\documentclass[twocolumn,preprintnumbers,amsmath,amssymb,aps,prb,superscriptaddress]{revtex4-2}

\pdfoutput=1
\usepackage[pdftex]{graphicx}
\usepackage[english]{babel}
\usepackage{soul}
\usepackage{cleveref}
\usepackage{bbold}
\usepackage{mathtools}
\usepackage{multirow}
\usepackage{colortbl}
\definecolor{kugray5}{RGB}{224,224,224}
\usepackage[latin1]{inputenc}
\usepackage{diagbox}
\usepackage{mciteplus}
\usepackage{appendix}

\usepackage{dcolumn}
\usepackage{bm}

\usepackage{color}
\usepackage{amstext}
\usepackage{braket}
\usepackage{mathdots}
\usepackage{tikz}
\usepackage{physics}
\usepackage{enumitem}

\usetikzlibrary{matrix,decorations.pathreplacing}

\begin{document}


\title{One-dimensional $2^n$-root topological insulators and superconductors}

\author{A. M. Marques}
\email{anselmomagalhaes@ua.pt}
\affiliation{Department of Physics $\&$ i3N, University of Aveiro, 3810-193 Aveiro, Portugal}

\author{L. Madail}
\affiliation{Department of Physics $\&$ i3N, University of Aveiro, 3810-193 Aveiro, Portugal}
\affiliation{International Iberian Nanotechnology Laboratory, 4715-310 Braga, Portugal
}

\author{R. G. Dias}
\affiliation{Department of Physics $\&$ i3N, University of Aveiro, 3810-193 Aveiro, Portugal}

\date{\today}


\begin{abstract}
Square-root topology is a recently emerged subfield describing a class of insulators and superconductors whose topological nature is only revealed upon squaring their Hamiltonians, \textit{i.e.}, the finite energy edge states of the starting square-root model inherit their topological features from the zero-energy edge states of a known topological insulator/superconductor present in the squared model.
Focusing on one-dimensional models, we show how this concept can be generalized to $2^n$-root topological insulators and superconductors, with $n$ any positive integer, whose rules of construction are systematized here.
Borrowing from graph theory, we introduce the concept of arborescence of $2^n$-root topological insulators/superconductors which connects the Hamiltonian of the starting model for any $n$, through a series of squaring operations followed by constant energy shifts, to the Hamiltonian of the known topological insulator/superconductor, identified as the source of its topological features.
Our work paves the way for an extension of $2^n$-root topology to higher-dimensional systems.
\end{abstract}

\pacs{74.25.Dw,74.25.Bt}

\maketitle
\section{Introduction}
\label{sec:intro}

Topological insulators \cite{Asboth2016} (TIs) are one of the most intensively studied topics in condensed matter in recent years.
Paradigmatic examples of one-dimensional (1D) TIs, such as the Su-Shrieffer-Heeger (SSH) model \cite{Su1979}, exhibit midgap zero-energy edge states under open boundary conditions (OBC), which can be related to a nontrivial and quantized topological index characterizing the bulk bands below that energy gap, in what is commonly referred to as the bulk-boundary correspondence \cite{Hasan2010}.

Inspired by Dirac's derivation of his eponymous equation from taking the square-root of the Klein-Gordon equation, Arkinstall \textit{et al.} \cite{Arkinstall2017} proposed a scheme of relating the topological properties of a given 1D model with finite energy edge states to those of its squared model, a conventional TI with zero-energy edge states.
Accordingly, these models came to be known as square-root TIs ($\sqrt{\text{TIs}}$). 
Later on it was realized \cite{Kremer2020} that the concept of $\sqrt{\text{TIs}}$ could be extended to bipartite models, since their squared versions appear in a block diagonal form, with one of them corresponding to the TI from which the topological features are inherited in the starting $\sqrt{\text{TI}}$.
Square-root topology was quickly extended to other 1D models \cite{Pelegri2019} including topological superconductors (TSs) and non-Hermitian systems \cite{Ezawa2020,Ke2020,Lin2021}, to higher-order TIs \cite{Song2020,Mizoguchi2020,Yan2020} [$d$-dimensional lattices hosting topological edge states in $(d-j)$-dimensions, with $j\geq 2$ \cite{Benalcazar2017,Benalcazar2017b,Pelegri2019c}], to topological semimetals \cite{Mizoguchi2021} and Chern insulators \cite{Ezawa2020}.

The general recipe for the construction of $\sqrt{\text{TIs}}$ and square-root TSs ($\sqrt{\text{TSs}}$) from their topological squared counterparts was developed by Ezawa \cite{Ezawa2020}.
It relies on the realization that, upon treating the tight-binding chain as a connected graph, one can construct the square-root versions of a given TI/TS by subdividing its tight-binding graph and renormalizing the resulting hopping parameters.
The main idea of this method is that subdivision of the tight-binding graph guarantees that it will become bipartite, even if it was not so before.
In turn, the bipartite property guarantees that the squared Hamiltonian can be written in a block diagonal form. 
Focusing on 1D models, we show here how a further elaboration of the method in [\onlinecite{Ezawa2020}] allows one to contruct TIs/TSs of root degree $2^n$ ($\sqrt[2^n]{\text{TIs}}$/$\sqrt[2^n]{\text{TSs}}$), with $n$ any positive integer.
Furthermore, since the distance between the original TI/TS and the constructed $\sqrt[2^n]{\text{TIs}}$/$\sqrt[2^n]{\text{TSs}}$, measured by the number of successive squaring operations that have to be applied to the Hamiltonian of the latter in order to get to the former, grows with $n$, we codified the relation between the two by introducing the ``arborescence of $\sqrt[2^n]{\text{TIs}}$/$\sqrt[2^n]{\text{TSs}}$'', a term taken from graph theory, that enables one to keep track of the original topological features which are inherited by the edge states present in the starting model.
In a recent work \cite{Dias2021}, we already identified a specific subclass of 1D linear  bipartite models, labeled sine-cosine models, which, after each squaring operation, retrieve a smaller self-similar version of themselves as one of the diagonal blocks, in what we described as a Matryoshka sequence.
The results analyzed there can be viewed as a notable subset appearing within the general framework of 1D $\sqrt[2^n]{\text{TIs}}$ that we draw here.

The rest of the paper is organized as follows.
In Sec.~\ref{sec:diamondchain}, we review the properties of a 1D $\sqrt{\text{TI}}$, namely the diamond chain with $\pi$-flux per plaquette, which we will take as our toy model from which $2^n$-root topology is derived.
In Sec.~\ref{sec:quarticroot}, we show how to construct the quartic-root TI from the $\sqrt{\text{TI}}$ introduced before, highlighting the relation between the edge states of the $\sqrt[4]{\text{TI}}$ and the topological state of the original TI.
In Sec.~\ref{sec:2nroottis}, we show how the method followed in the previous section to find the $\sqrt[4]{\text{TI}}$ can be replicated an arbitrary number of times to find the $\sqrt[2^n]{\text{TIs}}$ with a higher $n>2$ value.
The arborescence of $\sqrt[2^n]{\text{TIs}}$ is introduced here, as an intuitive way of relating any $\sqrt[2^n]{\text{TI}}$ with the original TI.
In Sec.~\ref{sec:2nrootss}, we derive the $\sqrt[2^n]{\text{TSs}}$ from the original TS, taken to correspond to the Kitaev chain \cite{Kitaev2001} mapped into its single-particle tight-binding analog model.
Finally, in Sec.~\ref{sec:conclusions} we present our conclusions.

\section{Diamond chain with $\pi$-flux per plaquette}
\label{sec:diamondchain}
\begin{figure*}[ht]
	\begin{centering}
	\includegraphics[width=0.9 \textwidth,height=9cm]{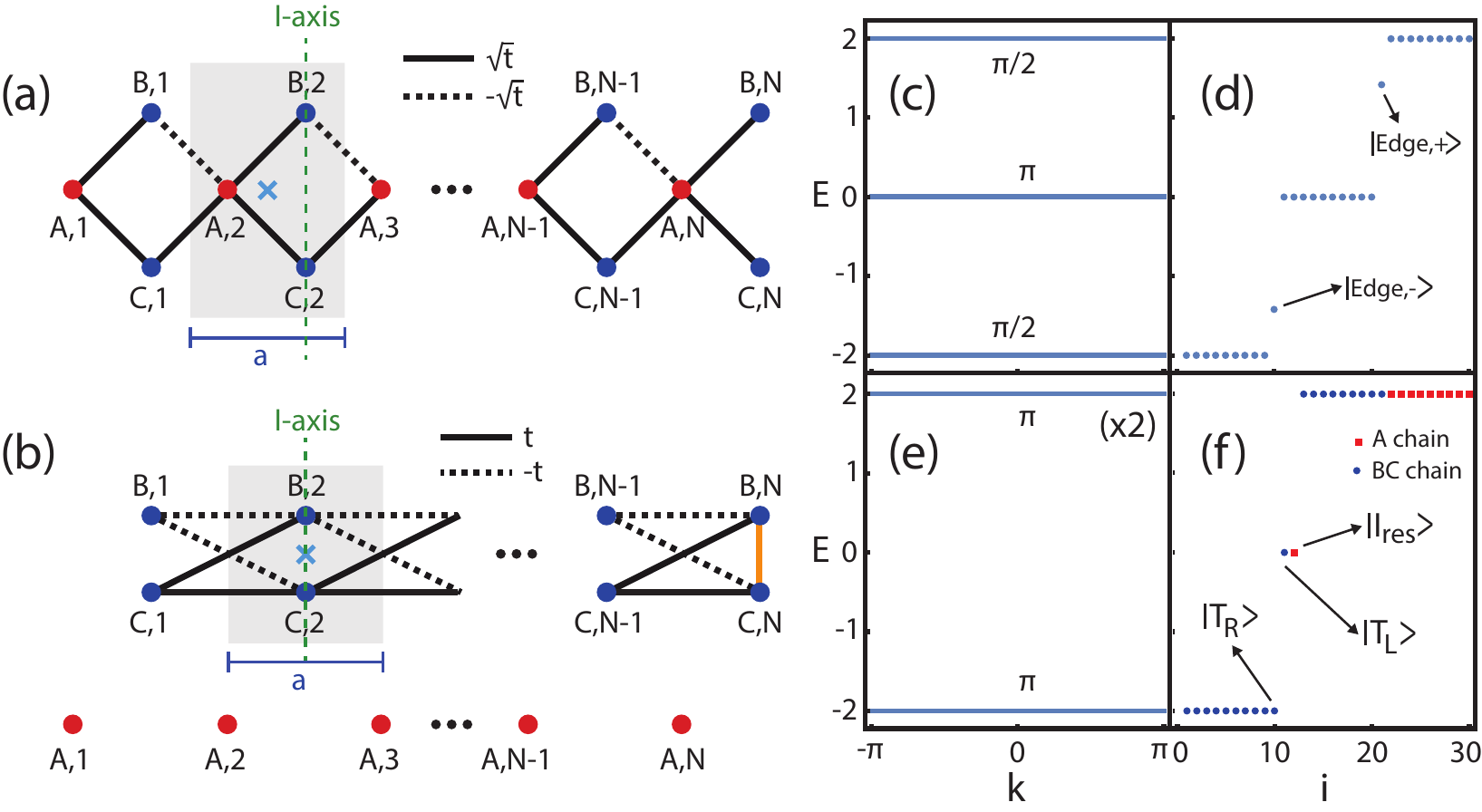} 
		\par\end{centering}
	\caption{(a) Diamond chain with $\pi$-flux per plaquette, corresponding to the $\sqrt{\text{TI}}$; (b) squared model of (a), yielding the Creutz ladder on top (the original TI) with a perturbed right end (extra orange hopping) and an independent residual chain of decoupled A sites at the bottom.
	Shaded regions represent the unit cell [only for the Creutz ladder in (b)], the blue cross its center and the vertical dashed green line the Inversion-axis.
   (c) Energy spectrum of the model at the left, with $t\equiv 1$, as a function of the momentum under PBC and (d) the state index for the same chain with OBC and $N=10$ unit cells.
(e),(f) Same as in (c),(d), respectively, but for the model in (b). 
The Zak phases for each band are indicated in the bulk spectra [cumulative for the degenerate bands in (e), with $\pi$ for the top band of the Creutz ladder and 0 for the band of decoupled A sites].}
	\label{fig:diamond}
\end{figure*}

Let us begin with a recapitulation of the properties of the diamond chain model with a $\pi$-flux per plaquette, depicted in Fig.~\ref{fig:diamond}(a), a known example of a 1D $\sqrt{\text{TI}}$ \cite{Kremer2020}. Under periodic boundary conditions (PBC), the square-root bulk Hamiltonian of this model, written in the $\{\ket{A(k)},\ket{B(k)},\ket{C(k)}\}$ basis, is given by
\begin{eqnarray}
	H_{\sqrt{\text{TI}}}&=&
	\sqrt{t}\begin{pmatrix}
	0&h^\dagger_{\sqrt{\text{TI}}}
	\\
	h_{\sqrt{\text{TI}}}&0
	\end{pmatrix},
	\label{eq:hamilt_diamond}
	\\
	h^\dagger_{\sqrt{\text{TI}}}&=&
	\begin{pmatrix}
	1-e^{-ik}&\ \ \ 1+e^{-ik}
	\end{pmatrix},
\end{eqnarray}
where the lattice constant is set to $a\equiv 1$ everywhere in this paper and the $k$-dependency is hidden in the terms and will remain so hereafter for convenience, except when deemed necessary.
The block antidiagonal form of (\ref{eq:hamilt_diamond}) indicates that the model is bipartite and, therefore, enjoys chiral-symmetry defined as $CH_{\sqrt{\text{TI}}}C^{-1}=-H_{\sqrt{\text{TI}}}$, with $C=\text{diag}(-1,1,1)$.
Diagonalization of the Hamiltonian in (\ref{eq:hamilt_diamond}) yields an all-bands flat energy spectrum \cite{Pelegri2020}, shown in Fig.~\ref{fig:diamond}(c), as a consequence of an Aharonov-Bohm caging effect induced by the $\pi$-flux in each plaquette \cite{Liberto2019,Pelegri2019b,Gligoric2020,Chang2021}.
Even though it would be more correct to speak of a \textit{square-root} energy spectrum, we will treat all $2^n$-root energy spectra in this paper as simply energy spectra.
In what follows, we set $t=1$.

Under open boundary conditions (OBC), there is a chiral pair of topological edge states with finite energies $E_{\text{edge}}^\pm=\pm\sqrt{2}$ [see in-gap states in the energy spectrum of an open chain in Fig.~\ref{fig:diamond}(d)], located around the end that terminates with a spinal A site.
A nonequivalent termination at both sites of the BC sublattice occurs at the oppposite end of the open chain, provided an integer number $N$ of unit cells is considered.
This feature is related to the fact that, even though the model is inversion ($\mathcal{I}$) symmetric, its $\mathcal{I}$-axis is shifted in relation to the center of the unit cell \cite{Marques2018,Pelegri2019,Madail2019} [see Fig.~\ref{fig:diamond}(a)].
As a consequence, the $\mathcal{I}$-symmetry operator becomes $k$-dependent \cite{Marques2019},
\begin{equation}
\mathcal{I}:\ \ \ \ P(k)H_{\sqrt{\text{TI}}}(k)P^{-1}(k)=H_{\sqrt{\text{TI}}}(-k),
\end{equation}
with $P(k)=\text{diag}(e^{ik},1,1)$.
The Zak phase of each band, a $Z_2$ topological index for 1D models \cite{Asboth2016}, can be expressed in this case as 
\begin{equation}
	\gamma_n=\arg(P_0^nP_\pi^n)-\int_{0}^{\pi}dk|u_{n,A}(k)|^2\ \ \mod 2\pi,
	\label{eq:zakphase}
\end{equation}
where $P_0^n,P_\pi^n=\pm 1$ are the parity eigenvalues of band $n$ at the $\mathcal{I}$-invariant momenta $k=0,\pi$, respectively, and $u_{n,A}(k)$ is the A-component of the corresponding eigenstate $\ket{u_n(k)}$.
The second term in (\ref{eq:zakphase}) appears as a consequence of the mismatch between the $\mathcal{I}$-axis and the center of the unit cell \cite{Marques2019}.
Since finite energy eigenstates in bipartite sublattices have half their weight in each sublattice, one can immediately see that $|u_{\pm,A}(k)|^2=1/2$ for all $k$, where $n=+(-)$ is the index of the top (bottom) band.
The first term in (\ref{eq:zakphase}) is quantized to either 0 or $\pi$ such that, modulo $2\pi$, one necessarily has $\gamma_\pm=\frac{\pi}{2}$.
It should be noted that this fractional quantization is an artifact of $\sqrt{\text{TIs}}$ where the correction term in (\ref{eq:zakphase}) goes over an entire sublattice (as will be shown to occur also for the $\sqrt{\text{TS}}$ analyzed later on).
However, this does not hold in general for all 1D $\sqrt{\text{TIs}}$, that is, the Zak phases of their bands can have arbitrary values between 0 and $2\pi$, which will be the case for the $\sqrt{\text{TI}}$ in Appendix~\ref{app:2nrootssh}, as shown elsewhere \cite{Marques2019}.
The states of the zero-energy flat band $n=0$, on the other hand, have no weight on the A sublattice \cite{Pelegri2020}, so the usual $\pi$-quantization holds for $\gamma_0$.

Upon squaring the square-root Hamiltonian in (\ref{eq:hamilt_diamond}) one arrives at
\begin{eqnarray}
H_{\sqrt{\text{TI}}}^2&=&
\begin{pmatrix}
H_{\text{res}}&0
\\
0&H_{\text{par}}
\end{pmatrix},
\\
H_{\text{res}}&=&h^\dagger_{\sqrt{\text{TI}}} h_{\sqrt{\text{TI}}}=2c_1,
\\
H_{\text{par}}&=&h_{\sqrt{\text{TI}}} h^\dagger_{\sqrt{\text{TI}}}=c_1\sigma_0+H_{\text{TI}},
\label{eq_hamilt_par}
\\
H_{\text{TI}}&=&\mathbf{d}\cdot\boldsymbol{\sigma},
\end{eqnarray} 
where $\sigma_0$ is the $2\times 2$ identity matrix, $\boldsymbol{\sigma}=(\sigma_x,\sigma_y,\sigma_z)$ is the vector of Pauli matrices, $\mathbf{d}=c_1(0,\sin{k},-\cos{k})$ and $c_1=2t$ is a constant energy shift. We label $H_{\text{res}}$ ($H_{\text{par}}$) as the residual (parent) Hamiltonian and $H_{\text{TI}}$ the Hamiltonian of the original TI.
The real-space representation of $H_{\sqrt{\text{TI}}}^2$ with a general energy downshift of $c_1$, such that $H_{\text{par}}\to H_{\text{TI}}$ and $H_{\text{res}}\to c_1$, is depicted in Fig.~\ref{fig:diamond}(b), where it can be seen that $H_{\text{TI}}$ models a Creutz ladder with $\pi$-flux in each triangle \cite{Creutz1999,Creutz2001,Zurita2020,Flannigan2020,Kuno2020,Kuno2020b} in the BC sublattice, known to host a nondecaying zero-energy topological edge state at each end for all $|t|\geq 0$ under OBC, and $H_{\text{res}}$ represents a chain of decoupled A sites.
Notice that the $\mathcal{I}$-axis of the Creutz ladder can be placed at the center of the unit cell, such that the Zak phases of both its bands recover the usual $\pi$-quantization [see Fig.~\ref{fig:diamond}(e)]. 
Through a suitable basis rotation, $H_{\text{TI}}$ can be transformed into the Hamiltonian of a fully dimerized SSH model in the topological phase \cite{Kremer2020}.

The edge spectrum of the open squared chain, downshifted by $c_1$, for $N=10$ complete unit cells is shown in Fig.~\ref{fig:diamond}(f). 
At first sight, one might be tempted to identify both midgap states with the topological states of the open Creutz ladder, our original TI.
There is, however, a subtle point that will prove to be of crucial importance throughout the rest of this paper: under OBC and for $N$ integer, the end sites of the starting square-root chain [see $\ket{A,1}$, $\ket{B,N}$, and $\ket{C,N}$ in Fig.~\ref{fig:diamond}(a)], have a lower coordination number than the rest of their equivalent bulk sites.
This will create an on-site energy offset upon squaring the Hamiltonian, akin to the offset in the self-energies for states with weight at the edge sites in second-order perturbation theory \cite{Marques2017},
\begin{eqnarray}
\bra{A,n}H_{\sqrt{\text{TI}}}^2\ket{A,n}&=&2c_1,\ \  n=,2,3,\dots,N,
\\
\bra{A,1}H_{\sqrt{\text{TI}}}^2\ket{A,1}&=&c_1,
\label{eq:impstate}
\\
\bra{\nu,n}H_{\sqrt{\text{TI}}}^2\ket{\nu,n}&=&c_1,\ \  n=,1,2,\dots,N-1,
\\
\bra{\nu,N}H_{\sqrt{\text{TI}}}^2\ket{\nu,N}&=&\frac{c_1}{2},
\end{eqnarray}
with $\nu=B,C$, and an extra hopping at the right edge of the Creutz ladder depicted in orange in Fig.~\ref{fig:diamond}(b),
\begin{equation}
	\bra{B,N}H_{\sqrt{\text{TI}}}^2\ket{C,N}=t=\frac{c_1}{2}.
\end{equation}
As a result, the left and right topological states of the Creutz ladder, written as $\ket{T_L}=\frac{1}{\sqrt{2}}(\ket{B,1}+\ket{C,1})$ and $\ket{T_R}=\frac{1}{\sqrt{2}}(\ket{B,N}-\ket{C,N})$, respectively, will have different energies, on account of the correction terms at the right edge,
\begin{eqnarray}
	H_{\sqrt{\text{TI}}}^2\ket{T_L}&=&c_1\ket{T_L},
	\label{eq:topostateleft}
	\\
	H_{\sqrt{\text{TI}}}^2\ket{T_R}&=&0.
	\label{eq:topostateright}
\end{eqnarray}
There is an impurity state at the left edge of the A chain, labeled $\ket{I_{\text{res}}}\equiv\ket{A,1}$, which is degenerate with $\ket{T_L}$ by comparing (\ref{eq:impstate}) and (\ref{eq:topostateleft}).
These are the two midgap states of Fig.~\ref{fig:diamond}(f), while $\ket{T_R}$ becomes degenerate with the bulk states of the lower band of the Creutz ladder, becoming a nontopological edge state due to the effect of the extra hopping term at the right end.
Returning now to the open square-root model, the finite energy states shown in Fig.~\ref{fig:diamond}(d) can be written, in the $\{\ket{A,1},\ket{B,1},\ket{C,1}\}$ basis, as
\begin{equation}
\ket{\text{Edge},\pm}=\frac{1}{2}
\begin{pmatrix}
\pm \sqrt{2}
\\
1
\\
1
\end{pmatrix}
=
\frac{1}{\sqrt{2}}
\begin{pmatrix}
\pm \ket{I_{\text{res}}}
\\
\ket{T_L}
\end{pmatrix},
\label{eq:ti2topoweight}
\end{equation}
that is, in relation to the squared model, each edge state of the starting square-root model can be said to be ``half'' topological and ``half'' impurity.
As we will see below, for $\sqrt[2^n]{\text{TIs}}$ the ratio of topological-to-impurity component for each edge state of the $2^n$-root model, relative to the original TI, will change in favor of the impurity component as $n$ increases.

\section{Quartic-root topological insulator}
\label{sec:quarticroot}

A $\sqrt[4]{\text{TI}}$ of our original TI [the Creutz ladder along the B and C sites of Fig.~\ref{fig:diamond}(b)] would have to be a given model that, upon squaring, would contain the $\sqrt{\text{TI}}$ model [the diamond chain of Fig.~\ref{fig:diamond}(a)] as one of its diagonal blocks, apart from an overall constant energy shift.
In other words, one has to ensure that the $\sqrt[4]{\text{TI}}$ chain: (i) keeps the same flux pattern as the $\sqrt{\text{TI}}$ model; (ii) is bipartite, and therefore has chiral symmetry and can be written in a block antidiagonal form; (iii) has a sublattice composed of the sites that will become the $\sqrt{\text{TI}}$ chain upon squaring; (iv) the sites in this sublattice have the same onsite energy upon squaring (constant energy shift).

\begin{figure}[ht]
	\begin{centering}
		\includegraphics[width=0.45 \textwidth]{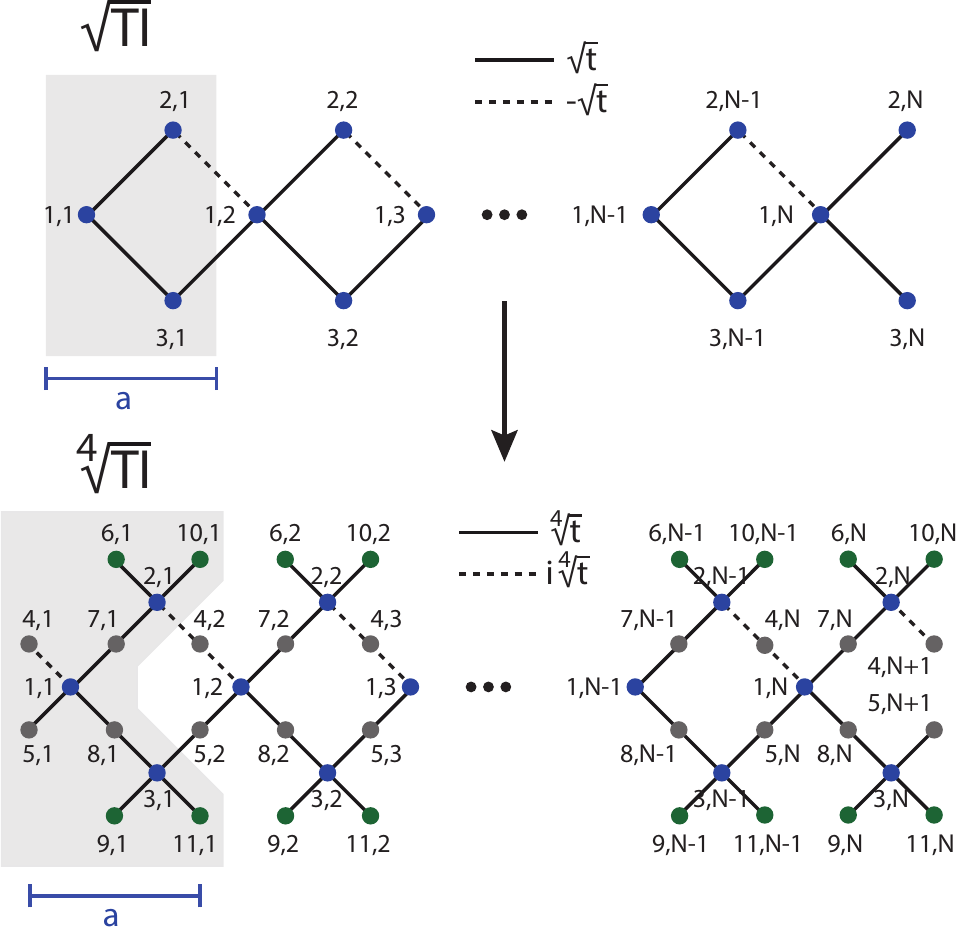} 
		\par\end{centering}
	\caption{Construction of the $\sqrt[4]{\text{TI}}$ from the $\sqrt{\text{TI}}$.
		The shaded regions indicate the respective unit cell.
	The gray sites appear from subdivision of the $\sqrt{\text{TI}}$ (a site is included in the middle of each link). The resulting hopping terms have, in relation to the corresponding ones at the $\sqrt{\text{TI}}$, the square root of the (constant) magnitude and halved phase factors.
Green extra sites are included in the $\sqrt[4]{\text{TI}}$ to keep the same coordination number for all sites of the blue sublattice, which implies that sites $\ket{4,N+1}$ and $\ket{5,N+1}$ have to be included at the right end under OBC.}
	\label{fig:quarticti}
\end{figure}

The first three conditions are met by following the method outlined by Ezawa\cite{Ezawa2020} which, in essence, treats the problem as one of graph theory: one constructs the split graph \cite{Ma2020} through subdivision of the original graph (chain) by adding nodes (sites) at the middle of each link (hopping term), as illustrated by the added gray sites in Fig.~\ref{fig:quarticti}, which guarantees that the higher-degree root TI is bipartite, with one of its sublattices corresponding to the sites of the starting model (the blue nodes in the $\sqrt[4]{\text{TI}}$ of Fig.~\ref{fig:quarticti} form a sublattice given by the sites of the $\sqrt{\text{TI}}$). 
In the process, the hopping terms are doubled in number and transformed as $\sqrt{t}e^{i\phi}\to\sqrt[4]{t}e^{i\phi/2}$, \textit{i.e.}, one takes the square-root of the magnitude and divides the phase by two, in order to keep the same $\pi$-flux per plaquette.

If we view the squaring of the Hamiltonian as a two-step quantum walk, where the on-site potential is given by the weighted sum of the paths that a particle can make from a given site to its connected neighbors in the first step and then hop back in the second step, it becomes clear that the last condition listed above is not met considering the blue and grey sites alone in Fig.~\ref{fig:quarticti} since the coordination number of the spinal sites $\ket{1,l}$ is four, where $\ket{j,l}$, with $j=1,2,\dots,11$ and $l=1,2,\dots,N$, is the state of a particle occupying site $j$ of unit cell $l$, while it is two for the same sublattice sites $\ket{2,l}$ and $\ket{3,l}$, resulting in
\begin{eqnarray}
\bra{1,l}H^2_{\sqrt[4]{\text{TI}}} \ket{1,l}&=&4\sqrt{t},
\label{eq:spinalsites}
\\
\bra{\beta,l}H^2_{\sqrt[4]{\text{TI}}}\ket{\beta,l}&=&2\sqrt{t},
\label{eq:topbottomsites}
\end{eqnarray}
with $\beta=2,3$.
As such, the on-site potentials at the relevant blue sublattice do not form a constant energy shift upon squaring.
\begin{figure}[ht]
	\begin{centering}
		\includegraphics[width=0.45 \textwidth]{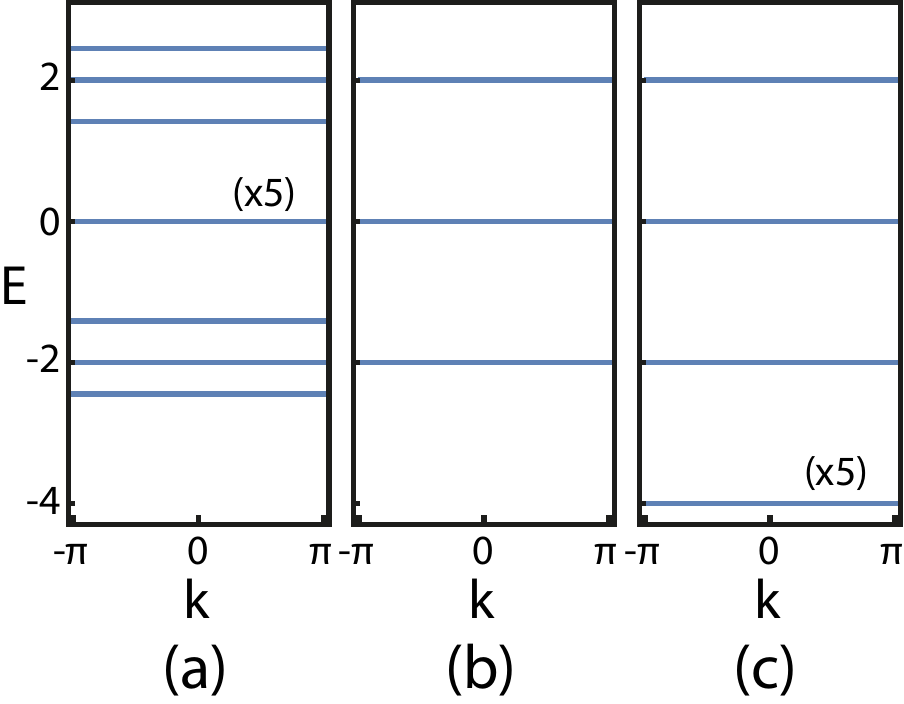} 
		\par\end{centering}
	\caption{Energy spectrum, with $t\equiv1$, as a function of the momentum for: (a) the $\sqrt[4]{\text{TI}}$ model given in (\ref{eq:hamilt_ti4}); (b),(c) the diagonal block of the $\sqrt[4]{\text{TI}}^2$ model corresponding to the $\sqrt{\text{TI}}$ model given in (\ref{eq:hamilt_diamond}) and to the $\sqrt{\text{res},2}^\prime$ model given in (\ref{eq:hamilt_res2}), respectively. The five-fold degenerate bands are indicated in the plots.}
	\label{fig:espectrumti4k}
\end{figure}
There are two fundamental methods that allow one to circumvent this limitation and impose a constant energy shift upon squaring the Hamiltonian: (i) by renormalizing some of the hopping terms, which is more economical at the level of the number of sites per unit cell, at the cost of requiring successive renormalizations of the hopping terms after each squaring operation; 
(ii) the magnitude of the hopping term  is kept constant (that is, $\sqrt[2^n]{t}$ for all hoppings in $\sqrt[2^n]{\text{TI}}$) and extra sites are added and connected to the sites of the relevant sublattice with lower coordination number, in order to compensate for the difference on their on-site energies upon squaring the Hamiltonian \footnote{An intermediate way that combines the distinguishing aspects of both fundamental methods (renormalized hopping terms and extra added sites) is also possible.}.
Since method (ii) is valid for a general model, while method (i) can only be applied to specific models such as the diamond chain, the sawtooth chain, etc., we will use method (ii) in the following sections of the main text and refer the reader to Appendix~\ref{app:alternativemethod} for further details on method (i), which is expected to become the easiest of the two to handle as the root degree of the $\sqrt[2^n]{\text{TI}}$ increases.
The effect on the energy spectrum of the extra green sites in the unit cell will simply be to originate the same number of extra zero-energy flat bands which, according to Lieb's theorem \cite{Lieb1989}, is given by the imbalance in the number of sites in each sublattice (eight for the gray and green sublattice and three for the blue sublattice, yielding a total of five zero-energy bands).
Application of method (ii) leads to the inclusion of the extra green sites depicted in Fig.~\ref{fig:quarticti}, such that (\ref{eq:topbottomsites}) becomes now
\begin{equation}
	\bra{\beta,l}H^2_{\sqrt[4]{\text{TI}}}\ket{\beta,l}=4\sqrt{t},
\end{equation}
yielding the same result as (\ref{eq:spinalsites}), as required.
\begin{figure*}[ht]
	\begin{centering}
		\includegraphics[width=0.9 \textwidth,height=9cm]{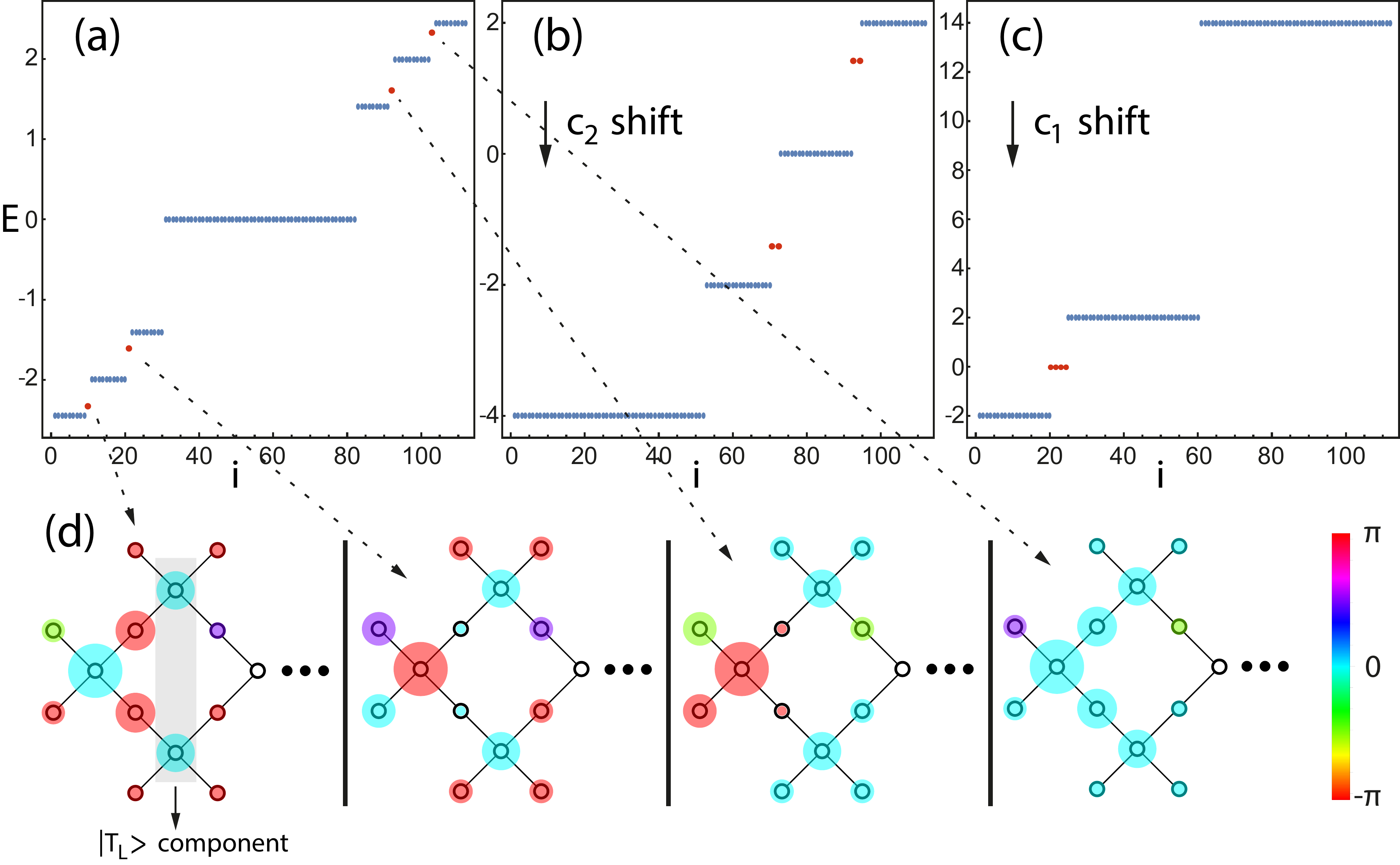} 
		\par\end{centering}
	\caption{Energy spectrum, in units of $t\equiv 1$, as a function of state index $i$, obtained from diagonalization of: (a) $H_{\sqrt[4]{\text{TI}}}$, the Hamiltonian of the open $\sqrt[4]{\text{TI}}$ chain at the bottom of Fig.~\ref{fig:quarticti} with $N=10$ complete unit cells plus extra sites 4 and 5 at unit cell $N+1$; (b) $H_{\sqrt[4]{\text{TI}}}^{2^\prime}=H_{\sqrt[4]{\text{TI}}}^{2}-c_2$, with $c_2=4\sqrt{t}$; (c) $H_{\sqrt[4]{\text{TI}}}^{4^\prime}=H_{\sqrt[4]{\text{TI}}}^{2^\prime}H_{\sqrt[4]{\text{TI}}}^{2^\prime}-c_1$, with $c_1=2t$. Bulk (edge) states are colored in blue (red). (d) Profile of the edge states in the $\sqrt[4]{\text{TI}}$ chain, where the radius of the circle represents the amplitude of the wavefunction at the respective site and the color represents its phase, coded by the color bar at the right.}
	\label{fig:espectrumti4r}
\end{figure*}

Under PBC, the bulk Hamiltonian of the $\sqrt[4]{\text{TI}}$, in the ordered $\{\ket{j(k)}\}$ basis, where $j=1,2,\dots,11$ refers to the $j^\text{th}$ component within the unit cell, has the form
\begin{eqnarray}
H_{\sqrt[4]{\text{TI}}}&=&
\sqrt[4]{t}\begin{pmatrix}
0&h^\dagger_{\sqrt[4]{\text{TI}}}
\\
h_{\sqrt[4]{\text{TI}}}&0
\end{pmatrix},
\label{eq:hamilt_ti4}
\\
h^\dagger_{\sqrt[4]{\text{TI}}}&=&
\begin{pmatrix}
i&1&0&1&1&0&0&0
\\
-ie^{ik}&0&1&1&0&0&1&0
\\
0&e^{ik}&0&0&1&1&0&1
\end{pmatrix},
\end{eqnarray}
whose squared version therefore becomes
\begin{eqnarray}
H_{\sqrt[4]{\text{TI}}}^2&=&
\begin{pmatrix}
H_{\sqrt{\text{par},2}}&0
\\
0&H_{\sqrt{\text{res},2}}
\end{pmatrix},
\label{eq:squaredti4}
\\
H_{\sqrt{\text{par},2}}&=&h_{\sqrt{\text{TI}}}^\dagger h_{\sqrt{\text{TI}}}=c_2I_3+H_{\sqrt{\text{TI}}},
\label{eq_hamilt_par}
\\
H_{\sqrt{\text{res},2}}&=&h_{\sqrt{\text{TI}}} h_{\sqrt{\text{TI}}}^\dagger=c_2I_8 +H_{\sqrt{\text{res},2}^\prime},
\label{eq:hamilt_res2}
\end{eqnarray} 
where $c_2=4\sqrt{t}$ is a constant energy shift, explicitly included also in $H_{\sqrt{\text{res},2}}$ to keep the energy spectrum of $H_{\sqrt{\text{TI}}}$ and $H_{\sqrt{\text{res},2}^\prime}$ leveled, $I_m$ is the $m\times m$ identity matrix and $H_{\sqrt{\text{TI}}}$ is given in (\ref{eq:hamilt_diamond}) with $\{\ket{1(k)},\ket{2(k)},\ket{3(k)}\}\to\{\ket{A(k)},\ket{B(k)},\ket{C(k)}\}$.
Diagonalization of $H_{\sqrt[4]{\text{TI}}}$, $H_{\sqrt{\text{TI}}}$ and $H_{\sqrt{\text{res},2}^\prime}$ yields the energy spectra of Fig.~\ref{fig:espectrumti4k}.
It can be seen that the $\sqrt{\text{TI}}$ and $\sqrt{\text{res},2}^\prime$ models share the same spectrum, apart from the five-fold degenerate lower band in the latter, corresponding to squaring and downshifting by $c_2$ the degenerate zero-energy bands of the $\sqrt[4]{\text{TI}}$ model, originated by the imbalance in the number of sublattice sites, as explained above.
As expected, the spectrum of Fig.~\ref{fig:espectrumti4k}(b) exactly matches the one on Fig.~\ref{fig:diamond}(c), since both of them correspond to the same $\sqrt{\text{TI}}$ model.

Finally, there is an ambiguity in the definition of the unit cell of $\sqrt[4]{\text{TI}}$, since sites $\ket{4,1}$ and $\ket{5,1}$ can be added either to the left, as in Fig.~\ref{fig:quarticti}, or to the right.
The ambiguity is somewhat resolved, however, when OBC are considered, in which case one has to add these two sites at the right of the last unit cell also (see sites $4$ and $5$ at the incomplete $N+1$ unit cell in Fig.~\ref{fig:quarticti}), in order to keep the coordination number at the blue sites of both edges the same as for the bulk blue sites.
Since the extra sites perturb the right edge physics, preventing in general the emergence of right edge states, one could argue that a direct bulk-edge correspondence is lost, since the lower number of edge states under OBC will not have a correspondence with the cummulative Zak phases of the bulk bands below the energy gap where each of them lies.
On the other hand, even without the extra sites at the right edge under OBC the usual bulk-edge correspondence is still broken, since the $\mathcal{I}$-axis does not cross the center of the unit cell of the $\sqrt[4]{\text{TI}}$, meaning that the Zak phases of the bulk bands are not $\pi$-quantized in general.

The energy spectrum of the open $\sqrt[4]{\text{TI}}$ chain at the bottom of Fig.~\ref{fig:quarticti}, for $N=10$ complete unit cells plus sites 4 and 5 of the $N+1$ unit cell, is shown in Fig.~\ref{fig:espectrumti4r}(a), where two chiral pairs of nondegenerate edge states with energies
\begin{eqnarray}
E_{\text{edge}}^{\sqrt[4]{\text{TI}}}&=&\pm\sqrt{E_{\text{edge}}^{\sqrt{\text{TI}}}+c_2} \nonumber
\\
&=&\pm\sqrt[4]{2t}\sqrt{2\sqrt{2}\pm1},
\end{eqnarray}
where $E_{\text{edge}}^{\sqrt{\text{TI}}}=\pm\sqrt{2t}$ [see edge states in Fig.~\ref{fig:diamond}(d)].
The spatial profile of these edge states is shown in Fig.~\ref{fig:espectrumti4r}(d).
After squaring this spectrum and taking out the constant shift $c_2$, we arrive at the spectrum of Fig.~\ref{fig:espectrumti4r}(b), where the edge states now appear in two doubly degenerates pairs and the $E=0$ states in Fig.~\ref{fig:espectrumti4r}(a) become the $E=-4\sqrt{t}=-c_2$ states in Fig.~\ref{fig:espectrumti4r}(b).
Lastly, this squared and shifted spectrum is in turn squared and downshifted by $c_1=2t$ in Fig.~\ref{fig:espectrumti4r}(c).
All four edge states are now degenerate at zero energy, with only one of them corresponding to the topological state $\ket{T_L}$ of the original TI model, defined above (\ref{eq:topostateleft}), while the other three are impurity states, one stemming from the $H_{\text{res}^\prime}=H_{\text{res}}-c_1$ [the $\ket{I_{\text{res}}}$ state defined below (\ref{eq:topostateright})] and the other two, labeled $\ket{I_{\text{res},2}^1}$ and $\ket{I_{\text{res},2}^2}$, from $H_{\text{res},2^\prime}=H_{\sqrt{\text{res},2}^\prime}^2-c_1$, with $H_{\sqrt{\text{res},2}^\prime}^2$ defined in (\ref{eq:hamilt_res2}), involving only the sites in the green and gray sublattice depicted at the bottom of Fig.~\ref{fig:quarticti}.

We define the ancestor Hamiltonian as the direct sum of all the terms contained in Fig.~\ref{fig:espectrumti4r}(c),
\begin{equation}
H_\text{anc}:=H_{\text{TI}}\oplus H_{\text{res}^\prime}\oplus H_{\text{res},2^\prime},
\end{equation}
which is the Hamiltonian describing the ancestor chain.
The four edge states of the $\sqrt[4]{\text{TI}}$ chain in Fig.~\ref{fig:espectrumti4r}(a), labeled $\ket{T_4,j}$, with $j=1,2,3,4$ in an increasing energy order, can be written as linear combinations of the degenerate edge states in Fig.~\ref{fig:espectrumti4r}(c) with equal weight in all of them, resulting in
\begin{equation}
|\braket{T_L}{T_4,j}|^2=\frac{1}{4},
\label{eq:ti4topoweight}
\end{equation}
that is, the edge states of the $\sqrt[4]{\text{TI}}$ can be said to be, in relation to the ancestor chain, ``one quarter'' topological [notice the same $\ket{T_L}$ component appearing in all edge states of Fig.~\ref{fig:espectrumti4r}(d)] and ``three quarters'' impurity, which
should be compared with the case of the edge states of the $\sqrt{\text{TI}}$ chain in (\ref{eq:ti2topoweight}).

\section{$2^n$-root topological insulators}
\label{sec:2nroottis}
\begin{figure*}[ht]
	\begin{centering}
		\includegraphics[width=0.8 \textwidth,height=11cm]{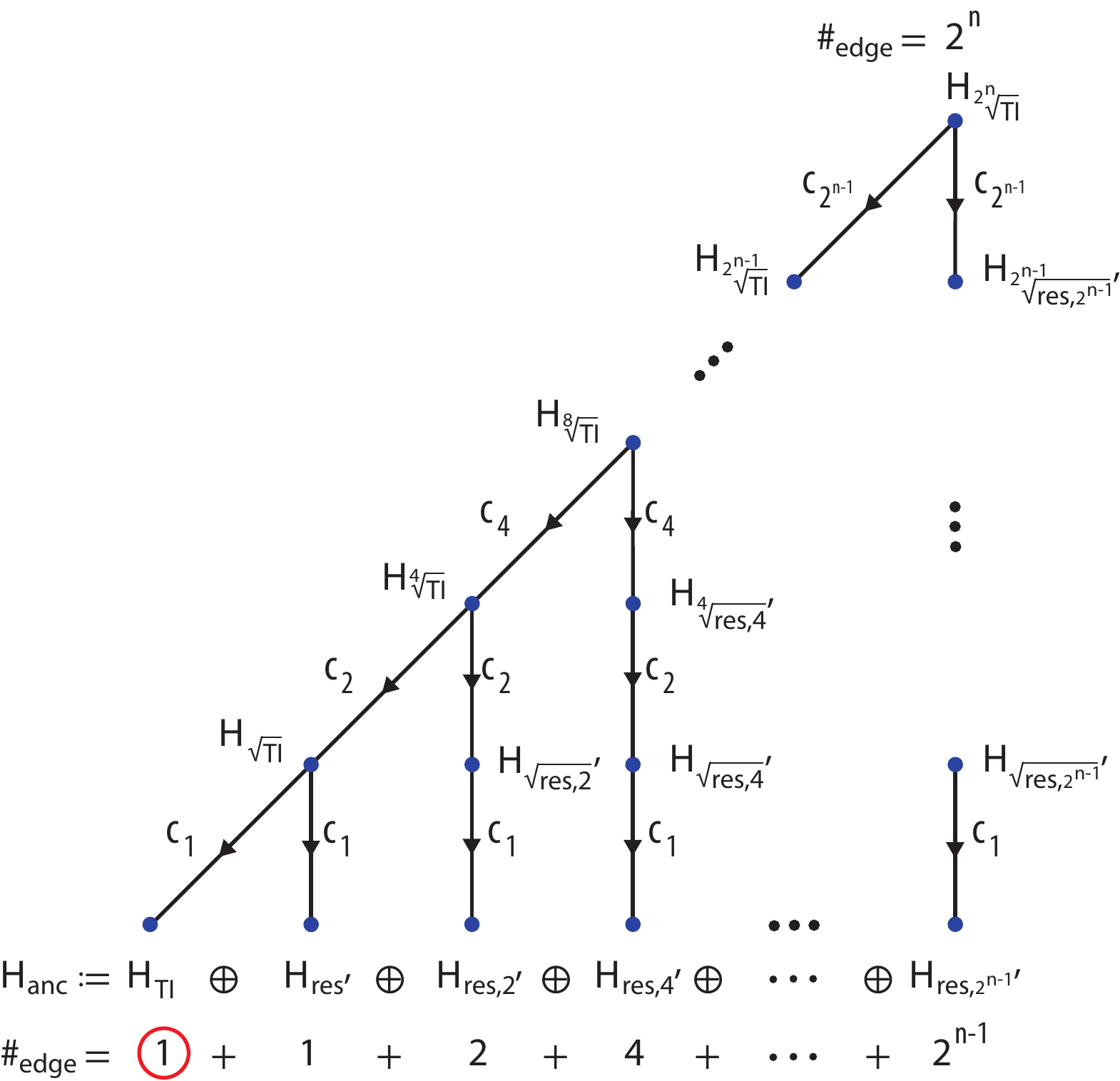} 
		\par\end{centering}
	\caption{Outwards directed rooted tree (arborescence) of $\sqrt[2^n]{\text{TIs}}$.
	The root node on top corresponds to the Hamiltonian of the starting $\sqrt[2^n]{\text{TI}}$ model.
From each level to the next the Hamiltonian is squared and shifted down in energy by $c_{2^{n-j}}$, with $j=1,2,\dots,n$ the level index. 
Each link selects one of the diagonal blocks of the resulting Hamiltonian, with the lower root degree TIs placed along the hypotenuse (each generating two children nodes from its block diagonal squared form) and the successive residual Hamiltonians placed at the right of those (each generating a single child node).
The process is stopped at the level where $H_\text{TI}$ is reached along the hypotenuse. The ancestor Hamiltonian $H_\text{anc}$ is defined as the direct sum over all leaf nodes.
$\#_{\text{edge}}$, indicated on the root and leaf nodes, is the number of edge states appearing from diagonalization of the Hamiltonian of the respective node.
In $H_\text{anc}$ only the red encircled edge state is topological, while all others are impurity states.}
	\label{fig:tree}
\end{figure*}
The procedure followed in the last section to find the $\sqrt[4]{\text{TI}}$ model from the $\sqrt{\text{TI}}$ can be readily generalized.
More concretely, one can find the $\sqrt[2^n]{\text{TI}}$, for any $n\in  \mathbb{N}$, from the $\sqrt[2^{n-1}]{\text{TI}}$ in four steps:
\\

1.	One starts by subdividing the graph made by the $\sqrt[2^{n-1}]{\text{TI}}$ chain, that is, a new node (site) is introduced at the middle of each link (hopping term). 
The resulting chain is therefore bipartite.
\\

2. The magnitude of the new hopping parameters is given by the square-root of the corresponding one in the $\sqrt[2^{n-1}]{\text{TI}}$ chain, while their phases are divided in half to keep the same flux pattern in the chain,
\begin{equation}
	e^{i\phi}\sqrt[2^{n-1}]{t}\to e^{i\phi/2}\sqrt[2^{n}]{t}.
\end{equation}
Note that each link in the $\sqrt[2^{n-1}]{\text{TI}}$ chain generates two after subdivision.
\\

3. One has to ensure that the same squared on-site potential appears at the sites in the sublattice that decouples from the other one upon squaring of the Hamiltonian to generate the  $\sqrt[2^{n-1}]{\text{TI}}$ chain.
First, one has to identify the sites of the relevant sublattice with the highest squared on-site potential (defined as $c_{2^{n-1}}$) and the others, with a lower squared on-site potential. To level the squared on-site potentials of the sites of these two subsets, one has to compensate for the difference of those in the latter subset by (i) renormalizing the hopping terms (the approach followed in Appendix~\ref{app:alternativemethod}), and/or (ii) introducing connections to extra sites, the approach followed in the previous sections. 
The phases of the extra hopping terms constitute an additional degree of freedom that one can use to guarantee that the relevant sublattice remains in itself bipartite upon squaring the model, as will be illustrated in the next section for a $\sqrt[4]{\text{TS}}$.
\\

4. This last step only applies under OBC. 
The extra sites  of the $\sqrt[2^{n}]{\text{TI}}$ that result from subdivision of the $\sqrt[2^{n-1}]{\text{TI}}$ create an ambiguity in the definition of the unit cell, since the outermost new sites can be placed either to the right or to the left in the unit cell.
We chose to place these sites at the leftmost region within the unit cell (see sites $\ket{4,1}$ and $\ket{5,1}$ of the $\sqrt[4]{\text{TI}}$ chain at the bottom of Fig.~\ref{fig:quarticti}).
However, under OBC these outermost sites have to be included also at the right of the last complete unit cell, in order to avoid a lower squared on-site potential at the rightmost sites of the relevant sublattice (see sites $\ket{4,N+1}$ and $\ket{5,N+1}$ at the bottom of Fig.~\ref{fig:quarticti}, keeping at four the coordination number of the two rightmost blue sites).
\\

By applying $n$ successive squaring operations to the Hamiltonian of the $\sqrt[2^n]{\text{TI}}$ chain, and taking out the constant energy shift given by the on-site potential energy at the relevant sublattice after each operation (in order to keep this sublattice bipartite), a cascade of block diagonal Hamiltonians appear, containing both the residual blocks and the lower degree roots of $\sqrt[2^n]{\text{TI}}$.
A way to visualize this is by constructing the outwards directed rooted tree (arborescence \cite{Deo2017}) of $\sqrt[2^n]{\text{TIs}}$, depicted in Fig.~\ref{fig:tree}.
The root node at the top represents the Hamiltonian of the starting $\sqrt[2^n]{\text{TI}}$ chain, hosting $\#_{\text{edge}}=2^n$ non-degenerate edge states.
Each link connects a Hamiltonian to one of its block diagonal terms upon squaring and shifting down by the constant energy term of the respective level, with $n+1$ levels in total.
Nodes along the hypotenuse correspond to a given root degree of the original TI and have two child nodes, since their squared Hamiltonians yield, apart from a constant energy shift, a residual block and the block with the lower root degree TI [see, e.g., the squared $\sqrt[4]{\text{TI}}$ model in (\ref{eq:squaredti4})].
Nodes living outside the hypotenuse, however, correspond to the successive residual Hamiltonians that one has to continue squaring and shifting down by the constant of the respective level, such that each originates only one child node.
The process stops when we reach the TI model along the hypotenuse.
We define the ancestor Hamiltonian as the direct sum of all leaf nodes,
\begin{equation}
	H_\text{anc}:=H_{\text{TI}}\oplus H_{\text{res}^\prime}\oplus H_{\text{res},2^\prime} \oplus \dots \oplus H_{\text{res},{2^{n-1}}^\prime}.
	\label{eq:ancestorhamilt}
\end{equation}
The number of zero-energy edge states $\#_{\text{edge}}$ originating from each of the terms in (\ref{eq:ancestorhamilt}) is indicated at the bottom of Fig.~\ref{fig:tree}.
Only the encircled one stemming from $H_\text{TI}$ is topological in nature [the $\ket{T_L}$ state defined in (\ref{eq:topostateleft})], while all others are impurity states of the residual Hamiltonians that appear due to the detuned on-site potentials at their respective left edge sites.
As expected, there is a total of $2^n$ degenerate zero-energy states in $H_\text{anc}$, the same as in the root node $H_{\sqrt[2^n]{\text{TI}}}$, with each of the $\ket{T_{2^n},j}$ states, with $j=1,2,\dots,2^n$ in an increasing energy order, appearing in a different energy gap of the latter.
If we define the projector onto the impurity subspace of the ancestor chain,
\begin{equation}
	\hat{I}=\sum\limits_{m=0}^n \sum\limits_{j=1}^{2^m} \ket{I_{\text{res,}2^m},j}\bra{I_{\text{res,}2^m},j},
\end{equation}
with $\ket{I_{\text{res,}2^0},1}\equiv\ket{I_{\text{res}}}$ defined below (\ref{eq:topostateright}), then we can determine the topological and impurity weights of the edge states of our starting $\sqrt[2^n]{\text{TI}}$ chain, relative to the edge states of the ancestor chain, as
\begin{eqnarray}
	|\braket{T_L}{T_{2^n},j}|^2&=&\frac{1}{2^n},
	\\
	\bra{T_{2^n},j}\hat{I}\ket{T_{2^n},j}&=&\frac{2^n-1}{2^n},
\end{eqnarray}
for all $j=1,2,\dots,2^n$.
We can conclude that, as $n$ increases, the weight of the starting edge states on the topological state of the ancestor chain gets diluted, in favor of an increasing weight over the sum of all impurity states.
Focusing on the subspace of $2^n$ edge states of the starting $\sqrt[2^n]{\text{TI}}$ chain, a method for the layered decomposition of each of these states in terms of the edge states of the successive residual chains plus the topological state of the original TI chain is presented in Appendix~\ref{app:holographic}.
Interestingly, this decomposition is shown there to be reminiscent of the holographic duality that defines the anti-de-Sitter (AdS)/conformal field theory (CFT) correspondence.

For the case of the model we have been considering, whose original TI is given by the Creutz ladder depicted in Fig.~\ref{fig:diamond}(b), the values of the constant energy shifts, corresponding to the on-site potential at the sites forming the diagonal block in the squared Hamiltonian yielding the lower root degree TI, can be readily found to be
\begin{eqnarray}
c_1&=&2t,
\\
c_{2^m}&=&4\sqrt[2^m]{t},\ \ m=1,2,\dots,n-1,
\end{eqnarray}
which, setting $t\equiv 1$, simplifies to $c_1=2$ and $c_{2^m}=4$.
The energy of the $2^n$ edge states coming from the diagonalization of the root node $H_{\sqrt[2^n]{\text{TI}}}$ can be obtained recursively through
\begin{eqnarray}
E_{\sqrt[2^n]{\text{TI}}}&=&\pm\sqrt{E_{\sqrt[2^{n-1}]{\text{TI}}}+c_{2^{n-1}}},
\label{eq:eneredgeti2n}
\\
E_\text{TI}&=&0.
\label{eq:eneredgeti}
\end{eqnarray}

\begin{figure}[ht]
	\begin{centering}
		\includegraphics[width=0.4 \textwidth]{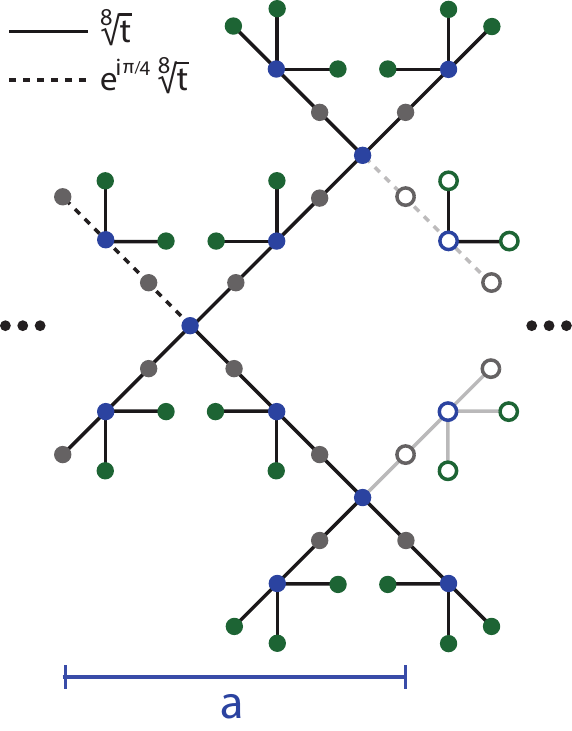} 
		\par\end{centering}
	\caption{43-sites unit cell of the $\sqrt[8]{\text{TI}}$ model. Blue sites form a sublattice yielding the $\sqrt[4]{\text{TI}}$ model at the bottom of Fig.~\ref{fig:quarticti} upon squaring, gray sites come from the subdivision of $\sqrt[4]{\text{TI}}$ and green sites are extra sites introduced to keep the coordination number constant for sites in the blue sublattice. The 10 open sites at the right belong to the next unit cell and have to be included at unit cell $N+1$ under OBC.}
	\label{fig:ti8ucell}
\end{figure}

As a final example, let us follow the four steps outlined above to construct the $\sqrt[8]{\text{TI}}$ model.
Its unit cell is depicted in Fig.~\ref{fig:ti8ucell}.
The blue sites form a sublattice corresponding to the unit cell of the $\sqrt[4]{\text{TI}}$ shown at the bottom of Fig.~\ref{fig:quarticti}, while the gray sites are a consequence of step 1, the subdivision of $\sqrt[4]{\text{TI}}$.
The magnitude of the new hopping parameters becomes $\sqrt[8]{t}$, while the phases of the dashed hopping terms are now $\frac{\pi}{4}$, that is, half the phase of the corresponding hoppings in the $\sqrt[4]{\text{TI}}$, in agreement with step 2, which keeps the same $\pi$-flux per plaquette pattern.
The green extra sites are introduced in order to keep the same coordination number at the sites in the blue sublattice such that, upon squaring $H_{\sqrt[8]{\text{TI}}}$, the on-site energy is the same for all blue sites, as required by step 3.
As a general rule of thumb, one identifies the sites within a unit cell that are positioned at the left of the spinal site as the set of sites that has to be included also at the right end under OBC, such that step 4 is fulfilled when, under OBC, the open sites in Fig.~\ref{fig:ti8ucell} are included in the incomplete $N+1$ unit cell for it to yield exactly, as one of its diagonal block and apart the $c_4$ energy shift, the $\sqrt[4]{\text{TI}}$ chain at the bottom of Fig.~\ref{fig:quarticti}.
\begin{figure*}[ht]
	\begin{centering}
		\includegraphics[width=0.95 \textwidth,height=5cm]{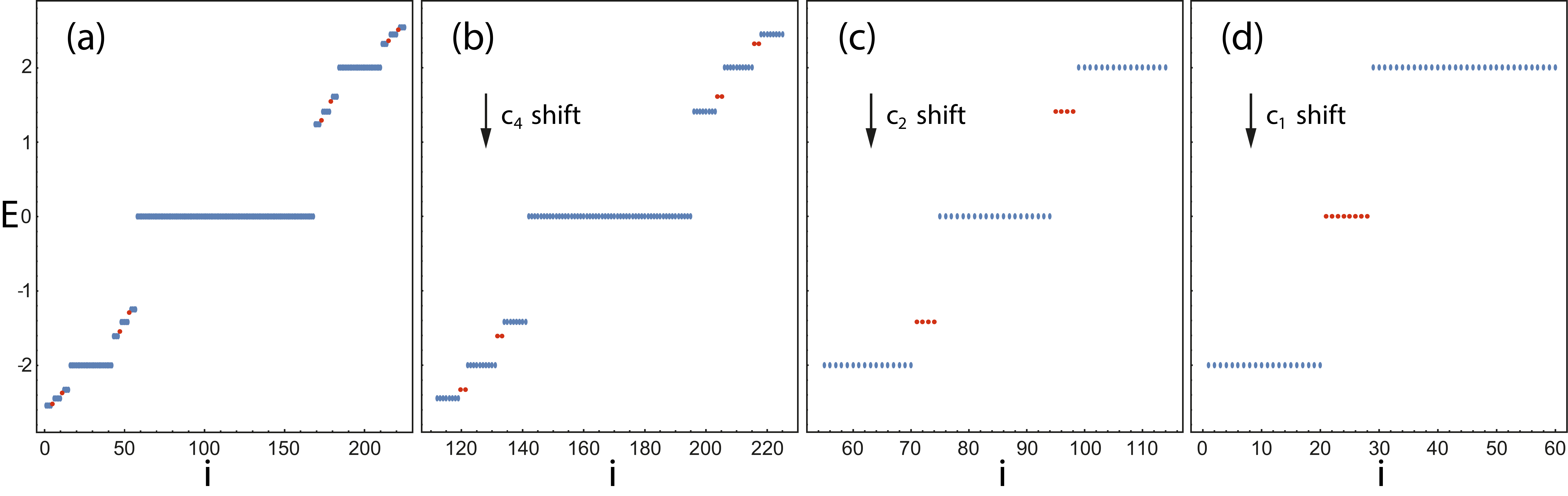} 
		\par\end{centering}
	\caption{Energy spectrum, in units of $t\equiv 1$, as a function of state index $i$, obtained from diagonalization of: (a) $H_{\sqrt[8]{\text{TI}}}$, the Hamiltonian of the open $\sqrt[8]{\text{TI}}$ chain with the unit cell of Fig.\ref{fig:ti8ucell}, with $N=5$ complete unit cells plus the 10 open extra sites at unit cell $N+1$; (b) $H_{\sqrt[8]{\text{TI}}}^{2^\prime}=H_{\sqrt[8]{\text{TI}}}^{2}-c_4$, with $c_4=4\sqrt[4]{t}$; (c) $H_{\sqrt[8]{\text{TI}}}^{4^\prime}=H_{\sqrt[8]{\text{TI}}}^{2^\prime}H_{\sqrt[8]{\text{TI}}}^{2^\prime}-c_2$, with $c_2=4\sqrt{t}$; (d) $H_{\sqrt[8]{\text{TI}}}^{8^\prime}=H_{\sqrt[8]{\text{TI}}}^{4^\prime}H_{\sqrt[8]{\text{TI}}}^{4^\prime}-c_1$, with $c_1=2t$. Bulk (edge) states are colored in blue (red). Only a partial spectrum is shown in (b)-(d) for clarity, with some high energy bulk bands, in absolute value, outside the range of the y-axis.}
	\label{fig:espectrumti8r}
\end{figure*}

The energy spectrum for the total Hamiltonian at each level of the arborescence of the $\sqrt[8]{\text{TI}}$, \textit{i.e.}, with $H_{\sqrt[8]{\text{TI}}}$ as the root node in Fig.~\ref{fig:tree}, for $N=5$ complete unit cells plus 10 extra sites at unit cell $N+1$, is shown in Fig.~\ref{fig:espectrumti8r}.
The progression shows how the eight nondegenerate edge states coming from diagonalization of $H_{\sqrt[8]{\text{TI}}}$ [in-gap red states in Fig.~\ref{fig:espectrumti8r}(a)] end up as the eight-fold degenerate zero-energy state coming from diagonalization of $H_{\text{anc}}$ [midgap states in Fig.~\ref{fig:espectrumti8r}(d)].

By applying steps 1 through 4, one can construct the $\sqrt[16]{\text{TI}}$ from the $\sqrt[8]{\text{TI}}$ in Fig.~\ref{fig:ti8ucell}.
We have been assuming in the main text that step 3 is satisfied by adding extra sites, while keeping the magnitude of the hopping terms the same everywhere, such that the number of sites per unit cell for the $\sqrt[2^n]{\text{TI}}$ follows the recurrence relation
\begin{eqnarray}
\#_{\text{uc}}^{2^n}&=&\#_{\text{uc}}^{2^{n-1}}+2^{2n-1}, \ \ n\geq 2,
\\
\#_{\text{uc}}^{2}&=&3,
\end{eqnarray}
meaning that one has $\#_{\text{uc}}^{16}=\#_{\text{uc}}^{2^4}=171$, $\#_{\text{uc}}^{32}=683$, etc.
In other words, as $n$ increases it becomes impractical to follow this method, with the one followed in Appendix~\ref{app:alternativemethod}, which relies on renormalizing the hopping parameters in place of adding extra sites, progressively gaining traction as the most tractable of the two.

It should be noted that the $\sqrt[2^n]{\text{TIs}}$ studied so far provide a practical way of generating 1D models with an all-bands-flat spectrum \cite{Danieli2020,Danieli2020b,Ichinose2021}, alternatively to a recent proposal of generating these spectra by taking advantage of a non-Abelian AB caging effect, related to the periodicity with which a $\pi$-flux appears in the plaquettes of the diamond chain \cite{Li2020,Mukherjee2020}.
Another type of $\sqrt[2^n]{\text{TI}}$, whose energy spectrum is in general dispersive, is analyzed in Appendix~\ref{app:2nrootssh}. This other type of $\sqrt[2^n]{\text{TI}}$ is labeled there as $\sqrt[2^n]{\text{SSH}}$, since the original TI is in this case taken to be the SSH model.

\section{$2^n$-root topological superconductors}
\label{sec:2nrootss}

In this section, we show how the same method for constructing $\sqrt[2^n]{\text{TIs}}$ can be applied for the construction of $\sqrt[2^n]{\text{TSs}}$, using the Kitaev model \cite{Kitaev2001} as the original TS, whose $\sqrt{\text{TS}}$ is already known \cite{Ezawa2020}.
We will go one step further and determine the $\sqrt[4]{\text{TS}}$, from which one can straightforwardly generalize for $n>2$.
\begin{figure*}[ht]
	\begin{centering}
		\includegraphics[width=0.95 \textwidth,height=8.3cm]{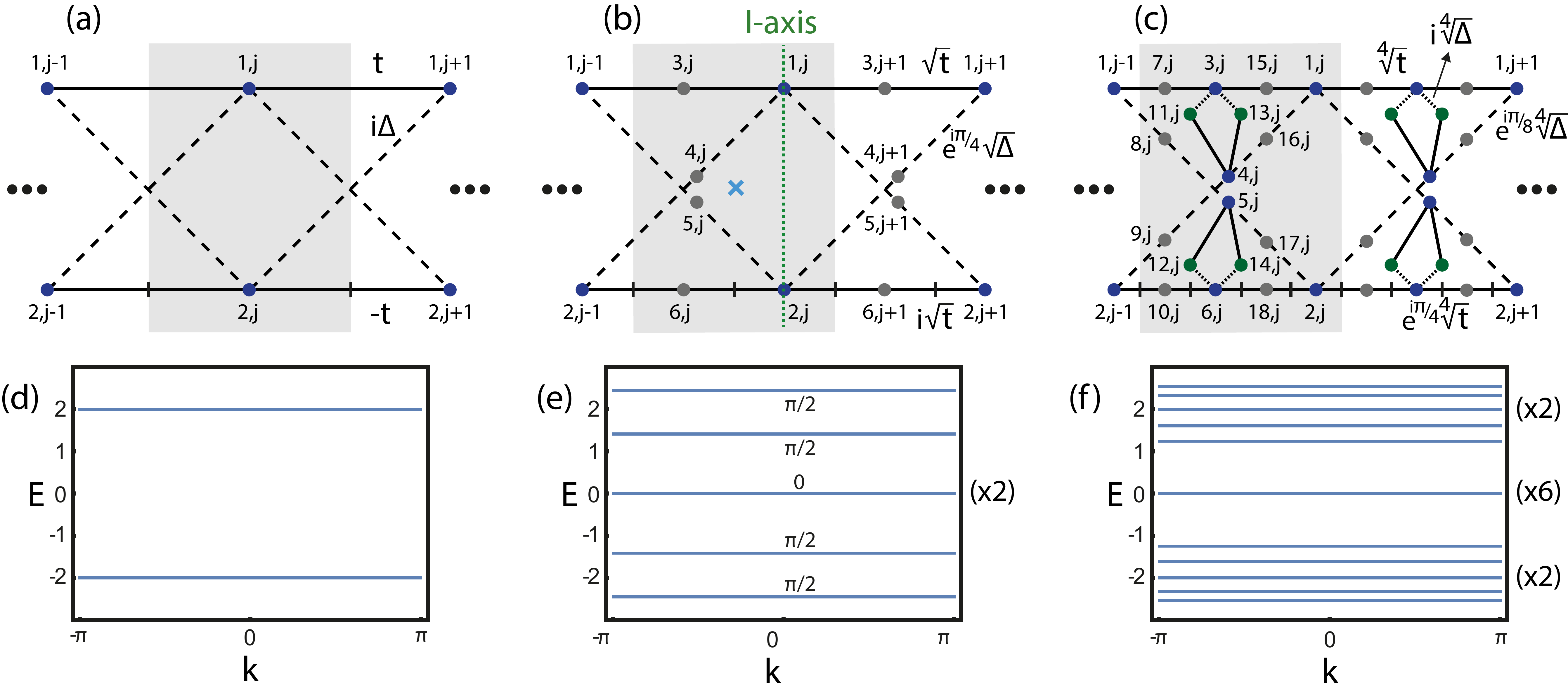} 
		\par\end{centering}
	\caption{Single-particle tight-binding analog representation of the: (a) Kitaev TS model; (b) $\sqrt{\text{TS}}$ model; (c) $\sqrt[4]{\text{TS}}$ model.
		Shaded regions represent the unit cell, and in (b) the blue cross marks its center and the vertical dashed green line the Inversion-axis.
	Gray sites come from the subdivision of the chain to the left, while the green sites in $\sqrt[4]{\text{TS}}$ were included to keep the on-site potential constant in the blue sublattice sites upon squaring $H_{\sqrt[4]{\text{TS}}}$.
(d)-(f) Energy spectrum as a function of the momentum for the respective models on top under PBC and with $\Delta=t\equiv 1$.
The Zak phase of each band is indicated in (e) (cumulative for the degenerate zero-energy bands).}
	\label{fig:ts4model}
\end{figure*}

The Hamiltonian of a periodic Kitaev chain, modeling a 1D quantum wire placed on top of a $p$-wave superconductor, reads as
\begin{equation}
	H_\text{TS}=\sum\limits_k(c^\dagger_k\ \ c_{-k})
	\begin{pmatrix}
	2t\cos k-\mu&2\Delta\sin k 
	\\
	2\Delta\sin k & -2t\cos k+\mu
	\end{pmatrix}
		\begin{pmatrix}
	c_k
	\\
	c^\dagger_{-k} 
	\end{pmatrix},
	\label{eq:hamiltkitaev}
\end{equation}
where $c_k$ is the particle annihilation operator acting on the state with momentum $k$, $\Delta$ is the superconducting pairing term induced in the chain by proximity effect with the superconductor, and $\mu$ the chemical potential.
Diagonalization of this Hamiltonian yields two symmetric flat bands for $t=\Delta$, as shown in Fig.~\ref{fig:ts4model}(d).
The presence of an annihilation operator in the Nambu pseudospinor $\mathbf{\Psi}_k=(c^\dagger_k\ \ c_{-k})$ prevents a real-space representation of the model by direct application of the inverse Fourier transform to (\ref{eq:hamiltkitaev}).
However, one can construct the single-particle tight-binding analog of the Kitaev chain through the mapping $\mathbf{\Psi}_k=(c^\dagger_k\ \ c_{-k})\to \mathbf{\Psi}_k=(c^\dagger_{1,k}\ \ c^\dagger_{2,k})$, that is, one treats each component of $\mathbf{\Psi}_k$ as if corresponding, e.g., to different atomic species or internal states.
Under this mapping, the real-space representation of the single-particle tight-binding analog model can be readily found to have the form of the Creutz ladder in Fig.~\ref{fig:ts4model}(a), with its ``electron channel'' at the top leg and its ``hole channel'' at the bottom leg.
Notice that the superconducting pairing is translated into an interleg crossed hopping term carrying a $\frac{\pi}{2}$ phase factor.
Thus, this system becomes amenable to the same four steps treatment outlined in the previous section for the construction of the $\sqrt[2^n]{\text{TI}}$.
We will further set $\mu=0$ below, otherwise the successive $\sqrt[2^n]{\text{TS}}$ models  would not be bipartite and would therefore lack chiral symmetry, such that their squared models could not be written in a block diagonal form and our method would not be applicable.
For convenience we will also set $t\equiv 1$ as the energy unit and keep $\Delta$ as our degree of freedom.

Subdivision of the graph of the Kitaev model in Fig.~\ref{fig:ts4model}(a), along with the corresponding changes to the magnitudes and phases of the hopping parameters (steps 1 and 2), is enough to generate the $\sqrt{\text{TS}}$ of Fig.~\ref{fig:ts4model}(b), whose bulk Hamiltonian, written in the ordered $\{\ket{j(k)}\}$ basis, where $j=1,2,\dots,6$ refers to the $j^{th}$ component within the unit cell, is given by
\begin{eqnarray}
\mathcal{H}_{\sqrt{\text{TS}}}&=&
\begin{pmatrix}
0&h^\dagger_{\sqrt{\text{TS}}}
\\
h_{\sqrt{\text{TS}}}&0
\end{pmatrix},
\label{eq:hamilt_ts2}
\\
h_{\sqrt{\text{TS}}}&=&
\begin{pmatrix}
\sqrt{t}(1+e^{-ik})&\ \ \ 0
\\
e^{-i\frac{\pi}{4}}\sqrt{\Delta}&\ \ \ e^{-i(k-\frac{\pi}{4})}\sqrt{\Delta}
\\
e^{-i(k-\frac{\pi}{4})}\sqrt{\Delta}&\ \ \ e^{-i\frac{\pi}{4}}\sqrt{\Delta}
\\
0&\ \ \ -i\sqrt{t}(1-e^{-ik}).
\end{pmatrix},
\end{eqnarray}
whose diagonalization, for $t=\Delta$, yields the energy spectrum in Fig.~\ref{fig:ts4model}(e), where the two-fold degenerate zero-energy band comes as a consequence of Lieb's theorem\cite{Lieb1989}, equaling in number the imbalance within the unit cell between sites in the gray sublattice (four) and in the blue sublattice (two).
Given the noncentered $\mathcal{I}$-axis within the unit cell, the Zak phase of band $n$ in (\ref{eq:zakphase}) can be rewritten as
\begin{equation}
\gamma_n=\arg(P_0^nP_\pi^n)-\sum\limits_{j=3}^6\int_{0}^{\pi}dk|u_{n,j}(k)|^2\ \ \mod 2\pi,
\label{eq:zakphasets2}
\end{equation}
since the correction term on the right-hand side goes over the components of the eigenstates in the gray sublattice \cite{Marques2019}.
Similarly to the case of the diamond chain analyzed in Sec.~\ref{sec:diamondchain}, the Zak phases of all finite energy bands of the $\sqrt{\text{TS}}$ yield $\frac{\pi}{2}$, that is, they are not $\pi$-quantized, as shown in Fig.~\ref{fig:ts4model}(e).

Let us suppose we want to construct the $\sqrt[4]{\text{TS}}$ from the $\sqrt{\text{TS}}$ in Fig.~\ref{fig:ts4model}(b), whose sites become now the relevant blue sublattice of $\sqrt[4]{\text{TS}}$.
We start by applying steps 1 and 2, subdivision of $\sqrt{\text{TS}}$ and subsequent renormalization of the magnitudes and phases of the hopping parameters (remember that the superconducting pairings were converted into hopping terms), leading to the model depicted in Fig.~\ref{fig:ts4model}(c) \textit{without} the green sites.
Upon squaring the Hamiltonian, the on-site potential in sites 1 and 2 in the blue sublattice can be easily seen to yield $c_2=2\sqrt{\Delta}+2\sqrt{t}$, while for sites 3 and 6 (4 and 5) it yields only $2\sqrt{t}$ ($2\sqrt{\Delta}$) because they are missing, in relation to sites 1 and 2, two $\sqrt[4]{\Delta}$ ($\sqrt[4]{t}$) hopping connections.
Since we want the same $c_2$ on-site potential at all sites in the blue sublattice in order to satisfy step 3, we introduce the extra green sites in Fig.~\ref{fig:ts4model}(c) with judiciously chosen hopping terms to sites 3, 4, 5 and 6 such that they exactly compensate for the difference in their respective on-site potential, upon squaring, in relation to $c_2$.
Notice the $\frac{\pi}{2}$ included in the $\sqrt[4]{\Delta}$ hoppings connected to the green sites, introducing a $\pi$-flux around these small loops.
This is required in order to prevent finite couplings between sites 3 and 4 (5 and 6) upon squaring the Hamiltonian (for instance, there are two two-step hopping processes between sites 3 and 4, with $\sqrt[4]{t\Delta}$ magnitude but symmetric $\pm\frac{\pi}{2}$ phase factors, thus canceling out), which would break the bipartite condition and, therefore, not yield the $\sqrt{\text{TS}}$ in Fig.~\ref{fig:ts4model}(b) as one of its diagonal blocks.
The bulk Hamiltonian of the $\sqrt[4]{\text{TS}}$, written in the ordered $\{\ket{j(k)}\}$ basis, where $j=1,2,\dots,18$ refers to the $j^{th}$ component within the unit cell, is given by
\begin{widetext}
	\begin{eqnarray}
	\mathcal{H}_{\sqrt[4]{\text{TS}}}&=&
	\begin{pmatrix}
	0&h^\dagger_{\sqrt[4]{\text{TS}}}
	\\
	h_{\sqrt[4]{\text{TI}}}&0
	\end{pmatrix},
	\label{eq:hamilt_ts4}
	\\
	h_{\sqrt[4]{\text{TS}}}&=&
	\begin{pmatrix}
    e^{-ik}\sqrt[4]{t}&0&\sqrt[4]{t}&0&0&0
    \\
    e^{-i(k-\frac{\pi}{8})}\sqrt[4]{\Delta}&0&0&0&e^{-i\frac{\pi}{8}}\sqrt[4]{\Delta}&0
    \\
    0& e^{-i(k-\frac{\pi}{8})}\sqrt[4]{\Delta}&0& e^{-i\frac{\pi}{8}}\sqrt[4]{\Delta}&0&0
    \\
    0&e^{-i(k-\frac{\pi}{4})}\sqrt[4]{t}&0&0&0&e^{-i\frac{\pi}{4}}\sqrt[4]{t}
    \\
    0&0&-i\sqrt[4]{\Delta}&\sqrt[4]{t}&0&0
    \\
    0&0&0&0&\sqrt[4]{t}&i\sqrt[4]{\Delta}
    \\
    0&0&i\sqrt[4]{\Delta}&\sqrt[4]{t}&0&0
    \\
    0&0&0&0&\sqrt[4]{t}&-i\sqrt[4]{\Delta}
    \\
    \sqrt[4]{t}&0&\sqrt[4]{t}&0&0&0
    \\
    e^{-i\frac{\pi}{8}}\sqrt[4]{\Delta}&0&0&e^{i\frac{\pi}{8}}\sqrt[4]{\Delta}&0&0
    \\
    0&e^{-i\frac{\pi}{8}}\sqrt[4]{\Delta}&0&0&e^{i\frac{\pi}{8}}\sqrt[4]{\Delta}&0
    \\
    0&e^{-i\frac{\pi}{4}}\sqrt[4]{t}&0&0&0&e^{i\frac{\pi}{4}}\sqrt[4]{t}
	\end{pmatrix},
	\end{eqnarray}
\end{widetext}
whose diagonalization, for $t=\Delta$, yields the energy spectrum in Fig.~\ref{fig:ts4model}(f), where the six-fold degenerate zero-energy band comes, similarly to the $\sqrt{\text{TS}}$, as a consequence of Lieb's theorem (gray and green sites belong to the same sublattice).
\begin{figure*}[ht]
	\begin{centering}
		\includegraphics[width=0.95 \textwidth,height=4.5cm]{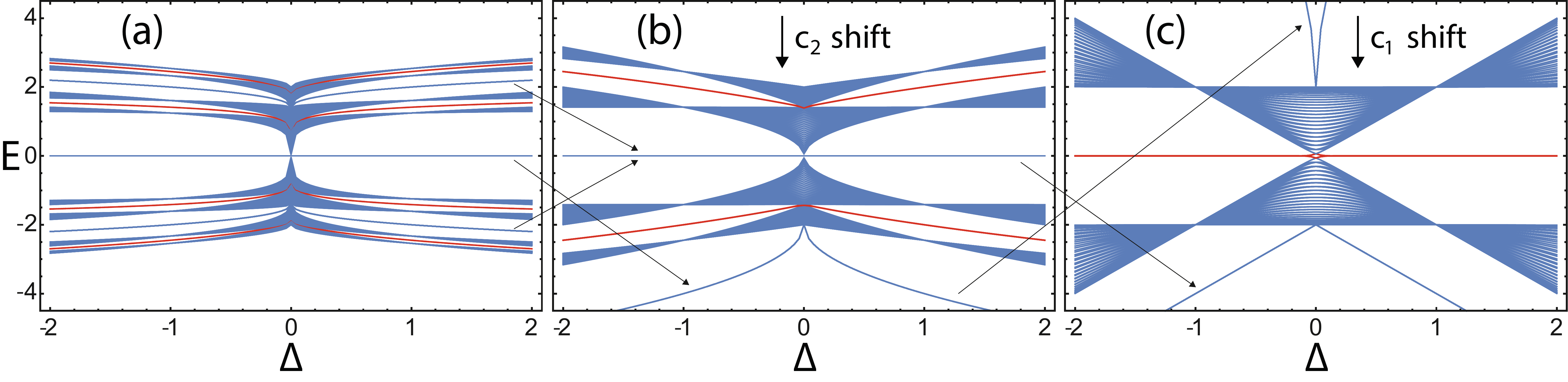} 
		\par\end{centering}
	\caption{Energy spectrum, in units of $t\equiv 1$, as a function of $\Delta$ obtained from diagonalization of: (a) $H_{\sqrt[4]{\text{TS}}}$, the Hamiltonian of the open $\sqrt[4]{\text{TS}}$ chain with the unit cell of Fig.~\ref{fig:ts4model}(c), with $N=50$ complete unit cells plus extra sites 3 to 18 at unit cell $N+1$; (b) $H_{\sqrt[4]{\text{TS}}}^{2^\prime}=H_{\sqrt[4]{\text{TS}}}^{2}-c_2$, with $c_2=2\sqrt{t}+2\sqrt{\Delta}$; (c) $H_{\sqrt[4]{\text{TS}}}^{4^\prime}=H_{\sqrt[4]{\text{TS}}}^{2^\prime}H_{\sqrt[4]{\text{TS}}}^{2^\prime}-c_1$, with $c_1=2t+2\Delta$. Bulk (edge) states are colored in blue (red). Each red edge band is two-fold degenerate in (a), four-fold degenerate in (b) and eight-fold degenerate in (c).}
	\label{fig:espectrumts4r}
\end{figure*}

As with the $\sqrt[2^n]{\text{TI}}$ studied above, there remains an ambiguity in the definition of the unit cell of the $\sqrt[2^n]{\text{TS}}$.
For instance, in the $\sqrt[4]{\text{TS}}$ of Fig.~\ref{fig:ts4model}(c), sites 3 through 18 have been placed at the left of sites 1 and 2 [those ultimately forming the original TS in Fig.~\ref{fig:ts4model}(a)] within the unit cell, but they could just as well been placed at the right.
If choosing one of the options dissipates the ambiguity, one must remember that step 4 requires that, under OBC, sites other than 1 and 2 have to appear at both ends.
Since we chose to grow the unit cell to the left, extra sites 3 through 18 have to be included at the incomplete unit cell $N+1$ at the right edge under OBC.
The energy spectrum of the open $\sqrt[4]{\text{TS}}$ with $N=50$ complete unit cell plus the 16 extra sites at unit cell $N+1$ is shown in Fig.~\ref{fig:espectrumts4r}(a).
Four doubly degenerate (red) bands of edge states are present at different energy gaps for finite $\Delta$.

Diagonalization of the squared Hamiltonian of the open $\sqrt[4]{\text{TS}}$ chain, followed by a $c_2=2\sqrt{t}+2\sqrt{\Delta}$ energy shift, yields the energy spectrum of Fig.~\ref{fig:espectrumts4r}(b), where the two red bands of edge states become four-fold degenerate.
Due to the $c_2$ shift, the six-fold degenerate band at $E=0$ of Fig.~\ref{fig:espectrumts4r}(a) becomes the lower $E=-c_2$ band in Fig.~\ref{fig:espectrumts4r}(b), which looses its flatness since $c_2$ depends on $\Delta$.

Upon squaring once more the Hamiltonian corresponding to the spectrum of Fig.~\ref{fig:espectrumts4r}(b) and subtracting a constant energy shift given by $c_1=2t+2\Delta$, one obtains the ancestor Hamiltonian $H_{\text{anc}}$ of the $\sqrt[4]{\text{TS}}$.
Its diagonalization yields the energy spectrum of Fig.~\ref{fig:espectrumts4r}(c).
The six-fold degenerate lowest energy band has $E=-c_1$ and the four-fold degenerate highest energy one has $E=c_2^2-c_1$.
The bands of edge states all collapse onto the eight-fold degenerate Majorana zero-energy red band.

For a finite $\Delta$, the $\sqrt[4]{\text{TS}}$ chain is in the topological phase and harbors $\#_{\text{edge}}=8$ edge states, double the number of the $\sqrt[4]{\text{TI}}$ chain [compare with Fig.~\ref{fig:espectrumti4r}(a)].
In general, one has $\#_{\text{edge}}=2^{n+1}$ edge states in the $\sqrt[2^n]{\text{TS}}$ chain.
The arborescence of $\sqrt[2^n]{\text{TSs}}$ is the same as in Fig.~\ref{fig:tree}, only doubling the number of edge states present in the root node and, consequently, also in all the leaf nodes.
Whereas the original TI we considered previously only contributed with one topological edge state to all $\sqrt[2^n]{\text{TI}}$, localized at the left edge, since the impurity hopping and on-site potential terms appearing at the right end send the right edge state to the bulk [see the Creutz ladder in Fig.~\ref{fig:diamond}(b) and energy spectrum of Fig.~\ref{fig:diamond}(f)], our original TS, the Kitaev chain depicted in Fig.~\ref{fig:ts4model}(a), contributes with one topological edge state at each end to all $\sqrt[2^n]{\text{TS}}$.
Note that these two oppositely located edge states would correspond, in the Majorana basis of $\gamma$ operators, to the two free $\gamma$ operators living at each end of the chain that combine to form a Majorana fermion.
In fact, it is because we want to keep this correspondence that, in contrast with  the right edge perturbed Creutz ladder in Fig.~\ref{fig:diamond}(b), the original ladder in Fig.~\ref{fig:ts4model}(a) is unperturbed, requiring that gray sites from the incomplete unit cell $N+1$ to be already present at the right edge at the square-root level shown in Fig.~\ref{fig:ts4model}(b).
This is why we find double the number of edge states here, in relation to the same level of the TI studied in the previous sections.

It is clear that a $\sqrt[8]{\text{TS}}$ can be constructed from the $\sqrt[4]{\text{TS}}$ in Fig.~\ref{fig:ts4model}(c) by application of steps 1 to 4 and, by further repetition of the process, any $\sqrt[2^n]{\text{TS}}$ can be constructed.
The energy of the edge states can be determined recursively, as in (\ref{eq:eneredgeti2n}), by changing only the degeneracy of the initial value in (\ref{eq:eneredgeti}),
\begin{eqnarray}
E_{\sqrt[2^n]{\text{TS}}}&=&\pm\sqrt{E_{\sqrt[2^{n-1}]{\text{TS}}}+c_{2^{n-1}}},
\label{eq:eneredgets2n}
\\
E_\text{TS}&=&0\ \ (\times 2),
\label{eq:eneredgets}
\end{eqnarray}
where the constant energy shifts are now given by
\begin{equation}
c_{2^m}=2\sqrt[2^m]{t}+2\sqrt[2^m]{\Delta},
\end{equation}
with $m=0,1,2,\dots,n-1$.
Analogously to the case of a $\sqrt[2^n]{\text{TI}}$, all $\#_{\text{edge}}=2^{n+1}$ edge states present in a $\sqrt[2^n]{\text{TS}}$ have a diluting weight in the two topological edge states of the TS as $n$ increases and, conversely, an increasing weight in the impurity edge states that stem from the residual Hamiltonian terms in $H_{\text{anc}}$.

Remarkably, for $\Delta=t$ the model of Fig.~\ref{fig:ts4model}(a) becomes the Creutz ladder in Fig.~\ref{fig:diamond}(b) \text{without} the right edge hopping and on-site potential perturbations.
This shows that modifications to the same original TI/TS can generate different types of $\sqrt[2^n]{\text{TIs}}$/$\sqrt[2^n]{\text{TIs}}$, given that before the diamond chain with $\pi$-flux per plaquette was identified as the square-root version of the right edge perturbed Creutz ladder combined with a chain of decoupled sites, whereas here we constructed the $\sqrt{\text{TS}}$ and the $\sqrt[4]{\text{TS}}$ from the unperturbed Creutz ladder only. 
Since it would be hard to guess \textit{a priori} that the very specific combination of Fig.~\ref{fig:diamond}(b) would be the one to generate a $\sqrt{\text{TI}}$, sometimes it is best in practice to start with a presumptive candidate for square-root topology, whose admission criteria are (i) being bipartite, (ii) having finite energy edge states and (iii) having nonquantized topological indices, and work from there in both directions, finding its original TI/TS along one and the higher root degree versions along the other.

\section{Conclusions}
\label{sec:conclusions}

We outlined here the procedure to go beyond square-root topology \cite{Ezawa2020}, towards topology with a root degree of any positive integer power of two ($2^n$-root topology, with $n\in \mathbb{N}$).
This was demonstrated for one-dimensional topological insulators and supercondutors, where the original models from which higher root degree versions were constructed were taken to be the Creutz ladder, for the former, which generated a family of extended diamond chains with a $\pi$-flux per plaquette, and the Kitaev chain mapped onto a single-particle tight-binding model, for the latter.
The connection between a $2^n$-root topological insulator/superconductor and its original topological insulator/superconductor was established through a graph shaped as an outwards directed rooted tree, where the Hamiltonian of the former is identified as the root node and the Hamiltonian of the latter as one of the leaf nodes.

Concerning the experimental realization of the $2^n$-root topological insulators studied here, several platforms appear as suitable candidates for their implementation, most notably photonic \cite{Baboux2016,Mukherjee2018,Zhang2019,XIa2020,Kremer2020,Jorg2020}, solid-state \cite{Huda2020} and optical lattices \cite{Tale2015}.
In what concerns the $2^n$-root topological superconductor model introduced here, on the other hand, its direct implementation in a superconducting system seems out-of-reach with current technology, as it would require some very fine local tuning of the phases of the superconducting and hopping terms.
Nevertheless, and as we explained in Sec.~\ref{sec:2nrootss}, one can simulate this model by translating it into its single-particle tight-binding analog model, which can be realized in the venues listed above or, as recently proposed \cite{Ezawa2019,Ezawa2020b}, in topoelectrical circuits \cite{Song2020,Li2020b,Olekhno2020,Wu2020,Yang2021}.

Our work invites future studies searching for deeper connections between a Hamiltonian and any of its positive integer powers.
We expect $2^n$-root topology to be generalizable to simply $n$-root topology.
In particular, odd-root topology would come as a natural extension of our results, since a block antidiagonal Hamiltonian will remain so when raised to any odd power (one should only be careful in taking out the constant energy shifts after each squaring operation).
In turn, raising this odd-power Hamiltonian to an even power will render it block diagonal.
For instance, one can treat six-root topology as a product of cubic-root and square-root topology.
It may be possible to construct, on this basis, a network of topological insulators such that, as with mathematical theorems, the set of those we elevate to an ``axiomatic'' status, like that implicitly granted to the Su-Schrieffer-Heeger model, would become a matter of convention.

The results concerning the extension of $2^n$-root topology to encompass two-dimensional Chern, weak and higher-order topological insulators, as well as topological semimetals, are being finalized and will be the subject of a forthcoming paper.

\section*{Acknowledgments}
\label{sec:acknowledments}

This work was developed within the scope of the Portuguese Institute for Nanostructures, Nanomodelling and Nanofabrication (i3N) projects UIDB/50025/2020 and UIDP/50025/2020 and funded by FCT - Portuguese Foundation for Science and Technology through the project PTDC/FIS-MAC/29291/2017. A.M.M. acknowledges financial support from the FCT through the work contract CDL-CTTRI-147-ARH/2018.
LM acknowledges financial support from the FCT through Grant No. SFRH/BD/150640/2020.
A.M.M. would like to thank Rui Am\'erico Costa for useful clarifications on graph theory.

\appendix

\section{Alternative construction of the $2^n$-root Creutz ladder}
\label{app:alternativemethod}

\begin{figure}[b]
	\centering
	\includegraphics[width=1\linewidth]{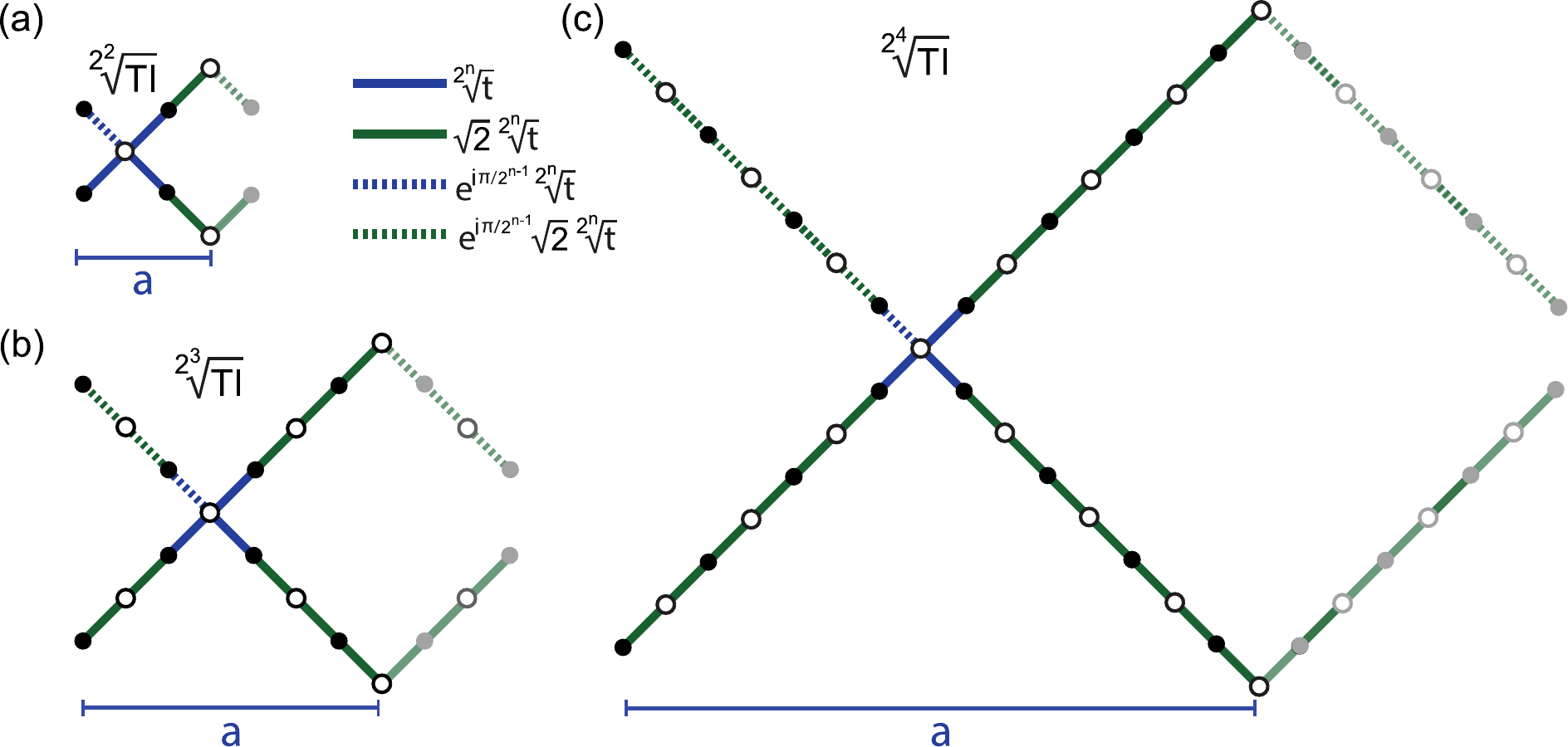}
	\caption{Unit cells of the $n=2,3,4$ $2^n$-root models using the renormalization method, constructed from the $\sqrt{\text{TI}}$ given by the diamond chain with $\pi$-flux of Fig.~\ref{fig:diamond}(a),  onwards to (a) $\sqrt[4]{\text{TI}}$ with 7-sites, (b) $\sqrt[8]{\text{TI}}$ with 15-sites and (c) $\sqrt[16]{\text{TI}}$ with 31-sites. Black sites appear from subdivision of the $\sqrt[n-1]{\text{TI}}$ (a site is included in the middle of each link). Gray sites belong to the next unit cell and have to be included at unit cell $N+1$ under OBC.}
	\label{fig:ap1fig1}
\end{figure}

In this section we introduce a more economic construction of the $2^n$-root	Creutz ladder by renormalization of the hopping parameters. This approach follows the same steps as those described in Sec.~\ref{sec:2nroottis} except that we adopt step 3(i) to ensure evened on-site energies upon squaring the Hamiltonian.

We start by introducing nodes in the middle of each link in the $\sqrt[2^{n-1}]{\text{TI}}$ where the magnitude of the new two-fold hopping terms is transformed as $\sqrt[2^{n-1}]{t}e^{i\phi/2^{n-2}}\to\sqrt[2^n]{t}e^{i\phi/2^{n-1}}$. We then identify the highest squared on-site potential $c_{2^{n-1}}$ which will determine the renormalization of the hopping terms connected to sites with lower squared on-site potential (often corresponds to the subset of lower coordination number). At this point, there is also an ambiguity in the definition of the unit cell under OBC (the outermost extra sites can be placed either to the right or to the left within the unit cell). In order to preserve an uniform squared on-site potential along the chain these outermost sites have to be included in both boundaries. 

This method was followed for the construction of the $2^n$-root Creutz ladder starting from its $\sqrt{\text{TI}}$, given by the diamond chain with $\pi$-flux of Fig.~\ref{fig:diamond}(a), and up to $n=4$, as depicted in Fig.~\ref{fig:ap1fig1}. White and black sites, respectively, set apart the old sublattice of the original model for $\sqrt[2^{n-1}]{\text{TI}}$ and the subset of extra sites added to the latter to model the $\sqrt[2^n]{\text{TI}}$ chain. 
\begin{figure*}[ht]
	\centering
	\includegraphics[width=1\linewidth]{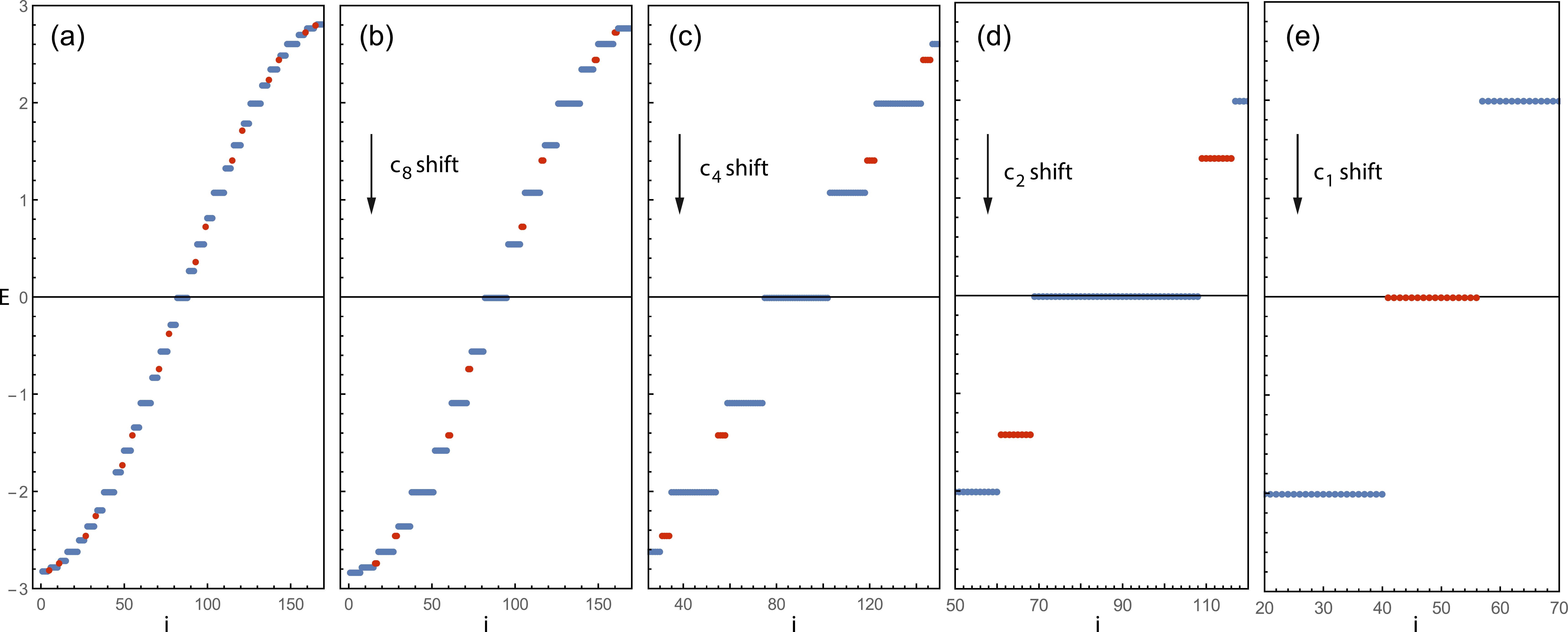}
	\caption{Energy spectrum, in units of $t\equiv 1$, as a function of state index $i$, obtained from diagonalization of (a) $H_{\sqrt[16]{\text{TI}}}$, the Hamiltonian of the open $\sqrt[16]{\text{TI}}$ chain obtained by the renormalization method with $N=5$ complete unit cells plus 14 open extra sites at unit cell $N+1$, (b) $H_{\sqrt[16]{\text{TI}}}^{2^\prime}=H_{\sqrt[16]{\text{TI}}}^{2}-c_8$, with $c_8=4\sqrt[8]{t}$, (c) $H_{\sqrt[16]{\text{TI}}}^{4^\prime}=H_{\sqrt[16]{\text{TI}}}^{2^\prime}H_{\sqrt[16]{\text{TI}}}^{2^\prime}-c_4$, with $c_4=4\sqrt[4]{t}$, (d) $H_{\sqrt[16]{\text{TI}}}^{8^\prime}=H_{\sqrt[16]{\text{TI}}}^{4^\prime}H_{\sqrt[16]{\text{TI}}}^{4^\prime}-c_2$, with $c_2=4\sqrt{t}$, and (e) $H_{\sqrt[16]{\text{TI}}}^{16^\prime}=H_{\sqrt[16]{\text{TI}}}^{8^\prime}H_{\sqrt[16]{\text{TI}}}^{8^\prime}-c_1$, with $c_1=2t$. Bulk (edge) states are colored in blue (red). In (b)-(d) the energy of all states was rescaled by a factor of $1/\sqrt{2}$, as explained in (\ref{eq:rescaling}). Only a partial spectrum is shown in (c)-(e) for clarity, with some high energy bulk bands, in absolute value, outside the range of the y-axis.}
	\label{fig:ap1fig2}
\end{figure*}
Predictably, the only hopping parameters that do not require renormalization are those connected to the spinal sites with highest coordination number with squared onsite potential of $c_{2^{n-1}}=4\sqrt[2^{n-1}]{t}$, provided that $n>1$. For the remainder, the renormalization factor of $\sqrt{2}$ ensures the same onsite potential in the ancestor subset upon squaring the Hamiltonian for all $\sqrt[2^{n}]{\text{TI}}$. 
It should be noticed that, in relation to the method 3(ii) followed in the main text, this method rescales all parameters upon squaring the Hamiltonian at each level by a factor of $\sqrt{2}$,
\begin{equation}
H_{\sqrt[2^i]{\text{TI}}}^2=
c_{2^{i-1}}I_{\sqrt[2^i]{\text{TI}}}+\sqrt{2}
\begin{pmatrix}
H_{\sqrt[2^{i-1}]{\text{TI}}}&0
\\
0&H_{\sqrt{\text{res},2^{i-1}}^\prime}
\end{pmatrix},
\label{eq:rescaling}
\end{equation} 
for $i=2,3,\dots,n$, where $I_{\sqrt[2^i]{\text{TI}}}$ is the identity matrix with the dimension of $H_{\sqrt[2^i]{\text{TI}}}$.
Accordingly, in order to keep the same energy scale a general rescaling given by $1/\sqrt{2}$ is applied to all Hamiltonians at the same level as $H_{\sqrt[2^{i-1}]{\text{TI}}}$ in Fig.~\ref{fig:tree}, that is, including the residual ones.

The number of sites per unit cell for the $\sqrt[2^n]{\text{TI}}$ follows the recurrence relation
\begin{eqnarray}
\#_{\text{uc}}^{2^n}&=&\#_{\text{uc}}^{2^{n-1}}+2^{n}, \ \ n\geq 2, \label{apEq1}
\\
\#_{\text{uc}}^{2}&=&3,
\end{eqnarray}
where we begin with the 3-site diamond chain unit cell. The closed form solution of (\ref{apEq1}) by iteration is explicitly written in terms of the initial condition $\#_{\text{uc}}^{2}$ as
\begin{equation}
\#_{\textit{(i)}}^{2^n}=\#_{\text{uc}}^{2}+\sum_{i=2}^{n}2^{i}=-1+2^{n+1},\label{apeq2}
\end{equation}
with $\sum_{i=2}^{n}2^{i}=-4+2^{n+1}$ being the general formula for this geometric progression of $n$ terms. 

We may probe the efficiency for both methods described in this report, namely (i) renormalization approach and (ii) the extra added sites method of Sec.~\ref{sec:2nroottis}, from the perspective of the $\sqrt[2^{n}]{\text{TI}}$-Hamiltonian recursive dimensions. For (\ref{apeq2}) the growth rate in order notation is $\mathcal{O}(2^n)$. For the alternative method, the solution for the recurrence relation yields
\begin{equation}
\#_{\textit{(ii)}}^{2^n}=\#_{\text{uc}}^{2}+\sum_{i=2}^{n}2^{2i-1}=\dfrac{1}{3}(1+2^{2n+3}),
\end{equation}
with a growth rate of $\mathcal{O}(2^{2n})$. At this point we can say that $\mathcal{O}(\#_{\textit{(ii)}}^{2^n})>\mathcal{O}(\#_{\textit{(i)}}^{2^n})$ and the ratio for the site increment as $n\gg1$ can be approximated to
\begin{equation}
\frac{\#_{\textit{(i)}}^{2^n}}{	\#_{\textit{(ii)}}^{2^n}}=\frac{3(-1+2^{n+1})}{1+2^{2n+3}}\sim\frac{3}{2^{n}},
\end{equation}
which is a good approximation for $n>4$ and asymptotically approaches zero in the limit $n\rightarrow+\infty$. Thus, for this criteria, the renormalization method is clearly more tractable in the realization of higher $\sqrt[2^{n}]{\text{TI}}$.

In Fig.~\ref{fig:ap1fig2} we show the energy spectrum of the $\sqrt[2^{4}]{\text{TI}}$ chain with $N=5$ unit cells with $\#_{\textit{(i)}}^{2^4}=31$ sites and of the $n=4$ successive squaring operations to its Hamiltonian, taking out the constant energy shift given by the on-site potential energy at the relevant sublattice after each operation (white sites in  Fig.~\ref{fig:ap1fig1}) and rescaling the energy by $1/\sqrt{2}$ in Figs.~\ref{fig:ap1fig2}(b)-\ref{fig:ap1fig2}(d), as discussed below (\ref{eq:rescaling}). As explained above, the outermost 14 sites were placed in the leftmost region within the unit cell and included also at the right-side of the last complete unit cell [see Fig.~\ref{fig:ap1fig1}(c)].

As expected, there is a total of $2^4=16$ degenerate zero-energy states [in the midgap of Fig.~\ref{fig:ap1fig2}(e)] arising from the diagonalization of $H_\text{anc}$ and the topological weight of the edge states of our starting $\sqrt[2^4]{\text{TI}}$ chain [in-gap red states in Fig.~\ref{fig:ap1fig2}(a)], relative to the edge states of the ancestor chain is given by the factor $|\braket{T_L}{T_{2^4},j}|^2=1/16$, with $\ket{T_L}$ defined above (\ref{eq:topostateleft}). Both the values of the constant energy shifts determined by the highest squared on-site potential and the energies of the $2^4$ edge states can be readily found to be equivalent to those described in the main text, thus validating this alternative method.

\section{Holographic duality in the subspace of edge states}
\label{app:holographic}
The $2^n$ edge states of the $\sqrt[2^n]{\text{TI}}$ can be decomposed in the residual and lower root-degree components of its squared model, by generalizing (\ref{eq:ti2topoweight}), as
\begin{widetext}
\begin{equation}
\begin{pmatrix}
\ket{T_{2^n},2^n+1-j}
\\
\ket{T_{2^n},j}
\end{pmatrix}
=
\frac{1}{\sqrt{2}}
\begin{pmatrix}
1&1
\\
1&-1
\end{pmatrix}
\begin{pmatrix}
\ket{T_{2^{n-1}},2^{n-1}+1-j}
\\
\ket{I_{\text{res,}2^{n-1}},2^{n-1}+1-j}
\end{pmatrix},\ \ \ \ j=1,2,\dots,2^{n-1},
\label{eq:transformationedgestates}
\end{equation} 
\end{widetext}
with the topological and impurity states coming from the residual block defined below (\ref{eq:ancestorhamilt}).
The two topological states at the left-hand side form a chiral pair with symmetric energies, $\epsilon_{2^n+1-j}=-\epsilon_j>0$ (as $j$ increases, the symmetric pairs approach the zero-energy level starting from $j=1$, the highest energy pair furthest away from zero-energy), while the edge states at the right-hand side are degenerate.
Inverting (\ref{eq:transformationedgestates}) and setting $l=n-1$ leads to
\begin{widetext}
	\begin{equation}
	\begin{pmatrix}
	\ket{T_{2^l},j}
	\\
	\ket{I_{\text{res,}2^l},j}
	\end{pmatrix}
	=
	\frac{1}{\sqrt{2}}
	\begin{pmatrix}
	1&1
	\\
	1&-1
	\end{pmatrix}
	\begin{pmatrix}
	\ket{T_{2^{l+1}},2^l+j}
	\\
	\ket{T_{2^{l+1}},2^l+1-j}
	\end{pmatrix},\ \ \ \ j=1,2,\dots,2^l,
	\label{eq:invtransformationedgestates}
	\end{equation} 
\end{widetext}
which is a local unitary transformation acting on pairs of edge states of the $\sqrt[2^{l+1}]{\text{TI}}$ chain.
\begin{figure}[ht]
	\centering
	\includegraphics[width=1\linewidth]{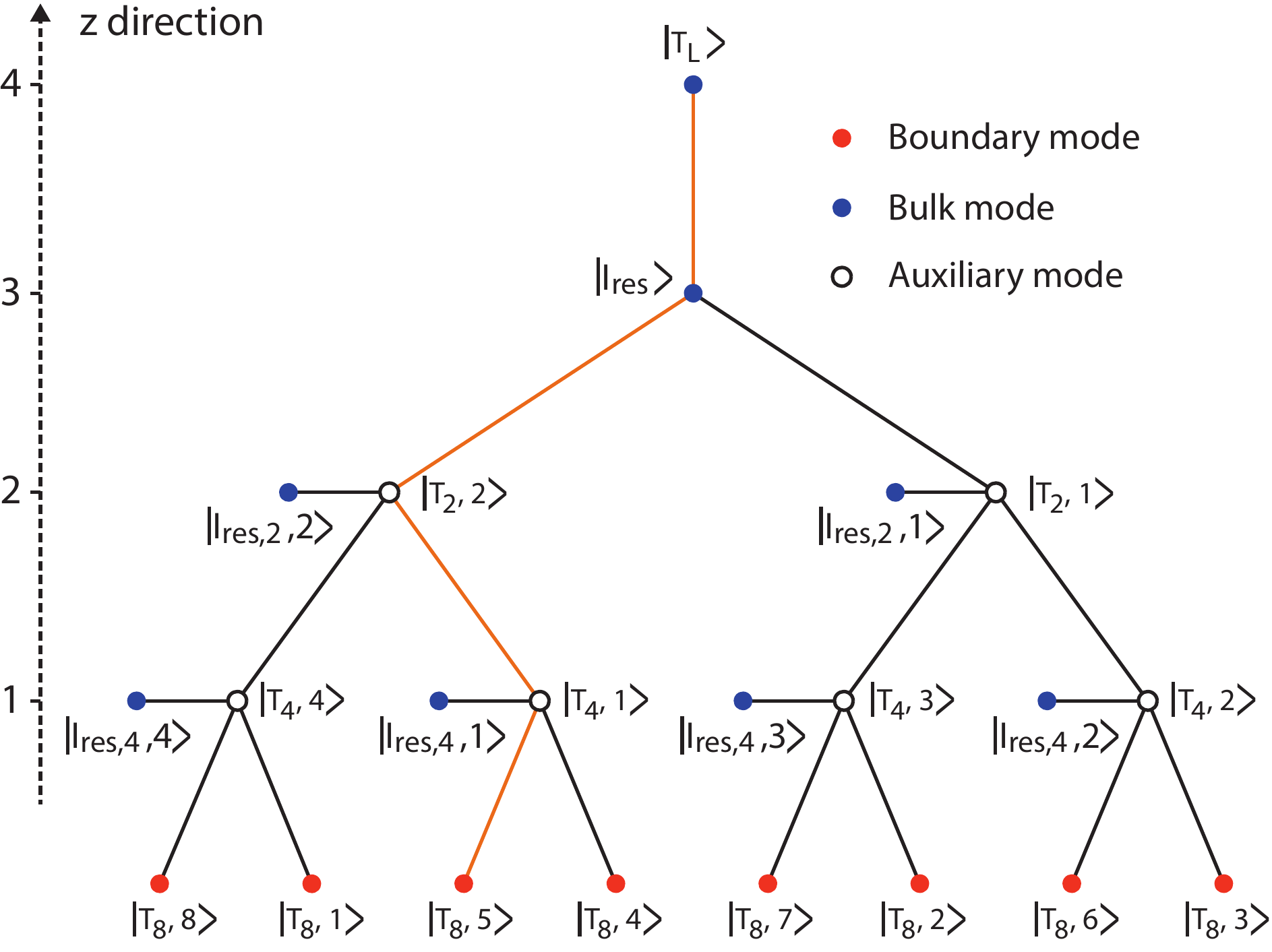}
	\caption{Resulting tensor network from application of the exact holographic mapping to the edge states of the $\sqrt[8]{\text{TI}}$ chain, represented here as the boundary modes from which the bulk modes at each level of the emergent $z$ direction are found, corresponding to the edge states of the successive residual chains plus the topological state of the original TI on top. Orange highlighted path illustrates the decomposition of the boundary mode $\ket{T_8,5}$ in terms of the bulk modes picked up along $z$, according to the rules explained in the text.}
	\label{fig:ap1fig5}
\end{figure}

An analogy can be drawn here with the AdS/CFT correspondence, also called holographic duality, if we iteratively apply (\ref{eq:invtransformationedgestates}) to the edge states of the topological block at each level, starting with a given $\sqrt[2^n]{\text{TI}}$ chain, until the edge state of the original TI is reached.
We identify the $2^n$ initial edge states as boundary modes of our ``CFT'' model.
Then, by applying (\ref{eq:invtransformationedgestates}) to the whole boundary we obtain $2^{n-1}$ residual edge states, which we retain at that level as our ``bulk modes'', and $2^{n-1}$ edge states from the lower root-degree TI block, which act as the ``auxiliary modes'' from which the next level is obtained by a subsequent application of (\ref{eq:invtransformationedgestates}). 
The tensor network constructed by this exact holographic mapping \cite{Qi2013,Lee2016,Hu2020} for the $\sqrt[8]{\text{TI}}$ is depicted in Fig.~\ref{fig:ap1fig5}, where after the last transformation both $\ket{I_{\text{res}}}\equiv\ket{I_{\text{res,}1},1}$ and $\ket{T_L}\equiv\ket{T_{1},1}$ are kept as bulk modes.
The indexation of the boundary sites is made according to the following rules: (i) the highest energy state is placed as the leftmost site; (ii) chiral pairs are adjacently placed, with the positive energy state of the two at the left; (iii) the disposition of the chiral pairs of sites is such that crossings between the solid lines are avoided for the entire tensor network.
The $z$ direction is the emergent  discrete dimension of our ``AdS'' space embedding the bulk model built by the layered decomposition of the boundary modes \cite{Gu2016} as $2^n\to 2^{n-1}+2^{n-2}+\dots+2+1+1$ .
\begin{figure}[ht]
	\centering
	\includegraphics[width=0.5\linewidth]{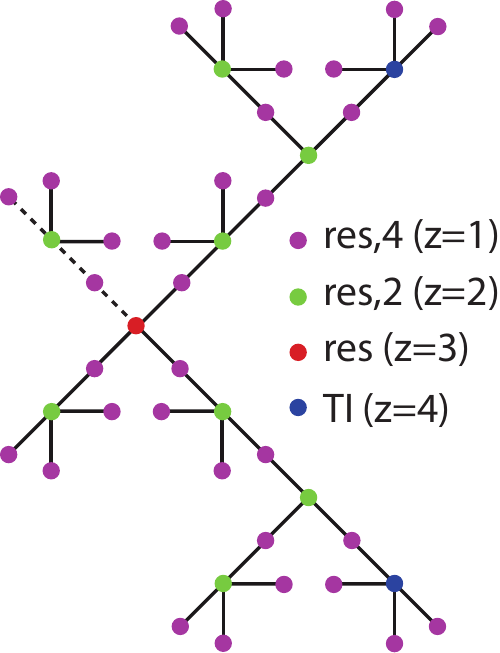}
	\caption{Hidden sublattices, distinguished by color, present in the unit cell of the $\sqrt[8]{\text{TI}}$ recovered here from Fig.~\ref{fig:ti8ucell}. The bulk modes at a given $z$ level in Fig.~\ref{fig:ap1fig5} only have finite weight on the respective sublattice depicted here.}
	\label{fig:ap1fig6}
\end{figure}
The number of levels of the network (the $z$ value of $\ket{T_L}$) has the physical meaning of counting the number of hidden sublattices of the $\sqrt[2^n]{\text{TI}}$, one corresponding to each residual chain and one to the original TI.
The four hidden sublattices of the $\sqrt[8]{\text{TI}}$ are represented by different colors on its unit cell in Fig.~\ref{fig:ap1fig6}.

The exact form of each of the $2^n$ boundary states can be written as a linear combination of all the bulk modes picked up on the path from the boundary to the top bulk mode, with the rule that (i) a ``+'' (``-'') sign is added whenever the lower level mode is the left (right) one, with the top $\ket{T_L}$ mode always with a ``+'' sign, and (ii) the weight of the state is halved from one level to the next, except from $z=n$ to $z=n+1$ on top, where it is kept constant.
For example, the boundary mode connected to $\ket{T_L}$ by the orange highlighted path in Fig.~\ref{fig:ap1fig6} is written as
\begin{equation}
\ket{T_8,5}=\frac{1}{2\sqrt{2}}\begin{pmatrix}
\ket{T_L}
\\
\ket{I_{\text{res}}}
\\
-\sqrt{2}\ket{I_{\text{res,}2},2}
\\
2\ket{I_{\text{res,}4},1}
\end{pmatrix},
\end{equation}
where it is implicit that each bulk mode only has weight on its respective hidden lattice of Fig.~\ref{fig:ap1fig6}.

It is clear from the normalization pre-factors that the bulk dual of a boundary mode has most of its weight concentrated in the ``ultraviolet'' region closer to the boundary, and becomes progressively diluted as it approaches the ``infrared'' region further from the boundary.

\section{The $2^n$-root SSH model}
\label{app:2nrootssh}
\begin{figure}[ht]
	\centering
	\includegraphics[width=1\linewidth]{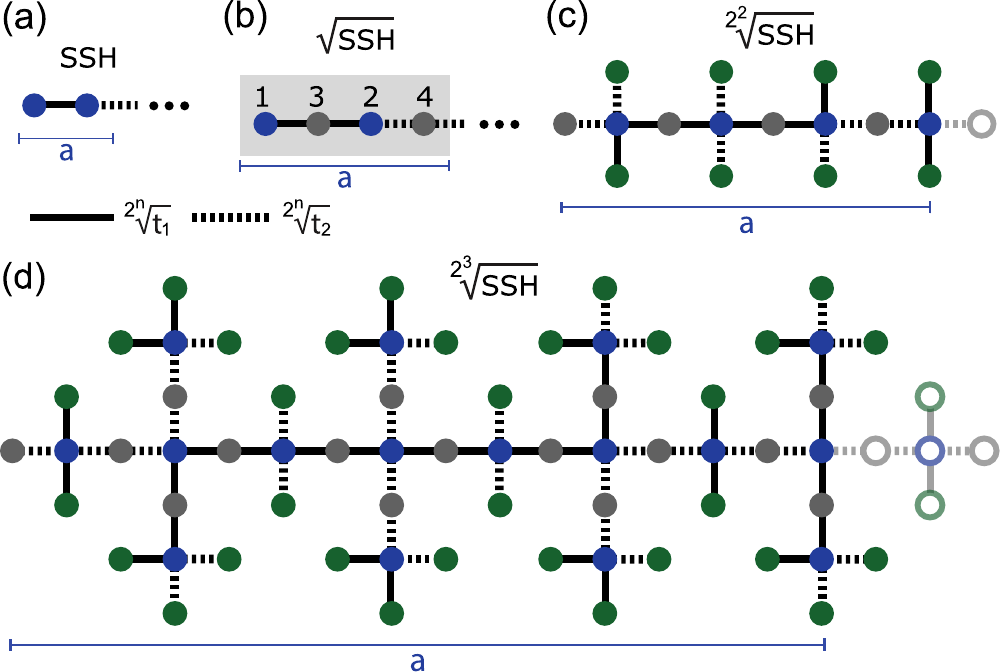}
	\caption{Unit cells of the $n=\{1,2,3\}$ $2^n$-root models using the renormalization method applied to the (a) SSH model with 2-sites, starting from (b) $\sqrt[2]{\text{SSH}}$ with 4-sites, to (c) $\sqrt[4]{\text{SSH}}$ with 16-sites and (d) $\sqrt[8]{\text{SSH}}$ with 64-sites. Blue sites form the sublattice from the previous $\sqrt[2^{n-1}]{\text{SSH}}$ chain, gray sites come from the subdivision of the latter, green sites are extra sites introduced to keep the coordination number constant for sites in the blue sublattice and white sites at the right belong to unit cell N + 1 under OBC.}
	\label{fig:ap2fig1}
\end{figure}
\begin{figure*}[ht]
	\centering
	\includegraphics[width=0.9\linewidth]{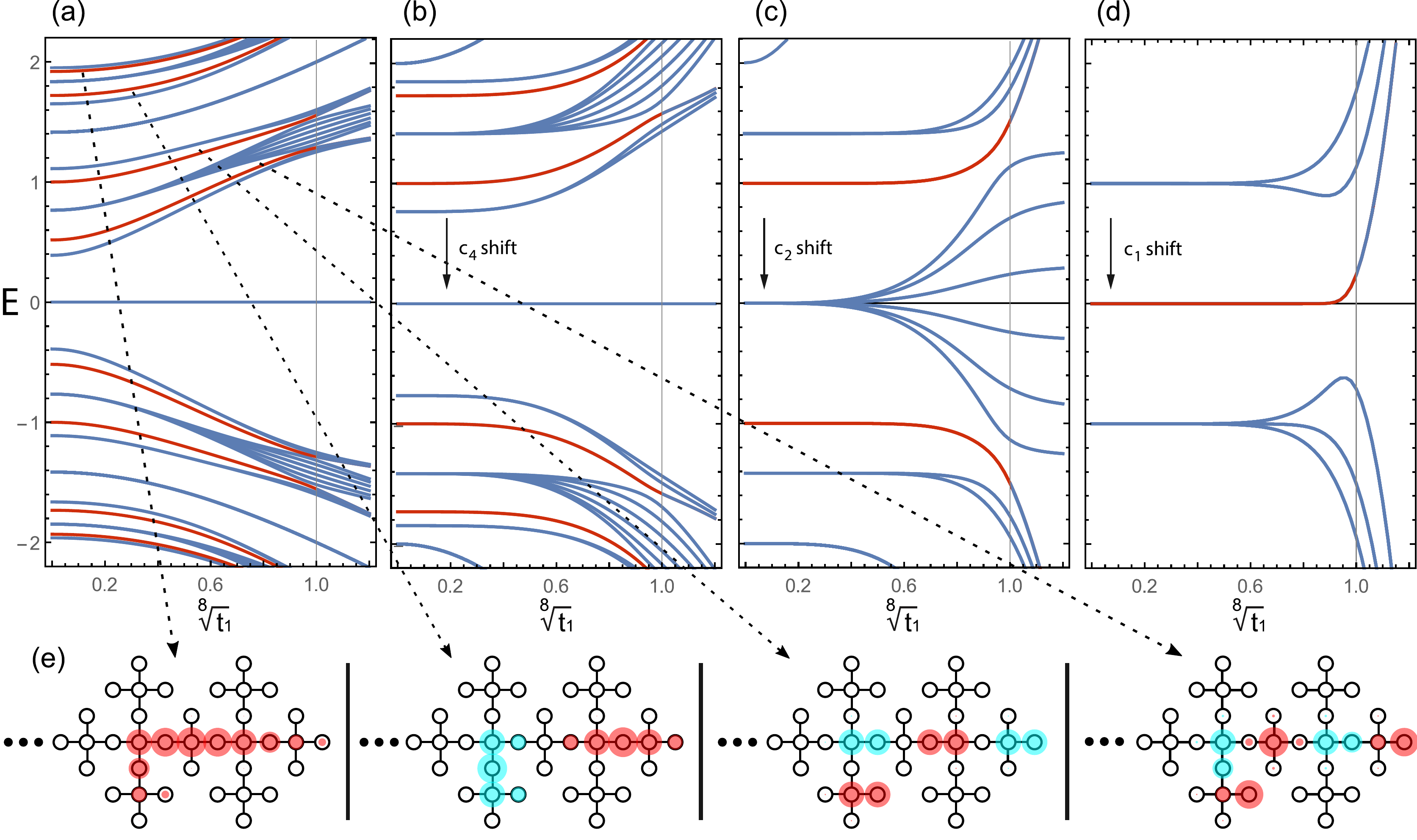}
	\caption{Energy spectrum, in units of $t_2\equiv 1$, as a function of a linear variation of $\sqrt[8]{t_1}$, obtained from diagonalization of (a) $H_{\sqrt[8]{\text{SSH}}}$, (b) $H_{\sqrt[8]{\text{TI}}}^{2^\prime}=H_{\sqrt[8]{\text{SSH}}}^{2}-c_4$, (c) $H_{\sqrt[8]{\text{SSH}}}^{4^\prime}=H_{\sqrt[8]{\text{SSH}}}^{2^\prime}H_{\sqrt[8]{\text{SSH}}}^{2^\prime}-c_2$, (d) $H_{\sqrt[8]{\text{SSH}}}^{8^\prime}=H_{\sqrt[8]{\text{SSH}}}^{4^\prime}H_{\sqrt[8]{\text{SSH}}}^{4^\prime}-c_1$. Bulk (edge) states are colored in blue (red). (e) Profile of the edge states in the $\sqrt[8]{\text{SSH}}$ chain, where the radius of the circle represents the amplitude of the wavefunction at the respective site and the color represents its phase, coded by the same color bar as in Fig.~\ref{fig:espectrumti4r}(d).}
	\label{fig:ap2fig2}
\end{figure*}
We will validate the proposed method for the construction of  energy dispersive $2^n$-root topological insulators using the well-known SSH model. We start with the already formulated $\sqrt{\text{SSH}}$ \cite{Ezawa2020,Lin2021}, corresponding to an SSH$_4$ model \cite{Maffei2018} whose solution both for PBC and OBC can be exactly found \cite{Eliashvili2017,Marques2019,Marques2020}, and recursively obtain $\sqrt[2^n]{\text{SSH}}$ for $n=\{2,3\}$ (see Fig.~\ref{fig:ap2fig1}).

Under PBC, the bulk Hamiltonian for the $\sqrt{\text{SSH}}$ model, written in the ordered $\{|j(k)\}$ basis, where $j = 1, 2,..., 4$ is the j$^{\text{th}}$ Fourier transformed component within the unit cell, following the indexation scheme of Fig.~\ref{fig:ap2fig1}(b), is given by
\begin{eqnarray}
\mathcal{H}_{\sqrt{\text{SSH}}}&=&
\begin{pmatrix}
0&h^\dagger_{\sqrt{\text{SSH}}}
\\
h_{\sqrt{\text{SSH}}}&0
\end{pmatrix},
\\
h^\dagger_{\sqrt{\text{SSH}}}&=&
\label{eq:srssh}
\begin{pmatrix}
\sqrt{t_1}&\sqrt{t_2}e^{-ik}
\\
\sqrt{t_1}&\sqrt{t_2}
\end{pmatrix}.
\end{eqnarray}

Considering OBC and assuming an integer number $N$ of unit cells, there is a topological nontrivial phase in the $t_1<t_2$ regime with two nondegenerate right-edge states of energy $E_{\text{edge}}^\pm=\pm\sqrt{t_1+t_2}$  [red curves in Fig.~\ref{fig:ap1fig2}(c)]. The resulting Hamiltonian from the squaring operation of $H_{\sqrt{\text{SSH}}}$ describes an SSH chain with alternating $\{t_1,t_2\}$ parameters and on-site local potentials $c_1=t_1+t_2$ (except at the perturbed left edge site, with on-site energy $c_1-t_2=t_1$, which sends the left-edge state to the bulk), plus an impurity chain with uniform $t=\sqrt{t_1t_2}$ and staggered on-site potentials $\{2t_1,2t_2\}$ (except at the perturbed right edge site, with on-site energy $t_2$, which generates a localized right-edge impurity state).
As such, the chiral-symmetric pair of right-edge states of the $\sqrt{\text{SSH}}$ chain can be written in the same form of (\ref{eq:ti2topoweight}), with $T_L\to T_R$ in this case.

For the construction of the higher root-degree SSH models from the steps described in the main text, we identify the sublattice with lower squared on-site potentials as the subset of sites of the previous $\sqrt[2^{n-1}]{\text{SSH}}$ model (blue in Fig.~\ref{fig:ap2fig1}). In each site, two extra connections were included (green in Fig.~\ref{fig:ap2fig1}), connected to the blue sites either by $\sqrt[2^n]{t_1}$ or $\sqrt[2^n]{t_2}$, such that each blue site is connected by two hoppings of each kind. This ensures a constant energy shift for each step $n$ defined as
\begin{eqnarray}
c_{2^{n-1}}&=&2\,(\sqrt[2^{n-1}]{t_1}+\sqrt[2^{n-1}]{t_2} ), \ \ n > 1,\\
c_{1}&=&t_1+t_2.
\end{eqnarray}
Recall that there are other options for leveling the squared on-site potentials, with our choice keeping only two different hopping parameters for all $n$. 
The squaring operation at each level reads as
\begin{eqnarray}
H_{\sqrt[2^n]{\text{SSH}}}^2&=&
\begin{pmatrix}
H_{\sqrt{\text{par},2^{n-1}}}&0
\\
0&H_{\sqrt{\text{res},2^{n-1}}}
\end{pmatrix},
\\
H_{\sqrt{\text{par},2^{n-1}}}&=&h_{\sqrt[2^n]{\text{SSH}}}^\dagger h_{\sqrt[2^n]{\text{SSH}}}= \nonumber
\\
&=&c_{2^{n-1}}I_{\text{par}}+H_{\sqrt[2^{n-1}]{\text{SSH}}},
\\
H_{\sqrt{\text{res},2^{n-1}}}&=&h_{\sqrt[2^n]{\text{SSH}}} h_{\sqrt[2^n]{\text{SSH}}}^\dagger = \nonumber
\\
&=&c_{2^{n-1}}I_{\text{res}} +H_{\sqrt{\text{res},2^{n-1}}^\prime},
\end{eqnarray} 
where $I_{\text{par(res)}}$ is the $m\times m$ identity matrix with $m=\dim ( H_{\sqrt{\text{par(res)},2^{n-1}}} )$.

In Fig.~\ref{fig:ap2fig2}, we show the energy spectrum at each level as a function of a linear variation of $\sqrt[8]{t_1}$, chosen for visual clarity reasons, starting with the root node $\sqrt[8]{\text{SSH}}$ with $N=3$ complete unit cells plus five extra sites at unit cell $N+1$. The progression shows how the eight nondegenerate edge states coming from diagonalization of $H_{\sqrt[8]{\text{SSH}}}$ [in-gap red curves in Fig.~\ref{fig:ap2fig2}(a)] end up as the eight-fold degenerate zero-energy state (in the topological regime $t_1<t_2$) coming from diagonalization of $H_{\text{anc}}$ [midgap red curve in Fig.~\ref{fig:ap2fig2}(d)]. Of these, only one corresponds to the topological state of the original perturbed SSH model. 

The spatial profile of four selected edge states of $H_{\sqrt[8]{\text{TI}}}$ are shown at the bottom of Fig.~\ref{fig:ap2fig2}. For the topological regime $t_1<t_2$, there are eight nondegenerate edge bands in Fig.~\ref{fig:ap2fig2}(a), which become four doubly degenerates edge bands in Fig.~\ref{fig:ap2fig2}(b), two four-fold degenerate edge bands in Fig.~\ref{fig:ap2fig2}(c) and, finally, a single eight-fold degenerate edge band in Fig.~\ref{fig:ap2fig2}(d). 
The energies of the edge states can be obtained recursively through
\begin{eqnarray}
E_{\text{edge}}^{\sqrt[2^n]{\text{TI}}}&=&\pm\sqrt{E_{\text{edge}}^{\sqrt[2^{n-1}]{\text{TI}}}+c_{2^{n-1}}}= 
\\
E_{\text{edge}}^{\text{TI}}&=&0.
\end{eqnarray}
These eight edge states of $\sqrt[8]{\text{SSH}}$, at any given $t_1<t_2$, can be written as linear combinations of the corresponding degenerate edge states in Fig.~\ref{fig:ap2fig2}(d), all with an equal weight of $1/8$ on the topological state of the SSH chain.
The generalization of these results for the $\sqrt[2^n]{\text{SSH}}$ model, with $n>3$, is straightforward, starting with the $n=4$ case constructed by applying the rules outlined in Section~\ref{sec:2nroottis} to the $\sqrt[8]{\text{SSH}}$ model in Fig.~\ref{fig:ap2fig1}(d).

\bibliography{sqrt1d}

\begin{thebibliography}{62}%
\makeatletter
\providecommand \@ifxundefined [1]{%
 \@ifx{#1\undefined}
}%
\providecommand \@ifnum [1]{%
 \ifnum #1\expandafter \@firstoftwo
 \else \expandafter \@secondoftwo
 \fi
}%
\providecommand \@ifx [1]{%
 \ifx #1\expandafter \@firstoftwo
 \else \expandafter \@secondoftwo
 \fi
}%
\providecommand \natexlab [1]{#1}%
\providecommand \enquote  [1]{``#1''}%
\providecommand \bibnamefont  [1]{#1}%
\providecommand \bibfnamefont [1]{#1}%
\providecommand \citenamefont [1]{#1}%
\providecommand \href@noop [0]{\@secondoftwo}%
\providecommand \href [0]{\begingroup \@sanitize@url \@href}%
\providecommand \@href[1]{\@@startlink{#1}\@@href}%
\providecommand \@@href[1]{\endgroup#1\@@endlink}%
\providecommand \@sanitize@url [0]{\catcode `\\12\catcode `\$12\catcode
  `\&12\catcode `\#12\catcode `\^12\catcode `\_12\catcode `\%12\relax}%
\providecommand \@@startlink[1]{}%
\providecommand \@@endlink[0]{}%
\providecommand \url  [0]{\begingroup\@sanitize@url \@url }%
\providecommand \@url [1]{\endgroup\@href {#1}{\urlprefix }}%
\providecommand \urlprefix  [0]{URL }%
\providecommand \Eprint [0]{\href }%
\providecommand \doibase [0]{https://doi.org/}%
\providecommand \selectlanguage [0]{\@gobble}%
\providecommand \bibinfo  [0]{\@secondoftwo}%
\providecommand \bibfield  [0]{\@secondoftwo}%
\providecommand \translation [1]{[#1]}%
\providecommand \BibitemOpen [0]{}%
\providecommand \bibitemStop [0]{}%
\providecommand \bibitemNoStop [0]{.\EOS\space}%
\providecommand \EOS [0]{\spacefactor3000\relax}%
\providecommand \BibitemShut  [1]{\csname bibitem#1\endcsname}%
\let\auto@bib@innerbib\@empty
\bibitem [{\citenamefont {Asb\'oth}\ \emph {et~al.}(2016)\citenamefont
  {Asb\'oth}, \citenamefont {Oroszl\'any},\ and\ \citenamefont
  {P\'alyi}}]{Asboth2016}%
  \BibitemOpen
  \bibfield  {author} {\bibinfo {author} {\bibfnamefont {J.~K.}\ \bibnamefont
  {Asb\'oth}}, \bibinfo {author} {\bibfnamefont {L.}~\bibnamefont
  {Oroszl\'any}},\ and\ \bibinfo {author} {\bibfnamefont {A.}~\bibnamefont
  {P\'alyi}},\ }\href@noop {} {\emph {\bibinfo {title} {A Short Course on
  Topological Insulators}}}\ (\bibinfo  {publisher} {Springer, Berlin},\
  \bibinfo {year} {2016})\BibitemShut {NoStop}%
\bibitem [{\citenamefont {Su}\ \emph {et~al.}(1979)\citenamefont {Su},
  \citenamefont {Schrieffer},\ and\ \citenamefont {Heeger}}]{Su1979}%
  \BibitemOpen
  \bibfield  {author} {\bibinfo {author} {\bibfnamefont {W.~P.}\ \bibnamefont
  {Su}}, \bibinfo {author} {\bibfnamefont {J.~R.}\ \bibnamefont {Schrieffer}},\
  and\ \bibinfo {author} {\bibfnamefont {A.~J.}\ \bibnamefont {Heeger}},\
  }\bibfield  {title} {\bibinfo {title} {Solitons in polyacetylene},\ }\href
  {https://doi.org/10.1103/PhysRevLett.42.1698} {\bibfield  {journal} {\bibinfo
   {journal} {Phys. Rev. Lett.}\ }\textbf {\bibinfo {volume} {42}},\ \bibinfo
  {pages} {1698} (\bibinfo {year} {1979})}\BibitemShut {NoStop}%
\bibitem [{\citenamefont {Hasan}\ and\ \citenamefont {Kane}(2010)}]{Hasan2010}%
  \BibitemOpen
  \bibfield  {author} {\bibinfo {author} {\bibfnamefont {M.~Z.}\ \bibnamefont
  {Hasan}}\ and\ \bibinfo {author} {\bibfnamefont {C.~L.}\ \bibnamefont
  {Kane}},\ }\bibfield  {title} {\bibinfo {title} {Colloquium: Topological
  insulators},\ }\href {https://doi.org/10.1103/RevModPhys.82.3045} {\bibfield
  {journal} {\bibinfo  {journal} {Rev. Mod. Phys.}\ }\textbf {\bibinfo {volume}
  {82}},\ \bibinfo {pages} {3045} (\bibinfo {year} {2010})}\BibitemShut
  {NoStop}%
\bibitem [{\citenamefont {Arkinstall}\ \emph {et~al.}(2017)\citenamefont
  {Arkinstall}, \citenamefont {Teimourpour}, \citenamefont {Feng},
  \citenamefont {El-Ganainy},\ and\ \citenamefont
  {Schomerus}}]{Arkinstall2017}%
  \BibitemOpen
  \bibfield  {author} {\bibinfo {author} {\bibfnamefont {J.}~\bibnamefont
  {Arkinstall}}, \bibinfo {author} {\bibfnamefont {M.~H.}\ \bibnamefont
  {Teimourpour}}, \bibinfo {author} {\bibfnamefont {L.}~\bibnamefont {Feng}},
  \bibinfo {author} {\bibfnamefont {R.}~\bibnamefont {El-Ganainy}},\ and\
  \bibinfo {author} {\bibfnamefont {H.}~\bibnamefont {Schomerus}},\ }\bibfield
  {title} {\bibinfo {title} {Topological tight-binding models from nontrivial
  square roots},\ }\href {https://doi.org/10.1103/PhysRevB.95.165109}
  {\bibfield  {journal} {\bibinfo  {journal} {Phys. Rev. B}\ }\textbf {\bibinfo
  {volume} {95}},\ \bibinfo {pages} {165109} (\bibinfo {year}
  {2017})}\BibitemShut {NoStop}%
\bibitem [{\citenamefont {Kremer}\ \emph {et~al.}(2020)\citenamefont {Kremer},
  \citenamefont {Petrides}, \citenamefont {Meyer}, \citenamefont {Heinrich},
  \citenamefont {Zilberberg},\ and\ \citenamefont {Szameit}}]{Kremer2020}%
  \BibitemOpen
  \bibfield  {author} {\bibinfo {author} {\bibfnamefont {M.}~\bibnamefont
  {Kremer}}, \bibinfo {author} {\bibfnamefont {I.}~\bibnamefont {Petrides}},
  \bibinfo {author} {\bibfnamefont {E.}~\bibnamefont {Meyer}}, \bibinfo
  {author} {\bibfnamefont {M.}~\bibnamefont {Heinrich}}, \bibinfo {author}
  {\bibfnamefont {O.}~\bibnamefont {Zilberberg}},\ and\ \bibinfo {author}
  {\bibfnamefont {A.}~\bibnamefont {Szameit}},\ }\bibfield  {title} {\bibinfo
  {title} {A square-root topological insulator with non-quantized indices
  realized with photonic aharonov-bohm cages},\ }\href
  {https://doi.org/10.1038/s41467-020-14692-4} {\bibfield  {journal} {\bibinfo
  {journal} {Nature Communications}\ }\textbf {\bibinfo {volume} {11}},\
  \bibinfo {pages} {907} (\bibinfo {year} {2020})}\BibitemShut {NoStop}%
\bibitem [{\citenamefont {Pelegr\'{\i}}\ \emph
  {et~al.}(2019{\natexlab{a}})\citenamefont {Pelegr\'{\i}}, \citenamefont
  {Marques}, \citenamefont {Dias}, \citenamefont {Daley}, \citenamefont
  {Ahufinger},\ and\ \citenamefont {Mompart}}]{Pelegri2019}%
  \BibitemOpen
  \bibfield  {author} {\bibinfo {author} {\bibfnamefont {G.}~\bibnamefont
  {Pelegr\'{\i}}}, \bibinfo {author} {\bibfnamefont {A.~M.}\ \bibnamefont
  {Marques}}, \bibinfo {author} {\bibfnamefont {R.~G.}\ \bibnamefont {Dias}},
  \bibinfo {author} {\bibfnamefont {A.~J.}\ \bibnamefont {Daley}}, \bibinfo
  {author} {\bibfnamefont {V.}~\bibnamefont {Ahufinger}},\ and\ \bibinfo
  {author} {\bibfnamefont {J.}~\bibnamefont {Mompart}},\ }\bibfield  {title}
  {\bibinfo {title} {Topological edge states with ultracold atoms carrying
  orbital angular momentum in a diamond chain},\ }\href
  {https://doi.org/10.1103/PhysRevA.99.023612} {\bibfield  {journal} {\bibinfo
  {journal} {Phys. Rev. A}\ }\textbf {\bibinfo {volume} {99}},\ \bibinfo
  {pages} {023612} (\bibinfo {year} {2019}{\natexlab{a}})}\BibitemShut
  {NoStop}%
\bibitem [{\citenamefont {Ezawa}(2020{\natexlab{a}})}]{Ezawa2020}%
  \BibitemOpen
  \bibfield  {author} {\bibinfo {author} {\bibfnamefont {M.}~\bibnamefont
  {Ezawa}},\ }\bibfield  {title} {\bibinfo {title} {Systematic construction of
  square-root topological insulators and superconductors},\ }\href
  {https://doi.org/10.1103/PhysRevResearch.2.033397} {\bibfield  {journal}
  {\bibinfo  {journal} {Phys. Rev. Research}\ }\textbf {\bibinfo {volume}
  {2}},\ \bibinfo {pages} {033397} (\bibinfo {year}
  {2020}{\natexlab{a}})}\BibitemShut {NoStop}%
\bibitem [{\citenamefont {{Ke}}\ \emph {et~al.}(2020)\citenamefont {{Ke}},
  \citenamefont {{Zhao}}, \citenamefont {{Fu}}, \citenamefont {{Liao}},
  \citenamefont {{Wang}},\ and\ \citenamefont {{Lu}}}]{Ke2020}%
  \BibitemOpen
  \bibfield  {author} {\bibinfo {author} {\bibfnamefont {S.}~\bibnamefont
  {{Ke}}}, \bibinfo {author} {\bibfnamefont {D.}~\bibnamefont {{Zhao}}},
  \bibinfo {author} {\bibfnamefont {J.}~\bibnamefont {{Fu}}}, \bibinfo {author}
  {\bibfnamefont {Q.}~\bibnamefont {{Liao}}}, \bibinfo {author} {\bibfnamefont
  {B.}~\bibnamefont {{Wang}}},\ and\ \bibinfo {author} {\bibfnamefont
  {P.}~\bibnamefont {{Lu}}},\ }\bibfield  {title} {\bibinfo {title}
  {Topological edge modes in non-hermitian photonic aharonov-bohm cages},\
  }\href {https://doi.org/10.1109/JSTQE.2020.3010586} {\bibfield  {journal}
  {\bibinfo  {journal} {IEEE Journal of Selected Topics in Quantum
  Electronics}\ }\textbf {\bibinfo {volume} {26}},\ \bibinfo {pages} {1}
  (\bibinfo {year} {2020})}\BibitemShut {NoStop}%
\bibitem [{\citenamefont {Lin}\ \emph {et~al.}(2021)\citenamefont {Lin},
  \citenamefont {Ke}, \citenamefont {Zhu},\ and\ \citenamefont {Li}}]{Lin2021}%
  \BibitemOpen
  \bibfield  {author} {\bibinfo {author} {\bibfnamefont {Z.}~\bibnamefont
  {Lin}}, \bibinfo {author} {\bibfnamefont {S.}~\bibnamefont {Ke}}, \bibinfo
  {author} {\bibfnamefont {X.}~\bibnamefont {Zhu}},\ and\ \bibinfo {author}
  {\bibfnamefont {X.}~\bibnamefont {Li}},\ }\bibfield  {title} {\bibinfo
  {title} {Square-root non-bloch topological insulators in non-hermitian ring
  resonators},\ }\href {https://doi.org/10.1364/OE.419852} {\bibfield
  {journal} {\bibinfo  {journal} {Opt. Express}\ }\textbf {\bibinfo {volume}
  {29}},\ \bibinfo {pages} {8462} (\bibinfo {year} {2021})}\BibitemShut
  {NoStop}%
\bibitem [{\citenamefont {Song}\ \emph {et~al.}(2020)\citenamefont {Song},
  \citenamefont {Yang}, \citenamefont {Cao},\ and\ \citenamefont
  {Yan}}]{Song2020}%
  \BibitemOpen
  \bibfield  {author} {\bibinfo {author} {\bibfnamefont {L.}~\bibnamefont
  {Song}}, \bibinfo {author} {\bibfnamefont {H.}~\bibnamefont {Yang}}, \bibinfo
  {author} {\bibfnamefont {Y.}~\bibnamefont {Cao}},\ and\ \bibinfo {author}
  {\bibfnamefont {P.}~\bibnamefont {Yan}},\ }\bibfield  {title} {\bibinfo
  {title} {Realization of the square-root higher-order topological insulator in
  electric circuits},\ }\href {https://doi.org/10.1021/acs.nanolett.0c03049}
  {\bibfield  {journal} {\bibinfo  {journal} {Nano Lett.}\ }\textbf {\bibinfo
  {volume} {20}},\ \bibinfo {pages} {7566} (\bibinfo {year}
  {2020})}\BibitemShut {NoStop}%
\bibitem [{\citenamefont {Mizoguchi}\ \emph {et~al.}(2020)\citenamefont
  {Mizoguchi}, \citenamefont {Kuno},\ and\ \citenamefont
  {Hatsugai}}]{Mizoguchi2020}%
  \BibitemOpen
  \bibfield  {author} {\bibinfo {author} {\bibfnamefont {T.}~\bibnamefont
  {Mizoguchi}}, \bibinfo {author} {\bibfnamefont {Y.}~\bibnamefont {Kuno}},\
  and\ \bibinfo {author} {\bibfnamefont {Y.}~\bibnamefont {Hatsugai}},\
  }\bibfield  {title} {\bibinfo {title} {Square-root higher-order topological
  insulator on a decorated honeycomb lattice},\ }\href
  {https://doi.org/10.1103/PhysRevA.102.033527} {\bibfield  {journal} {\bibinfo
   {journal} {Phys. Rev. A}\ }\textbf {\bibinfo {volume} {102}},\ \bibinfo
  {pages} {033527} (\bibinfo {year} {2020})}\BibitemShut {NoStop}%
\bibitem [{\citenamefont {Yan}\ \emph {et~al.}(2020)\citenamefont {Yan},
  \citenamefont {Huang}, \citenamefont {Luo}, \citenamefont {Lu}, \citenamefont
  {Deng},\ and\ \citenamefont {Liu}}]{Yan2020}%
  \BibitemOpen
  \bibfield  {author} {\bibinfo {author} {\bibfnamefont {M.}~\bibnamefont
  {Yan}}, \bibinfo {author} {\bibfnamefont {X.}~\bibnamefont {Huang}}, \bibinfo
  {author} {\bibfnamefont {L.}~\bibnamefont {Luo}}, \bibinfo {author}
  {\bibfnamefont {J.}~\bibnamefont {Lu}}, \bibinfo {author} {\bibfnamefont
  {W.}~\bibnamefont {Deng}},\ and\ \bibinfo {author} {\bibfnamefont
  {Z.}~\bibnamefont {Liu}},\ }\bibfield  {title} {\bibinfo {title} {Acoustic
  square-root topological states},\ }\href
  {https://doi.org/10.1103/PhysRevB.102.180102} {\bibfield  {journal} {\bibinfo
   {journal} {Phys. Rev. B}\ }\textbf {\bibinfo {volume} {102}},\ \bibinfo
  {pages} {180102(R)} (\bibinfo {year} {2020})}\BibitemShut {NoStop}%
\bibitem [{\citenamefont {Benalcazar}\ \emph
  {et~al.}(2017{\natexlab{a}})\citenamefont {Benalcazar}, \citenamefont
  {Bernevig},\ and\ \citenamefont {Hughes}}]{Benalcazar2017}%
  \BibitemOpen
  \bibfield  {author} {\bibinfo {author} {\bibfnamefont {W.~A.}\ \bibnamefont
  {Benalcazar}}, \bibinfo {author} {\bibfnamefont {B.~A.}\ \bibnamefont
  {Bernevig}},\ and\ \bibinfo {author} {\bibfnamefont {T.~L.}\ \bibnamefont
  {Hughes}},\ }\bibfield  {title} {\bibinfo {title} {Quantized electric
  multipole insulators},\ }\href {https://doi.org/10.1126/science.aah6442}
  {\bibfield  {journal} {\bibinfo  {journal} {Science}\ }\textbf {\bibinfo
  {volume} {357}},\ \bibinfo {pages} {61} (\bibinfo {year}
  {2017}{\natexlab{a}})}\BibitemShut {NoStop}%
\bibitem [{\citenamefont {Benalcazar}\ \emph
  {et~al.}(2017{\natexlab{b}})\citenamefont {Benalcazar}, \citenamefont
  {Bernevig},\ and\ \citenamefont {Hughes}}]{Benalcazar2017b}%
  \BibitemOpen
  \bibfield  {author} {\bibinfo {author} {\bibfnamefont {W.~A.}\ \bibnamefont
  {Benalcazar}}, \bibinfo {author} {\bibfnamefont {B.~A.}\ \bibnamefont
  {Bernevig}},\ and\ \bibinfo {author} {\bibfnamefont {T.~L.}\ \bibnamefont
  {Hughes}},\ }\bibfield  {title} {\bibinfo {title} {Electric multipole
  moments, topological multipole moment pumping, and chiral hinge states in
  crystalline insulators},\ }\href {https://doi.org/10.1103/PhysRevB.96.245115}
  {\bibfield  {journal} {\bibinfo  {journal} {Phys. Rev. B}\ }\textbf {\bibinfo
  {volume} {96}},\ \bibinfo {pages} {245115} (\bibinfo {year}
  {2017}{\natexlab{b}})}\BibitemShut {NoStop}%
\bibitem [{\citenamefont {Pelegr\'{\i}}\ \emph
  {et~al.}(2019{\natexlab{b}})\citenamefont {Pelegr\'{\i}}, \citenamefont
  {Marques}, \citenamefont {Ahufinger}, \citenamefont {Mompart},\ and\
  \citenamefont {Dias}}]{Pelegri2019c}%
  \BibitemOpen
  \bibfield  {author} {\bibinfo {author} {\bibfnamefont {G.}~\bibnamefont
  {Pelegr\'{\i}}}, \bibinfo {author} {\bibfnamefont {A.~M.}\ \bibnamefont
  {Marques}}, \bibinfo {author} {\bibfnamefont {V.}~\bibnamefont {Ahufinger}},
  \bibinfo {author} {\bibfnamefont {J.}~\bibnamefont {Mompart}},\ and\ \bibinfo
  {author} {\bibfnamefont {R.~G.}\ \bibnamefont {Dias}},\ }\bibfield  {title}
  {\bibinfo {title} {Second-order topological corner states with ultracold
  atoms carrying orbital angular momentum in optical lattices},\ }\href
  {https://doi.org/10.1103/PhysRevB.100.205109} {\bibfield  {journal} {\bibinfo
   {journal} {Phys. Rev. B}\ }\textbf {\bibinfo {volume} {100}},\ \bibinfo
  {pages} {205109} (\bibinfo {year} {2019}{\natexlab{b}})}\BibitemShut
  {NoStop}%
\bibitem [{\citenamefont {Mizoguchi}\ \emph {et~al.}(2021)\citenamefont
  {Mizoguchi}, \citenamefont {Yoshida},\ and\ \citenamefont
  {Hatsugai}}]{Mizoguchi2021}%
  \BibitemOpen
  \bibfield  {author} {\bibinfo {author} {\bibfnamefont {T.}~\bibnamefont
  {Mizoguchi}}, \bibinfo {author} {\bibfnamefont {T.}~\bibnamefont {Yoshida}},\
  and\ \bibinfo {author} {\bibfnamefont {Y.}~\bibnamefont {Hatsugai}},\ }\href
  {https://doi.org/10.1103/PhysRevB.103.045136} {\bibfield  {journal} {\bibinfo
   {journal} {Phys. Rev. B}\ }\textbf {\bibinfo {volume} {103}},\ \bibinfo
  {pages} {045136} (\bibinfo {year} {2021})}\BibitemShut {NoStop}%
\bibitem [{\citenamefont {Dias}\ and\ \citenamefont
  {Marques}(2021)}]{Dias2021}%
  \BibitemOpen
  \bibfield  {author} {\bibinfo {author} {\bibfnamefont {R.~G.}\ \bibnamefont
  {Dias}}\ and\ \bibinfo {author} {\bibfnamefont {A.~M.}\ \bibnamefont
  {Marques}},\ }\bibfield  {title} {\bibinfo {title} {Matryoshka approach to
  sine-cosine topological models},\ }\href
  {https://doi.org/10.1103/PhysRevB.103.245112} {\bibfield  {journal} {\bibinfo
   {journal} {Phys. Rev. B}\ }\textbf {\bibinfo {volume} {103}},\ \bibinfo
  {pages} {245112} (\bibinfo {year} {2021})}\BibitemShut {NoStop}%
\bibitem [{\citenamefont {Kitaev}(2001)}]{Kitaev2001}%
  \BibitemOpen
  \bibfield  {author} {\bibinfo {author} {\bibfnamefont {A.~Y.}\ \bibnamefont
  {Kitaev}},\ }\bibfield  {title} {\bibinfo {title} {Unpaired majorana fermions
  in quantum wires},\ }\href {https://doi.org/10.1070/1063-7869/44/10s/s29}
  {\bibfield  {journal} {\bibinfo  {journal} {Physics-Uspekhi}\ }\textbf
  {\bibinfo {volume} {44}},\ \bibinfo {pages} {131} (\bibinfo {year}
  {2001})}\BibitemShut {NoStop}%
\bibitem [{\citenamefont {Pelegr\'{\i}}\ \emph {et~al.}(2020)\citenamefont
  {Pelegr\'{\i}}, \citenamefont {Marques}, \citenamefont {Ahufinger},
  \citenamefont {Mompart},\ and\ \citenamefont {Dias}}]{Pelegri2020}%
  \BibitemOpen
  \bibfield  {author} {\bibinfo {author} {\bibfnamefont {G.}~\bibnamefont
  {Pelegr\'{\i}}}, \bibinfo {author} {\bibfnamefont {A.~M.}\ \bibnamefont
  {Marques}}, \bibinfo {author} {\bibfnamefont {V.}~\bibnamefont {Ahufinger}},
  \bibinfo {author} {\bibfnamefont {J.}~\bibnamefont {Mompart}},\ and\ \bibinfo
  {author} {\bibfnamefont {R.~G.}\ \bibnamefont {Dias}},\ }\bibfield  {title}
  {\bibinfo {title} {Interaction-induced topological properties of two bosons
  in flat-band systems},\ }\href
  {https://doi.org/10.1103/PhysRevResearch.2.033267} {\bibfield  {journal}
  {\bibinfo  {journal} {Phys. Rev. Research}\ }\textbf {\bibinfo {volume}
  {2}},\ \bibinfo {pages} {033267} (\bibinfo {year} {2020})}\BibitemShut
  {NoStop}%
\bibitem [{\citenamefont {Di~Liberto}\ \emph {et~al.}(2019)\citenamefont
  {Di~Liberto}, \citenamefont {Mukherjee},\ and\ \citenamefont
  {Goldman}}]{Liberto2019}%
  \BibitemOpen
  \bibfield  {author} {\bibinfo {author} {\bibfnamefont {M.}~\bibnamefont
  {Di~Liberto}}, \bibinfo {author} {\bibfnamefont {S.}~\bibnamefont
  {Mukherjee}},\ and\ \bibinfo {author} {\bibfnamefont {N.}~\bibnamefont
  {Goldman}},\ }\bibfield  {title} {\bibinfo {title} {Nonlinear dynamics of
  aharonov-bohm cages},\ }\href {https://doi.org/10.1103/PhysRevA.100.043829}
  {\bibfield  {journal} {\bibinfo  {journal} {Phys. Rev. A}\ }\textbf {\bibinfo
  {volume} {100}},\ \bibinfo {pages} {043829} (\bibinfo {year}
  {2019})}\BibitemShut {NoStop}%
\bibitem [{\citenamefont {Pelegr\'{\i}}\ \emph
  {et~al.}(2019{\natexlab{c}})\citenamefont {Pelegr\'{\i}}, \citenamefont
  {Marques}, \citenamefont {Dias}, \citenamefont {Daley}, \citenamefont
  {Mompart},\ and\ \citenamefont {Ahufinger}}]{Pelegri2019b}%
  \BibitemOpen
  \bibfield  {author} {\bibinfo {author} {\bibfnamefont {G.}~\bibnamefont
  {Pelegr\'{\i}}}, \bibinfo {author} {\bibfnamefont {A.~M.}\ \bibnamefont
  {Marques}}, \bibinfo {author} {\bibfnamefont {R.~G.}\ \bibnamefont {Dias}},
  \bibinfo {author} {\bibfnamefont {A.~J.}\ \bibnamefont {Daley}}, \bibinfo
  {author} {\bibfnamefont {J.}~\bibnamefont {Mompart}},\ and\ \bibinfo {author}
  {\bibfnamefont {V.}~\bibnamefont {Ahufinger}},\ }\bibfield  {title} {\bibinfo
  {title} {Topological edge states and aharanov-bohm caging with ultracold
  atoms carrying orbital angular momentum},\ }\href
  {https://doi.org/10.1103/PhysRevA.99.023613} {\bibfield  {journal} {\bibinfo
  {journal} {Phys. Rev. A}\ }\textbf {\bibinfo {volume} {99}},\ \bibinfo
  {pages} {023613} (\bibinfo {year} {2019}{\natexlab{c}})}\BibitemShut
  {NoStop}%
\bibitem [{\citenamefont {Gligori\ifmmode~\acute{c}\else \'{c}\fi{}}\ \emph
  {et~al.}(2020)\citenamefont {Gligori\ifmmode~\acute{c}\else \'{c}\fi{}},
  \citenamefont {Leykam},\ and\ \citenamefont {Maluckov}}]{Gligoric2020}%
  \BibitemOpen
  \bibfield  {author} {\bibinfo {author} {\bibfnamefont {G.}~\bibnamefont
  {Gligori\ifmmode~\acute{c}\else \'{c}\fi{}}}, \bibinfo {author}
  {\bibfnamefont {D.}~\bibnamefont {Leykam}},\ and\ \bibinfo {author}
  {\bibfnamefont {A.}~\bibnamefont {Maluckov}},\ }\bibfield  {title} {\bibinfo
  {title} {Influence of different disorder types on aharonov-bohm caging in the
  diamond chain},\ }\href {https://doi.org/10.1103/PhysRevA.101.023839}
  {\bibfield  {journal} {\bibinfo  {journal} {Phys. Rev. A}\ }\textbf {\bibinfo
  {volume} {101}},\ \bibinfo {pages} {023839} (\bibinfo {year}
  {2020})}\BibitemShut {NoStop}%
\bibitem [{\citenamefont {Chang}\ \emph {et~al.}(2021)\citenamefont {Chang},
  \citenamefont {Gundogdu}, \citenamefont {Leykam}, \citenamefont {Angelakis},
  \citenamefont {Kou}, \citenamefont {Flach},\ and\ \citenamefont
  {Maluckov}}]{Chang2021}%
  \BibitemOpen
  \bibfield  {author} {\bibinfo {author} {\bibfnamefont {N.}~\bibnamefont
  {Chang}}, \bibinfo {author} {\bibfnamefont {S.}~\bibnamefont {Gundogdu}},
  \bibinfo {author} {\bibfnamefont {D.}~\bibnamefont {Leykam}}, \bibinfo
  {author} {\bibfnamefont {D.~G.}\ \bibnamefont {Angelakis}}, \bibinfo {author}
  {\bibfnamefont {S.}~\bibnamefont {Kou}}, \bibinfo {author} {\bibfnamefont
  {S.}~\bibnamefont {Flach}},\ and\ \bibinfo {author} {\bibfnamefont
  {A.}~\bibnamefont {Maluckov}},\ }\bibfield  {title} {\bibinfo {title}
  {Nonlinear bloch wave dynamics in photonic aharonov-bohm cages},\ }\href
  {https://doi.org/10.1063/5.0037767} {\bibfield  {journal} {\bibinfo
  {journal} {APL Photonics}\ }\textbf {\bibinfo {volume} {6}},\ \bibinfo
  {pages} {030801} (\bibinfo {year} {2021})}\BibitemShut {NoStop}%
\bibitem [{\citenamefont {Marques}\ and\ \citenamefont
  {Dias}(2018)}]{Marques2018}%
  \BibitemOpen
  \bibfield  {author} {\bibinfo {author} {\bibfnamefont {A.~M.}\ \bibnamefont
  {Marques}}\ and\ \bibinfo {author} {\bibfnamefont {R.~G.}\ \bibnamefont
  {Dias}},\ }\bibfield  {title} {\bibinfo {title} {Topological bound states in
  interacting su{\textendash}schrieffer{\textendash}heeger rings},\ }\href
  {https://doi.org/10.1088/1361-648x/aacd7c} {\bibfield  {journal} {\bibinfo
  {journal} {Journal of Physics: Condensed Matter}\ }\textbf {\bibinfo {volume}
  {30}},\ \bibinfo {pages} {305601} (\bibinfo {year} {2018})}\BibitemShut
  {NoStop}%
\bibitem [{\citenamefont {Madail}\ \emph {et~al.}(2019)\citenamefont {Madail},
  \citenamefont {Flannigan}, \citenamefont {Marques}, \citenamefont {Daley},\
  and\ \citenamefont {Dias}}]{Madail2019}%
  \BibitemOpen
  \bibfield  {author} {\bibinfo {author} {\bibfnamefont {L.}~\bibnamefont
  {Madail}}, \bibinfo {author} {\bibfnamefont {S.}~\bibnamefont {Flannigan}},
  \bibinfo {author} {\bibfnamefont {A.~M.}\ \bibnamefont {Marques}}, \bibinfo
  {author} {\bibfnamefont {A.~J.}\ \bibnamefont {Daley}},\ and\ \bibinfo
  {author} {\bibfnamefont {R.~G.}\ \bibnamefont {Dias}},\ }\bibfield  {title}
  {\bibinfo {title} {Enhanced localization and protection of topological edge
  states due to geometric frustration},\ }\href
  {https://doi.org/10.1103/PhysRevB.100.125123} {\bibfield  {journal} {\bibinfo
   {journal} {Phys. Rev. B}\ }\textbf {\bibinfo {volume} {100}},\ \bibinfo
  {pages} {125123} (\bibinfo {year} {2019})}\BibitemShut {NoStop}%
\bibitem [{\citenamefont {Marques}\ and\ \citenamefont
  {Dias}(2019)}]{Marques2019}%
  \BibitemOpen
  \bibfield  {author} {\bibinfo {author} {\bibfnamefont {A.~M.}\ \bibnamefont
  {Marques}}\ and\ \bibinfo {author} {\bibfnamefont {R.~G.}\ \bibnamefont
  {Dias}},\ }\bibfield  {title} {\bibinfo {title} {One-dimensional topological
  insulators with noncentered inversion symmetry axis},\ }\href
  {https://doi.org/10.1103/PhysRevB.100.041104} {\bibfield  {journal} {\bibinfo
   {journal} {Phys. Rev. B}\ }\textbf {\bibinfo {volume} {100}},\ \bibinfo
  {pages} {041104(R)} (\bibinfo {year} {2019})}\BibitemShut {NoStop}%
\bibitem [{\citenamefont {Creutz}(1999)}]{Creutz1999}%
  \BibitemOpen
  \bibfield  {author} {\bibinfo {author} {\bibfnamefont {M.}~\bibnamefont
  {Creutz}},\ }\bibfield  {title} {\bibinfo {title} {End states, ladder
  compounds, and domain-wall fermions},\ }\href
  {https://doi.org/10.1103/PhysRevLett.83.2636} {\bibfield  {journal} {\bibinfo
   {journal} {Phys. Rev. Lett.}\ }\textbf {\bibinfo {volume} {83}},\ \bibinfo
  {pages} {2636} (\bibinfo {year} {1999})}\BibitemShut {NoStop}%
\bibitem [{\citenamefont {Creutz}(2001)}]{Creutz2001}%
  \BibitemOpen
  \bibfield  {author} {\bibinfo {author} {\bibfnamefont {M.}~\bibnamefont
  {Creutz}},\ }\bibfield  {title} {\bibinfo {title} {Aspects of chiral symmetry
  and the lattice},\ }\href {https://doi.org/10.1103/RevModPhys.73.119}
  {\bibfield  {journal} {\bibinfo  {journal} {Rev. Mod. Phys.}\ }\textbf
  {\bibinfo {volume} {73}},\ \bibinfo {pages} {119} (\bibinfo {year}
  {2001})}\BibitemShut {NoStop}%
\bibitem [{\citenamefont {Zurita}\ \emph {et~al.}(2020)\citenamefont {Zurita},
  \citenamefont {Creffield},\ and\ \citenamefont {Platero}}]{Zurita2020}%
  \BibitemOpen
  \bibfield  {author} {\bibinfo {author} {\bibfnamefont {J.}~\bibnamefont
  {Zurita}}, \bibinfo {author} {\bibfnamefont {C.~E.}\ \bibnamefont
  {Creffield}},\ and\ \bibinfo {author} {\bibfnamefont {G.}~\bibnamefont
  {Platero}},\ }\bibfield  {title} {\bibinfo {title} {Topology and interactions
  in the photonic creutz and creutz-hubbard ladders},\ }\href
  {https://doi.org/https://doi.org/10.1002/qute.201900105} {\bibfield
  {journal} {\bibinfo  {journal} {Adv. Quantum Technol.}\ }\textbf {\bibinfo
  {volume} {3}},\ \bibinfo {pages} {1900105} (\bibinfo {year}
  {2020})}\BibitemShut {NoStop}%
\bibitem [{\citenamefont {Flannigan}\ and\ \citenamefont
  {Daley}(2020)}]{Flannigan2020}%
  \BibitemOpen
  \bibfield  {author} {\bibinfo {author} {\bibfnamefont {S.}~\bibnamefont
  {Flannigan}}\ and\ \bibinfo {author} {\bibfnamefont {A.~J.}\ \bibnamefont
  {Daley}},\ }\bibfield  {title} {\bibinfo {title} {Enhanced repulsively bound
  atom pairs in topological optical lattice ladders},\ }\href
  {https://doi.org/10.1088/2058-9565/abb028} {\bibfield  {journal} {\bibinfo
  {journal} {Quantum Science and Technology}\ }\textbf {\bibinfo {volume}
  {5}},\ \bibinfo {pages} {045017} (\bibinfo {year} {2020})}\BibitemShut
  {NoStop}%
\bibitem [{\citenamefont {Kuno}(2020)}]{Kuno2020}%
  \BibitemOpen
  \bibfield  {author} {\bibinfo {author} {\bibfnamefont {Y.}~\bibnamefont
  {Kuno}},\ }\bibfield  {title} {\bibinfo {title} {Extended flat band,
  entanglement, and topological properties in a creutz ladder},\ }\href
  {https://doi.org/10.1103/PhysRevB.101.184112} {\bibfield  {journal} {\bibinfo
   {journal} {Phys. Rev. B}\ }\textbf {\bibinfo {volume} {101}},\ \bibinfo
  {pages} {184112} (\bibinfo {year} {2020})}\BibitemShut {NoStop}%
\bibitem [{\citenamefont {Kuno}\ \emph {et~al.}(2020)\citenamefont {Kuno},
  \citenamefont {Mizoguchi},\ and\ \citenamefont {Hatsugai}}]{Kuno2020b}%
  \BibitemOpen
  \bibfield  {author} {\bibinfo {author} {\bibfnamefont {Y.}~\bibnamefont
  {Kuno}}, \bibinfo {author} {\bibfnamefont {T.}~\bibnamefont {Mizoguchi}},\
  and\ \bibinfo {author} {\bibfnamefont {Y.}~\bibnamefont {Hatsugai}},\
  }\bibfield  {title} {\bibinfo {title} {Interaction-induced doublons and
  embedded topological subspace in a complete flat-band system},\ }\href
  {https://doi.org/10.1103/PhysRevA.102.063325} {\bibfield  {journal} {\bibinfo
   {journal} {Phys. Rev. A}\ }\textbf {\bibinfo {volume} {102}},\ \bibinfo
  {pages} {063325} (\bibinfo {year} {2020})}\BibitemShut {NoStop}%
\bibitem [{\citenamefont {Marques}\ and\ \citenamefont
  {Dias}(2017)}]{Marques2017}%
  \BibitemOpen
  \bibfield  {author} {\bibinfo {author} {\bibfnamefont {A.~M.}\ \bibnamefont
  {Marques}}\ and\ \bibinfo {author} {\bibfnamefont {R.~G.}\ \bibnamefont
  {Dias}},\ }\bibfield  {title} {\bibinfo {title} {Multihole edge states in
  su-schrieffer-heeger chains with interactions},\ }\href
  {https://doi.org/10.1103/PhysRevB.95.115443} {\bibfield  {journal} {\bibinfo
  {journal} {Phys. Rev. B}\ }\textbf {\bibinfo {volume} {95}},\ \bibinfo
  {pages} {115443} (\bibinfo {year} {2017})}\BibitemShut {NoStop}%
\bibitem [{\citenamefont {Ma}\ \emph {et~al.}(2020)\citenamefont {Ma},
  \citenamefont {Xu}, \citenamefont {Chiu}, \citenamefont {Regnault},
  \citenamefont {Houck}, \citenamefont {Song},\ and\ \citenamefont
  {Bernevig}}]{Ma2020}%
  \BibitemOpen
  \bibfield  {author} {\bibinfo {author} {\bibfnamefont {D.-S.}\ \bibnamefont
  {Ma}}, \bibinfo {author} {\bibfnamefont {Y.}~\bibnamefont {Xu}}, \bibinfo
  {author} {\bibfnamefont {C.~S.}\ \bibnamefont {Chiu}}, \bibinfo {author}
  {\bibfnamefont {N.}~\bibnamefont {Regnault}}, \bibinfo {author}
  {\bibfnamefont {A.~A.}\ \bibnamefont {Houck}}, \bibinfo {author}
  {\bibfnamefont {Z.}~\bibnamefont {Song}},\ and\ \bibinfo {author}
  {\bibfnamefont {B.~A.}\ \bibnamefont {Bernevig}},\ }\bibfield  {title}
  {\bibinfo {title} {Spin-orbit-induced topological flat bands in line and
  split graphs of bipartite lattices},\ }\href
  {https://doi.org/10.1103/PhysRevLett.125.266403} {\bibfield  {journal}
  {\bibinfo  {journal} {Phys. Rev. Lett.}\ }\textbf {\bibinfo {volume} {125}},\
  \bibinfo {pages} {266403} (\bibinfo {year} {2020})}\BibitemShut {NoStop}%
\bibitem [{Note1()}]{Note1}%
  \BibitemOpen
  \bibinfo {note} {An intermediate way that combines the distinguishing aspects
  of both fundamental methods (renormalized hopping terms and extra added
  sites) is also possible.}\BibitemShut {Stop}%
\bibitem [{\citenamefont {Lieb}(1989)}]{Lieb1989}%
  \BibitemOpen
  \bibfield  {author} {\bibinfo {author} {\bibfnamefont {E.~H.}\ \bibnamefont
  {Lieb}},\ }\bibfield  {title} {\bibinfo {title} {Two theorems on the hubbard
  model},\ }\href {https://doi.org/10.1103/PhysRevLett.62.1201} {\bibfield
  {journal} {\bibinfo  {journal} {Phys. Rev. Lett.}\ }\textbf {\bibinfo
  {volume} {62}},\ \bibinfo {pages} {1201} (\bibinfo {year}
  {1989})}\BibitemShut {NoStop}%
\bibitem [{\citenamefont {Deo}(2017)}]{Deo2017}%
  \BibitemOpen
  \bibfield  {author} {\bibinfo {author} {\bibfnamefont {N.}~\bibnamefont
  {Deo}},\ }\href {https://books.google.pt/books?id=DSBMDgAAQBAJ} {\emph
  {\bibinfo {title} {Graph Theory with Applications to Engineering and Computer
  Science}}}\ (\bibinfo  {publisher} {Dover Publications, London},\ \bibinfo
  {year} {2017})\BibitemShut {NoStop}%
\bibitem [{\citenamefont {Danieli}\ \emph
  {et~al.}(2020{\natexlab{a}})\citenamefont {Danieli}, \citenamefont
  {Andreanov}, \citenamefont {Mithun},\ and\ \citenamefont
  {Flach}}]{Danieli2020}%
  \BibitemOpen
  \bibfield  {author} {\bibinfo {author} {\bibfnamefont {C.}~\bibnamefont
  {Danieli}}, \bibinfo {author} {\bibfnamefont {A.}~\bibnamefont {Andreanov}},
  \bibinfo {author} {\bibfnamefont {T.}~\bibnamefont {Mithun}},\ and\ \bibinfo
  {author} {\bibfnamefont {S.}~\bibnamefont {Flach}},\ }\href@noop {} {\bibinfo
  {title} {Nonlinear caging in all-bands-flat lattices}} (\bibinfo {year}
  {2020}{\natexlab{a}}),\ \Eprint {https://arxiv.org/abs/2004.11871}
  {arXiv:2004.11871 [cond-mat.quant-gas]} \BibitemShut {NoStop}%
\bibitem [{\citenamefont {Danieli}\ \emph
  {et~al.}(2020{\natexlab{b}})\citenamefont {Danieli}, \citenamefont
  {Andreanov}, \citenamefont {Mithun},\ and\ \citenamefont
  {Flach}}]{Danieli2020b}%
  \BibitemOpen
  \bibfield  {author} {\bibinfo {author} {\bibfnamefont {C.}~\bibnamefont
  {Danieli}}, \bibinfo {author} {\bibfnamefont {A.}~\bibnamefont {Andreanov}},
  \bibinfo {author} {\bibfnamefont {T.}~\bibnamefont {Mithun}},\ and\ \bibinfo
  {author} {\bibfnamefont {S.}~\bibnamefont {Flach}},\ }\href@noop {} {\bibinfo
  {title} {Quantum caging in interacting many-body all-bands-flat lattices}}
  (\bibinfo {year} {2020}{\natexlab{b}}),\ \Eprint
  {https://arxiv.org/abs/2004.11880} {arXiv:2004.11880 [cond-mat.quant-gas]}
  \BibitemShut {NoStop}%
\bibitem [{\citenamefont {Ichinose}\ \emph {et~al.}(2021)\citenamefont
  {Ichinose}, \citenamefont {Orito},\ and\ \citenamefont
  {Kuno}}]{Ichinose2021}%
  \BibitemOpen
  \bibfield  {author} {\bibinfo {author} {\bibfnamefont {I.}~\bibnamefont
  {Ichinose}}, \bibinfo {author} {\bibfnamefont {T.}~\bibnamefont {Orito}},\
  and\ \bibinfo {author} {\bibfnamefont {Y.}~\bibnamefont {Kuno}},\ }\href@noop
  {} {\bibinfo {title} {Flat-band full localization and symmetry-protected
  topological phase on bilayer lattice systems}} (\bibinfo {year} {2021}),\
  \Eprint {https://arxiv.org/abs/2102.10986} {arXiv:2102.10986
  [cond-mat.dis-nn]} \BibitemShut {NoStop}%
\bibitem [{\citenamefont {Li}\ \emph {et~al.}(2020{\natexlab{a}})\citenamefont
  {Li}, \citenamefont {Xue}, \citenamefont {Gong},\ and\ \citenamefont
  {Hu}}]{Li2020}%
  \BibitemOpen
  \bibfield  {author} {\bibinfo {author} {\bibfnamefont {S.}~\bibnamefont
  {Li}}, \bibinfo {author} {\bibfnamefont {Z.-Y.}\ \bibnamefont {Xue}},
  \bibinfo {author} {\bibfnamefont {M.}~\bibnamefont {Gong}},\ and\ \bibinfo
  {author} {\bibfnamefont {Y.}~\bibnamefont {Hu}},\ }\bibfield  {title}
  {\bibinfo {title} {Non-abelian aharonov-bohm caging in photonic lattices},\
  }\href {https://doi.org/10.1103/PhysRevA.102.023524} {\bibfield  {journal}
  {\bibinfo  {journal} {Phys. Rev. A}\ }\textbf {\bibinfo {volume} {102}},\
  \bibinfo {pages} {023524} (\bibinfo {year} {2020}{\natexlab{a}})}\BibitemShut
  {NoStop}%
\bibitem [{\citenamefont {Mukherjee}\ \emph {et~al.}(2020)\citenamefont
  {Mukherjee}, \citenamefont {Nandy}, \citenamefont {Sil},\ and\ \citenamefont
  {Chakrabarti}}]{Mukherjee2020}%
  \BibitemOpen
  \bibfield  {author} {\bibinfo {author} {\bibfnamefont {A.}~\bibnamefont
  {Mukherjee}}, \bibinfo {author} {\bibfnamefont {A.}~\bibnamefont {Nandy}},
  \bibinfo {author} {\bibfnamefont {S.}~\bibnamefont {Sil}},\ and\ \bibinfo
  {author} {\bibfnamefont {A.}~\bibnamefont {Chakrabarti}},\ }\bibfield
  {title} {\bibinfo {title} {Engineering topological phase transition and
  aharonov{\textendash}bohm caging in a flux-staggered lattice},\ }\href
  {https://doi.org/10.1088/1361-648x/abbc9a} {\bibfield  {journal} {\bibinfo
  {journal} {Journal of Physics: Condensed Matter}\ }\textbf {\bibinfo {volume}
  {33}},\ \bibinfo {pages} {035502} (\bibinfo {year} {2020})}\BibitemShut
  {NoStop}%
\bibitem [{\citenamefont {Baboux}\ \emph {et~al.}(2016)\citenamefont {Baboux},
  \citenamefont {Ge}, \citenamefont {Jacqmin}, \citenamefont {Biondi},
  \citenamefont {Galopin}, \citenamefont {Lema\^{\i}tre}, \citenamefont
  {Le~Gratiet}, \citenamefont {Sagnes}, \citenamefont {Schmidt}, \citenamefont
  {T\"ureci}, \citenamefont {Amo},\ and\ \citenamefont {Bloch}}]{Baboux2016}%
  \BibitemOpen
  \bibfield  {author} {\bibinfo {author} {\bibfnamefont {F.}~\bibnamefont
  {Baboux}}, \bibinfo {author} {\bibfnamefont {L.}~\bibnamefont {Ge}}, \bibinfo
  {author} {\bibfnamefont {T.}~\bibnamefont {Jacqmin}}, \bibinfo {author}
  {\bibfnamefont {M.}~\bibnamefont {Biondi}}, \bibinfo {author} {\bibfnamefont
  {E.}~\bibnamefont {Galopin}}, \bibinfo {author} {\bibfnamefont
  {A.}~\bibnamefont {Lema\^{\i}tre}}, \bibinfo {author} {\bibfnamefont
  {L.}~\bibnamefont {Le~Gratiet}}, \bibinfo {author} {\bibfnamefont
  {I.}~\bibnamefont {Sagnes}}, \bibinfo {author} {\bibfnamefont
  {S.}~\bibnamefont {Schmidt}}, \bibinfo {author} {\bibfnamefont {H.~E.}\
  \bibnamefont {T\"ureci}}, \bibinfo {author} {\bibfnamefont {A.}~\bibnamefont
  {Amo}},\ and\ \bibinfo {author} {\bibfnamefont {J.}~\bibnamefont {Bloch}},\
  }\bibfield  {title} {\bibinfo {title} {Bosonic condensation and
  disorder-induced localization in a flat band},\ }\href
  {https://doi.org/10.1103/PhysRevLett.116.066402} {\bibfield  {journal}
  {\bibinfo  {journal} {Phys. Rev. Lett.}\ }\textbf {\bibinfo {volume} {116}},\
  \bibinfo {pages} {066402} (\bibinfo {year} {2016})}\BibitemShut {NoStop}%
\bibitem [{\citenamefont {Mukherjee}\ \emph {et~al.}(2018)\citenamefont
  {Mukherjee}, \citenamefont {Di~Liberto}, \citenamefont {\"Ohberg},
  \citenamefont {Thomson},\ and\ \citenamefont {Goldman}}]{Mukherjee2018}%
  \BibitemOpen
  \bibfield  {author} {\bibinfo {author} {\bibfnamefont {S.}~\bibnamefont
  {Mukherjee}}, \bibinfo {author} {\bibfnamefont {M.}~\bibnamefont
  {Di~Liberto}}, \bibinfo {author} {\bibfnamefont {P.}~\bibnamefont
  {\"Ohberg}}, \bibinfo {author} {\bibfnamefont {R.~R.}\ \bibnamefont
  {Thomson}},\ and\ \bibinfo {author} {\bibfnamefont {N.}~\bibnamefont
  {Goldman}},\ }\bibfield  {title} {\bibinfo {title} {Experimental observation
  of aharonov-bohm cages in photonic lattices},\ }\href
  {https://doi.org/10.1103/PhysRevLett.121.075502} {\bibfield  {journal}
  {\bibinfo  {journal} {Phys. Rev. Lett.}\ }\textbf {\bibinfo {volume} {121}},\
  \bibinfo {pages} {075502} (\bibinfo {year} {2018})}\BibitemShut {NoStop}%
\bibitem [{\citenamefont {Zhang}\ \emph {et~al.}(2019)\citenamefont {Zhang},
  \citenamefont {Teimourpour}, \citenamefont {Arkinstall}, \citenamefont {Pan},
  \citenamefont {Miao}, \citenamefont {Schomerus}, \citenamefont {El-Ganainy},\
  and\ \citenamefont {Feng}}]{Zhang2019}%
  \BibitemOpen
  \bibfield  {author} {\bibinfo {author} {\bibfnamefont {Z.}~\bibnamefont
  {Zhang}}, \bibinfo {author} {\bibfnamefont {M.~H.}\ \bibnamefont
  {Teimourpour}}, \bibinfo {author} {\bibfnamefont {J.}~\bibnamefont
  {Arkinstall}}, \bibinfo {author} {\bibfnamefont {M.}~\bibnamefont {Pan}},
  \bibinfo {author} {\bibfnamefont {P.}~\bibnamefont {Miao}}, \bibinfo {author}
  {\bibfnamefont {H.}~\bibnamefont {Schomerus}}, \bibinfo {author}
  {\bibfnamefont {R.}~\bibnamefont {El-Ganainy}},\ and\ \bibinfo {author}
  {\bibfnamefont {L.}~\bibnamefont {Feng}},\ }\bibfield  {title} {\bibinfo
  {title} {Experimental realization of multiple topological edge states in a 1d
  photonic lattice},\ }\href
  {https://doi.org/https://doi.org/10.1002/lpor.201800202} {\bibfield
  {journal} {\bibinfo  {journal} {Laser \& Photonics Reviews}\ }\textbf
  {\bibinfo {volume} {13}},\ \bibinfo {pages} {1800202} (\bibinfo {year}
  {2019})}\BibitemShut {NoStop}%
\bibitem [{\citenamefont {Xia}\ \emph {et~al.}(2020)\citenamefont {Xia},
  \citenamefont {Danieli}, \citenamefont {Yan}, \citenamefont {Li},
  \citenamefont {Xia}, \citenamefont {Ma}, \citenamefont {Lu}, \citenamefont
  {Song}, \citenamefont {Tang}, \citenamefont {Flach},\ and\ \citenamefont
  {Chen}}]{XIa2020}%
  \BibitemOpen
  \bibfield  {author} {\bibinfo {author} {\bibfnamefont {S.}~\bibnamefont
  {Xia}}, \bibinfo {author} {\bibfnamefont {C.}~\bibnamefont {Danieli}},
  \bibinfo {author} {\bibfnamefont {W.}~\bibnamefont {Yan}}, \bibinfo {author}
  {\bibfnamefont {D.}~\bibnamefont {Li}}, \bibinfo {author} {\bibfnamefont
  {S.}~\bibnamefont {Xia}}, \bibinfo {author} {\bibfnamefont {J.}~\bibnamefont
  {Ma}}, \bibinfo {author} {\bibfnamefont {H.}~\bibnamefont {Lu}}, \bibinfo
  {author} {\bibfnamefont {D.}~\bibnamefont {Song}}, \bibinfo {author}
  {\bibfnamefont {L.}~\bibnamefont {Tang}}, \bibinfo {author} {\bibfnamefont
  {S.}~\bibnamefont {Flach}},\ and\ \bibinfo {author} {\bibfnamefont
  {Z.}~\bibnamefont {Chen}},\ }\bibfield  {title} {\bibinfo {title}
  {Observation of quincunx-shaped and dipole-like flatband states in photonic
  rhombic lattices without band-touching},\ }\href
  {https://doi.org/10.1063/1.5131501} {\bibfield  {journal} {\bibinfo
  {journal} {APL Photonics}\ }\textbf {\bibinfo {volume} {5}},\ \bibinfo
  {pages} {016107} (\bibinfo {year} {2020})}\BibitemShut {NoStop}%
\bibitem [{\citenamefont {J\"{o}rg}\ \emph {et~al.}(2020)\citenamefont
  {J\"{o}rg}, \citenamefont {Queralt\'{o}}, \citenamefont {Kremer},
  \citenamefont {Pelegr\'{\i}}, \citenamefont {Schulz}, \citenamefont
  {Szameit}, \citenamefont {von Freymann}, \citenamefont {Mompart},\ and\
  \citenamefont {Ahufinger}}]{Jorg2020}%
  \BibitemOpen
  \bibfield  {author} {\bibinfo {author} {\bibfnamefont {C.}~\bibnamefont
  {J\"{o}rg}}, \bibinfo {author} {\bibfnamefont {G.}~\bibnamefont
  {Queralt\'{o}}}, \bibinfo {author} {\bibfnamefont {M.}~\bibnamefont
  {Kremer}}, \bibinfo {author} {\bibfnamefont {G.}~\bibnamefont
  {Pelegr\'{\i}}}, \bibinfo {author} {\bibfnamefont {J.}~\bibnamefont
  {Schulz}}, \bibinfo {author} {\bibfnamefont {A.}~\bibnamefont {Szameit}},
  \bibinfo {author} {\bibfnamefont {G.}~\bibnamefont {von Freymann}}, \bibinfo
  {author} {\bibfnamefont {J.}~\bibnamefont {Mompart}},\ and\ \bibinfo {author}
  {\bibfnamefont {V.}~\bibnamefont {Ahufinger}},\ }\bibfield  {title} {\bibinfo
  {title} {Artificial gauge field switching using orbital angular momentum
  modes in optical waveguides},\ }\href
  {https://doi.org/10.1038/s41377-020-00385-6} {\bibfield  {journal} {\bibinfo
  {journal} {Light: Science \& Applications}\ }\textbf {\bibinfo {volume}
  {9}},\ \bibinfo {pages} {150} (\bibinfo {year} {2020})}\BibitemShut {NoStop}%
\bibitem [{\citenamefont {Huda}\ \emph {et~al.}(2020)\citenamefont {Huda},
  \citenamefont {Kezilebieke},\ and\ \citenamefont {Liljeroth}}]{Huda2020}%
  \BibitemOpen
  \bibfield  {author} {\bibinfo {author} {\bibfnamefont {M.~N.}\ \bibnamefont
  {Huda}}, \bibinfo {author} {\bibfnamefont {S.}~\bibnamefont {Kezilebieke}},\
  and\ \bibinfo {author} {\bibfnamefont {P.}~\bibnamefont {Liljeroth}},\
  }\bibfield  {title} {\bibinfo {title} {Designer flat bands in
  quasi-one-dimensional atomic lattices},\ }\href
  {https://doi.org/10.1103/PhysRevResearch.2.043426} {\bibfield  {journal}
  {\bibinfo  {journal} {Phys. Rev. Research}\ }\textbf {\bibinfo {volume}
  {2}},\ \bibinfo {pages} {043426} (\bibinfo {year} {2020})}\BibitemShut
  {NoStop}%
\bibitem [{\citenamefont {Taie}\ \emph {et~al.}(2015)\citenamefont {Taie},
  \citenamefont {Ozawa}, \citenamefont {Ichinose}, \citenamefont {Nishio},
  \citenamefont {Nakajima},\ and\ \citenamefont {Takahashi}}]{Tale2015}%
  \BibitemOpen
  \bibfield  {author} {\bibinfo {author} {\bibfnamefont {S.}~\bibnamefont
  {Taie}}, \bibinfo {author} {\bibfnamefont {H.}~\bibnamefont {Ozawa}},
  \bibinfo {author} {\bibfnamefont {T.}~\bibnamefont {Ichinose}}, \bibinfo
  {author} {\bibfnamefont {T.}~\bibnamefont {Nishio}}, \bibinfo {author}
  {\bibfnamefont {S.}~\bibnamefont {Nakajima}},\ and\ \bibinfo {author}
  {\bibfnamefont {Y.}~\bibnamefont {Takahashi}},\ }\bibfield  {title} {\bibinfo
  {title} {Coherent driving and freezing of bosonic matter wave in an optical
  lieb lattice},\ }\bibfield  {journal} {\bibinfo  {journal} {Science
  Advances}\ }\textbf {\bibinfo {volume} {1}},\ \href
  {https://doi.org/10.1126/sciadv.1500854} {10.1126/sciadv.1500854} (\bibinfo
  {year} {2015})\BibitemShut {NoStop}%
\bibitem [{\citenamefont {Ezawa}(2019)}]{Ezawa2019}%
  \BibitemOpen
  \bibfield  {author} {\bibinfo {author} {\bibfnamefont {M.}~\bibnamefont
  {Ezawa}},\ }\bibfield  {title} {\bibinfo {title} {Braiding of majorana-like
  corner states in electric circuits and its non-hermitian generalization},\
  }\href {https://doi.org/10.1103/PhysRevB.100.045407} {\bibfield  {journal}
  {\bibinfo  {journal} {Phys. Rev. B}\ }\textbf {\bibinfo {volume} {100}},\
  \bibinfo {pages} {045407} (\bibinfo {year} {2019})}\BibitemShut {NoStop}%
\bibitem [{\citenamefont {Ezawa}(2020{\natexlab{b}})}]{Ezawa2020b}%
  \BibitemOpen
  \bibfield  {author} {\bibinfo {author} {\bibfnamefont {M.}~\bibnamefont
  {Ezawa}},\ }\bibfield  {title} {\bibinfo {title} {Non-abelian braiding of
  majorana-like edge states and topological quantum computations in electric
  circuits},\ }\href {https://doi.org/10.1103/PhysRevB.102.075424} {\bibfield
  {journal} {\bibinfo  {journal} {Phys. Rev. B}\ }\textbf {\bibinfo {volume}
  {102}},\ \bibinfo {pages} {075424} (\bibinfo {year}
  {2020}{\natexlab{b}})}\BibitemShut {NoStop}%
\bibitem [{\citenamefont {Li}\ \emph {et~al.}(2020{\natexlab{b}})\citenamefont
  {Li}, \citenamefont {Wu}, \citenamefont {Huang}, \citenamefont {Lu},
  \citenamefont {Li}, \citenamefont {Deng},\ and\ \citenamefont
  {Liu}}]{Li2020b}%
  \BibitemOpen
  \bibfield  {author} {\bibinfo {author} {\bibfnamefont {Z.}~\bibnamefont
  {Li}}, \bibinfo {author} {\bibfnamefont {J.}~\bibnamefont {Wu}}, \bibinfo
  {author} {\bibfnamefont {X.}~\bibnamefont {Huang}}, \bibinfo {author}
  {\bibfnamefont {J.}~\bibnamefont {Lu}}, \bibinfo {author} {\bibfnamefont
  {F.}~\bibnamefont {Li}}, \bibinfo {author} {\bibfnamefont {W.}~\bibnamefont
  {Deng}},\ and\ \bibinfo {author} {\bibfnamefont {Z.}~\bibnamefont {Liu}},\
  }\bibfield  {title} {\bibinfo {title} {Bound state in the continuum in
  topological inductor-capacitor circuit},\ }\href
  {https://doi.org/10.1063/5.0011719} {\bibfield  {journal} {\bibinfo
  {journal} {Applied Physics Letters}\ }\textbf {\bibinfo {volume} {116}},\
  \bibinfo {pages} {263501} (\bibinfo {year} {2020}{\natexlab{b}})}\BibitemShut
  {NoStop}%
\bibitem [{\citenamefont {Olekhno}\ \emph {et~al.}(2020)\citenamefont
  {Olekhno}, \citenamefont {Kretov}, \citenamefont {Stepanenko}, \citenamefont
  {Ivanova}, \citenamefont {Yaroshenko}, \citenamefont {Puhtina}, \citenamefont
  {Filonov}, \citenamefont {Cappello}, \citenamefont {Matekovits},\ and\
  \citenamefont {Gorlach}}]{Olekhno2020}%
  \BibitemOpen
  \bibfield  {author} {\bibinfo {author} {\bibfnamefont {N.~A.}\ \bibnamefont
  {Olekhno}}, \bibinfo {author} {\bibfnamefont {E.~I.}\ \bibnamefont {Kretov}},
  \bibinfo {author} {\bibfnamefont {A.~A.}\ \bibnamefont {Stepanenko}},
  \bibinfo {author} {\bibfnamefont {P.~A.}\ \bibnamefont {Ivanova}}, \bibinfo
  {author} {\bibfnamefont {V.~V.}\ \bibnamefont {Yaroshenko}}, \bibinfo
  {author} {\bibfnamefont {E.~M.}\ \bibnamefont {Puhtina}}, \bibinfo {author}
  {\bibfnamefont {D.~S.}\ \bibnamefont {Filonov}}, \bibinfo {author}
  {\bibfnamefont {B.}~\bibnamefont {Cappello}}, \bibinfo {author}
  {\bibfnamefont {L.}~\bibnamefont {Matekovits}},\ and\ \bibinfo {author}
  {\bibfnamefont {M.~A.}\ \bibnamefont {Gorlach}},\ }\bibfield  {title}
  {\bibinfo {title} {Topological edge states of interacting photon pairs
  emulated in a topolectrical circuit},\ }\href
  {https://doi.org/10.1038/s41467-020-14994-7} {\bibfield  {journal} {\bibinfo
  {journal} {Nature Communications}\ }\textbf {\bibinfo {volume} {11}},\
  \bibinfo {pages} {1436} (\bibinfo {year} {2020})}\BibitemShut {NoStop}%
\bibitem [{\citenamefont {Wu}\ \emph {et~al.}(2020)\citenamefont {Wu},
  \citenamefont {Huang}, \citenamefont {Lu}, \citenamefont {Wu}, \citenamefont
  {Deng}, \citenamefont {Li},\ and\ \citenamefont {Liu}}]{Wu2020}%
  \BibitemOpen
  \bibfield  {author} {\bibinfo {author} {\bibfnamefont {J.}~\bibnamefont
  {Wu}}, \bibinfo {author} {\bibfnamefont {X.}~\bibnamefont {Huang}}, \bibinfo
  {author} {\bibfnamefont {J.}~\bibnamefont {Lu}}, \bibinfo {author}
  {\bibfnamefont {Y.}~\bibnamefont {Wu}}, \bibinfo {author} {\bibfnamefont
  {W.}~\bibnamefont {Deng}}, \bibinfo {author} {\bibfnamefont {F.}~\bibnamefont
  {Li}},\ and\ \bibinfo {author} {\bibfnamefont {Z.}~\bibnamefont {Liu}},\
  }\bibfield  {title} {\bibinfo {title} {Observation of corner states in
  second-order topological electric circuits},\ }\href
  {https://doi.org/10.1103/PhysRevB.102.104109} {\bibfield  {journal} {\bibinfo
   {journal} {Phys. Rev. B}\ }\textbf {\bibinfo {volume} {102}},\ \bibinfo
  {pages} {104109} (\bibinfo {year} {2020})}\BibitemShut {NoStop}%
\bibitem [{\citenamefont {Yang}\ \emph {et~al.}(2021)\citenamefont {Yang},
  \citenamefont {Zhu}, \citenamefont {Hang},\ and\ \citenamefont
  {Chong}}]{Yang2021}%
  \BibitemOpen
  \bibfield  {author} {\bibinfo {author} {\bibfnamefont {Y.}~\bibnamefont
  {Yang}}, \bibinfo {author} {\bibfnamefont {D.}~\bibnamefont {Zhu}}, \bibinfo
  {author} {\bibfnamefont {Z.}~\bibnamefont {Hang}},\ and\ \bibinfo {author}
  {\bibfnamefont {Y.}~\bibnamefont {Chong}},\ }\bibfield  {title} {\bibinfo
  {title} {Observation of antichiral edge states in a circuit lattice},\ }\href
  {https://doi.org/10.1007/s11433-021-1675-0} {\bibfield  {journal} {\bibinfo
  {journal} {Science China Physics, Mechanics \& Astronomy}\ }\textbf {\bibinfo
  {volume} {64}},\ \bibinfo {pages} {257011} (\bibinfo {year}
  {2021})}\BibitemShut {NoStop}%
\bibitem [{\citenamefont {Qi}(2013)}]{Qi2013}%
  \BibitemOpen
  \bibfield  {author} {\bibinfo {author} {\bibfnamefont {X.-L.}\ \bibnamefont
  {Qi}},\ }\href@noop {} {\bibinfo {title} {Exact holographic mapping and
  emergent space-time geometry}} (\bibinfo {year} {2013}),\ \Eprint
  {https://arxiv.org/abs/1309.6282} {arXiv:1309.6282 [hep-th]} \BibitemShut
  {NoStop}%
\bibitem [{\citenamefont {Lee}\ and\ \citenamefont {Qi}(2016)}]{Lee2016}%
  \BibitemOpen
  \bibfield  {author} {\bibinfo {author} {\bibfnamefont {C.~H.}\ \bibnamefont
  {Lee}}\ and\ \bibinfo {author} {\bibfnamefont {X.-L.}\ \bibnamefont {Qi}},\
  }\bibfield  {title} {\bibinfo {title} {Exact holographic mapping in free
  fermion systems},\ }\href {https://doi.org/10.1103/PhysRevB.93.035112}
  {\bibfield  {journal} {\bibinfo  {journal} {Phys. Rev. B}\ }\textbf {\bibinfo
  {volume} {93}},\ \bibinfo {pages} {035112} (\bibinfo {year}
  {2016})}\BibitemShut {NoStop}%
\bibitem [{\citenamefont {Hu}\ \emph {et~al.}(2020)\citenamefont {Hu},
  \citenamefont {Li}, \citenamefont {Wang},\ and\ \citenamefont
  {You}}]{Hu2020}%
  \BibitemOpen
  \bibfield  {author} {\bibinfo {author} {\bibfnamefont {H.-Y.}\ \bibnamefont
  {Hu}}, \bibinfo {author} {\bibfnamefont {S.-H.}\ \bibnamefont {Li}}, \bibinfo
  {author} {\bibfnamefont {L.}~\bibnamefont {Wang}},\ and\ \bibinfo {author}
  {\bibfnamefont {Y.-Z.}\ \bibnamefont {You}},\ }\bibfield  {title} {\bibinfo
  {title} {Machine learning holographic mapping by neural network
  renormalization group},\ }\href
  {https://doi.org/10.1103/PhysRevResearch.2.023369} {\bibfield  {journal}
  {\bibinfo  {journal} {Phys. Rev. Research}\ }\textbf {\bibinfo {volume}
  {2}},\ \bibinfo {pages} {023369} (\bibinfo {year} {2020})}\BibitemShut
  {NoStop}%
\bibitem [{\citenamefont {Gu}\ \emph {et~al.}(2016)\citenamefont {Gu},
  \citenamefont {Lee}, \citenamefont {Wen}, \citenamefont {Cho}, \citenamefont
  {Ryu},\ and\ \citenamefont {Qi}}]{Gu2016}%
  \BibitemOpen
  \bibfield  {author} {\bibinfo {author} {\bibfnamefont {Y.}~\bibnamefont
  {Gu}}, \bibinfo {author} {\bibfnamefont {C.~H.}\ \bibnamefont {Lee}},
  \bibinfo {author} {\bibfnamefont {X.}~\bibnamefont {Wen}}, \bibinfo {author}
  {\bibfnamefont {G.~Y.}\ \bibnamefont {Cho}}, \bibinfo {author} {\bibfnamefont
  {S.}~\bibnamefont {Ryu}},\ and\ \bibinfo {author} {\bibfnamefont {X.-L.}\
  \bibnamefont {Qi}},\ }\bibfield  {title} {\bibinfo {title} {Holographic
  duality between $(2+1)$-dimensional quantum anomalous hall state and
  $(3+1)$-dimensional topological insulators},\ }\href
  {https://doi.org/10.1103/PhysRevB.94.125107} {\bibfield  {journal} {\bibinfo
  {journal} {Phys. Rev. B}\ }\textbf {\bibinfo {volume} {94}},\ \bibinfo
  {pages} {125107} (\bibinfo {year} {2016})}\BibitemShut {NoStop}%
\bibitem [{\citenamefont {Maffei}\ \emph {et~al.}(2018)\citenamefont {Maffei},
  \citenamefont {Dauphin}, \citenamefont {Cardano}, \citenamefont
  {Lewenstein},\ and\ \citenamefont {Massignan}}]{Maffei2018}%
  \BibitemOpen
  \bibfield  {author} {\bibinfo {author} {\bibfnamefont {M.}~\bibnamefont
  {Maffei}}, \bibinfo {author} {\bibfnamefont {A.}~\bibnamefont {Dauphin}},
  \bibinfo {author} {\bibfnamefont {F.}~\bibnamefont {Cardano}}, \bibinfo
  {author} {\bibfnamefont {M.}~\bibnamefont {Lewenstein}},\ and\ \bibinfo
  {author} {\bibfnamefont {P.}~\bibnamefont {Massignan}},\ }\bibfield  {title}
  {\bibinfo {title} {Topological characterization of chiral models through
  their long time dynamics},\ }\href {https://doi.org/10.1088/1367-2630/aa9d4c}
  {\bibfield  {journal} {\bibinfo  {journal} {New Journal of Physics}\ }\textbf
  {\bibinfo {volume} {20}},\ \bibinfo {pages} {013023} (\bibinfo {year}
  {2018})}\BibitemShut {NoStop}%
\bibitem [{\citenamefont {Eliashvili}\ \emph {et~al.}(2017)\citenamefont
  {Eliashvili}, \citenamefont {Kereselidze}, \citenamefont {Tsitsishvili},\
  and\ \citenamefont {Tsitsishvili}}]{Eliashvili2017}%
  \BibitemOpen
  \bibfield  {author} {\bibinfo {author} {\bibfnamefont {M.}~\bibnamefont
  {Eliashvili}}, \bibinfo {author} {\bibfnamefont {D.}~\bibnamefont
  {Kereselidze}}, \bibinfo {author} {\bibfnamefont {G.}~\bibnamefont
  {Tsitsishvili}},\ and\ \bibinfo {author} {\bibfnamefont {M.}~\bibnamefont
  {Tsitsishvili}},\ }\bibfield  {title} {\bibinfo {title} {Edge states of a
  periodic chain with four-band energy spectrum},\ }\href
  {https://doi.org/10.7566/JPSJ.86.074712} {\bibfield  {journal} {\bibinfo
  {journal} {Journal of the Physical Society of Japan}\ }\textbf {\bibinfo
  {volume} {86}},\ \bibinfo {pages} {074712} (\bibinfo {year}
  {2017})}\BibitemShut {NoStop}%
\bibitem [{\citenamefont {Marques}\ and\ \citenamefont
  {Dias}(2020)}]{Marques2020}%
  \BibitemOpen
  \bibfield  {author} {\bibinfo {author} {\bibfnamefont {A.~M.}\ \bibnamefont
  {Marques}}\ and\ \bibinfo {author} {\bibfnamefont {R.~G.}\ \bibnamefont
  {Dias}},\ }\bibfield  {title} {\bibinfo {title} {Analytical solution of open
  crystalline linear 1d tight-binding models},\ }\href
  {https://doi.org/10.1088/1751-8121/ab6a6e} {\bibfield  {journal} {\bibinfo
  {journal} {Journal of Physics A: Mathematical and Theoretical}\ }\textbf
  {\bibinfo {volume} {53}},\ \bibinfo {pages} {075303} (\bibinfo {year}
  {2020})}\BibitemShut {NoStop}%
\end{thebibliography}%

\end{document}